\documentstyle[12pt,epsf,a4]{report}
\begin{document}
%%%%%%%%%%%%%%%%%%%%%%%%%% MACROS
\newlength{\dinwidth}
\newlength{\dinmargin}
\setlength{\dinwidth}{24.0cm}
\textheight24.0cm \textwidth16.4cm
\setlength{\dinmargin}{\dinwidth}
\addtolength{\dinmargin}{-\textwidth}
\setlength{\dinmargin}{0.5\dinmargin}
%wdb\oddsidemargin  -1.0in
%\parindent0cm
\oddsidemargin -3.8cm
\addtolength{\oddsidemargin}{\dinmargin}
\setlength{\evensidemargin}{\oddsidemargin}
\setlength{\marginparwidth}{0.9\dinmargin}
\setlength{\fboxrule}{3.0pt}
%setlength{\linethickness}{3.0pt}
\marginparsep 8pt \marginparpush 5pt
\topmargin -38pt
\headheight 12pt
\def\aii{\alpha_i^{-1}}
\def\rZ{{\rm Z}}
\def\rW{{\rm W}}
\def\rG{{\rm GUT}}
\def\rS{{\rm SUSY}}
\def\rH{{\rm Higgs}}
\def\rF{{\rm Fam}}
\def\MG{M_\rG}
\renewcommand{\floatpagefraction}{0.05}
\renewcommand{\textfraction}{0.05}
\newcommand{\mc}{Monte Carlo }
\newcommand{\mcs}{Monte Carlos }
\newcommand{\brem}{brems\-strah\-lung }
\newcommand{\bq}{\begin{equation}}
\newcommand{\eq}{\end{equation}}
\newcommand{\ba}{\begin{array}}
\newcommand{\ea}{\end{array}}
\newcommand{\bqa}{\begin{eqnarray}}
\newcommand{\eqa}{\end{eqnarray}}
\newcommand{\nn}{\nonumber \\}
\newcommand{\mpmm}{\mu^{+}\mu^{-}}
\newcommand{\tptm}{\tau^{+}\tau^{-}}
\newcommand{\sq}{^{2}}
\newcommand{\etal}{\it et al.\rm}
\newcommand{\ra}{\rightarrow}
\newcommand{\lnf}{{\ifmmode \Lambda^{(N_f)} \else $\Lambda^{(N_f)}$\fi}}
\newcommand{\ms}{{\ifmmode \overline{MS} \else $\overline{MS}$\fi}}
\newcommand{\dr}{{\ifmmode \overline{DR} \else $\overline{DR}$\fi}}
\newcommand{\lms}{{\ifmmode \Lambda^{(5)}_{\overline{MS}}
                 \else $\Lambda^{(5)}_{\overline{MS}}$\fi}}
\newcommand{\lam}{{\ifmmode \Lambda \else $\Lambda$\fi}}
\newcommand{\gev}{{\ifmmode {\rm GeV} \else ${\rm GeV}$\fi}}
\newcommand{\gevc}{{\ifmmode {\rm GeV/c^2} \else ${\rm GeV/c^2}$\fi}}
\newcommand{\tev}{{\ifmmode {\rm TeV} \else ${\rm TeV}$\fi}}
\newcommand{\tevc}{{\ifmmode {\rm TeV/c^2} \else ${\rm TeV/c^2}$\fi}}
\newcommand{\lp}{{\ifmmode L^+  \else $L^+$\fi}}
\newcommand{\lm}{{\ifmmode L^-  \else $L^-$\fi}}
\newcommand{\mlp}{{\ifmmode M(L^-)  \else $M(L^-)$\fi}}
\newcommand{\mlz}{{\ifmmode M(L^0)  \else $M(L^0)$\fi}}
\newcommand{\lz}{{\ifmmode L^0     \else $L^0$\fi}}
\newcommand{\ev}{{\ifmmode GeV/c^2       else $GeV/c^2$\fi}}
\newcommand{\tri}{{\ifmmode \triangleup  \else $\triangleup$\fi}}
\newcommand{\unl}{{\ifmmode U_{lL^0}  \else $U_{lL^0}$\fi}}
\newcommand{\gL}{{\ifmmode g_L  \else $g_{L}$\fi}}
\newcommand{\gR}{{\ifmmode g_R  \else $g_{R}$\fi}}
\newcommand{\gumu}{{\ifmmode \gamma^{\mu}  \else $\gamma^{\mu}$\fi}}
\newcommand{\gunu}{{\ifmmode \gamma^{\nu}  \else $\gamma^{\nu}$\fi}}
\newcommand{\gdmu}{{\ifmmode \gamma_{\mu}  \else $\gamma_{\mu}$\fi}}
\newcommand{\gdnu}{{\ifmmode \gamma_{\nu}  \else $\gamma_{\nu}$\fi}}
\newcommand{\stw}{{\ifmmode\sin^2\theta_W  \else $\sin^{2}\theta_{W}$
\fi}}
\newcommand{\sws}{{\ifmmode \;\sin^2\theta_W  \else $\;\sin^{2}\theta_{W}$
\fi}}
\newcommand{\cws}{{\ifmmode \;\cos^2\theta_W  \else $\;\cos^{2}\theta_{W}$
\fi}}
\newcommand{\sw}{{\ifmmode \;\sin\theta_W  \else $\sin\theta_{W}$
\fi}}
\newcommand{\cw}{{\ifmmode \;\cos\theta_W  \else $\;\cos\theta_{W}$
\fi}}
\newcommand{\tw}{{\ifmmode \;\tan\theta_W  \else $\;\tan\theta_{W}$
\fi}}
\newcommand{\qq}{{\ifmmode q\overline{q} \else $q\overline{q}$\fi}}
\newcommand{\lR}{{\ifmmode l_R  \else $l_R$\fi}}
\newcommand{\lL}{{\ifmmode l_L  \else &l_L$\fi}}
\newcommand{\nt}{{\ifmmode \nu_{\tau} \else $\nu_{\tau}$\fi}}
\newcommand{\nuR}{{\ifmmode \nu_R  \else $\nu_R$\fi}}
\newcommand{\nuL}{{\ifmmode \nu_L  \else $\nu_L$\fi}}
\newcommand{\qR}{{\ifmmode g_R  \else $q_R$\fi}}
\newcommand{\qL}{{\ifmmode q_L  \else $q_L$\fi}}
\newcommand{\qRp}{{\ifmmode q_R'  \else $q_{R}$'\fi}}
\newcommand{\qLp}{{\ifmmode q_L'  \else $q_{L}$'\fi}}
\newcommand{\est}{{\ifmmode e^{\bf \ast}  \else $e^{\bf \ast}$\fi}}
\newcommand{\lst}{{\ifmmode l^{\bf \ast}  \else $l^{\bf \ast}$\fi}}
\newcommand{\must}{{\ifmmode \mu^{\bf \ast}  \else $\mu^{\bf \ast}$\fi}}
\newcommand{\taust}{{\ifmmode \tau^{\bf \ast}  \else $\tau^{\bf \ast}$
\fi}}
\newcommand{\pperp}{{\ifmmode p_t  \else $p_t$\fi}}
\newcommand{\et}{{\ifmmode E_t  \else $E_t$\fi}}
\newcommand{\xt}{{\ifmmode x_t  \else $x_t$\fi}}
\newcommand{\smumu}{{\ifmmode \sigma_{\mu\mu}  \else $\sigma_{\mu\mu}$
\fi}}
\newcommand{\eg}{{\ifmmode e\gamma  \else $e\gamma$\fi}}
\newcommand{\epem}{{\ifmmode e^+e^-  \else $e^+e^-$\fi}}
\newcommand{\lplm}{{\ifmmode L^+L^-  \else $L^+L^-$\fi}}
\newcommand{\pp}{{\ifmmode p\overline p  \else $p\overline p$\fi}}
\newcommand{\llz}{{\ifmmode L^0\overline{L}^0 \else
$L^0\overline{L}^0$\fi}}
\newcommand{\epemt}{{\ifmmode e^+e^- \to  \else $e^+e^- \to$\fi}}
\newcommand{\eb}{{\ifmmode E_{beam}  \else $E_{beam}$\fi}}
\newcommand{\ip}{{\ifmmode pb^{-1}  \else $pb^{-1}$\fi}}
\newcommand{\upm}{{\ifmmode ^{\pm}  \else $^{\pm}$\fi}}
\newcommand{\de}{{\ifmmode ^{\circ}  \else $^{\circ}$ \fi}}
\newcommand{\appr}{{\ifmmode \sim \else $\sim$ \fi}}
\newcommand{\corresp}{{\ifmmode \stackrel{\wedge}{=}
                      \else   $\stackrel{\wedge}{=}$ \fi}}
\newcommand{\sqrts}{{\ifmmode \sqrt{s} \else $\sqrt{s}$\fi}}
\newcommand{\zz}{{\ifmmode Z^0  \else $Z^0$\fi}}
\newcommand{\mz}{{\ifmmode M_{Z}  \else $M_{Z}$\fi}}
\newcommand{\mzs}{{\ifmmode M_{Z}^2  \else $M_{Z}^2$\fi}}
\newcommand{\mw}{{\ifmmode M_{W}  \else $M_{W}$\fi}}
\newcommand{\mws}{{\ifmmode M_{W}^2  \else $M_{W}^2$\fi}}
\newcommand{\mh}{{\ifmmode M_{Higgs}  \else $M_{Higgs}$\fi}}
\newcommand{\gt}{{\ifmmode \Gamma_{tot} \else $\Gamma_{tot}$\fi}}
\newcommand{\msusy}{{\ifmmode M_{SUSY}  \else $M_{SUSY}$\fi}}
\newcommand{\msusys}{{\ifmmode M_{SUSY}^2  \else $M_{SUSY}^2$\fi}}
\newcommand{\su}{{\ifmmode SU(3)_C\otimes\- SU(2)_L\otimes\- U(1)_Y  
                    \else $SU(3)_C\otimes SU(2)_L\otimes U(1)_Y$\fi}}
\newcommand{\suthree}{{\ifmmode SU(3)_C  \else $SU(3)_C$\fi}}
\newcommand{\sutwo}{{\ifmmode  SU(2)_L\otimes U(1)_Y 
                     \else $SU(2)_L\otimes U(1)_Y$\fi}}
\newcommand{\taup} {{\ifmmode \tau_{proton} \else $\tau_{proton}$\fi}}
\newcommand{\agut}{{\ifmmode \alpha_{GUT}  \else $\alpha_{GUT}$\fi}}
\newcommand{\mgut}{{\ifmmode M_{GUT}  \else $M_{GUT}$\fi}}
\newcommand{\mguts}{{\ifmmode M_{GUT}^2  \else $M_{GUT}^2$\fi}}
\newcommand{\mze} {{\ifmmode m_0        \else $m_0$\fi}}
\newcommand{\mha}{{\ifmmode m_{1/2}    \else $m_{1/2}$\fi}}
\newcommand{\mb} {{\ifmmode m_{b}    \else $m_{b}$\fi}}
\newcommand{\mt} {{\ifmmode m_{t}    \else $m_{t}$\fi}}
\newcommand{\mts} {{\ifmmode m_{t}^2    \else $m_{t}^2$\fi}}
\newcommand{\tb} {{\ifmmode \tan\beta  \else $\tan\beta$\fi}}
\newcommand {\rb}[1]{\raisebox{1.5ex}[-1.5ex]{#1}}

\hyphenation{multi-pli-ci-ties}
\hyphenation{cor-rections}
\hyphenation{pa-ra-me-ters}
\newcommand{\mtau}{{\ifmmode m_{\tau}  \else $m_{\tau}$\fi}}
\newcommand{\dpp}{{\ifmmode \delta_{pert} \else $\delta_{pert}$\fi}}
\newcommand{\dnp}{{\ifmmode\delta_{non-pert}\else$\delta_{non-pert}$\fi}}
\newcommand{\dew}{{\ifmmode \delta_{\rm EW}\else $\delta_{\rm EW}$\fi}}
\newcommand{\rt}{{\ifmmode R_{\tau}  \else
                 $R_{\tau} $\fi}}
\newcommand{\rz}{{\ifmmode R_{Z}  \else
                 $R_{Z} $\fi}}
\newcommand{\into}{\rightarrow}
\newcommand{\SM}{Standard Model}
\newcommand{\swb}{{\ifmmode \sin^2\theta_{\overline{MS}}
                     \else $\sin^2\theta_{\overline{MS}}$\fi}}
\newcommand{\cwb}{{\ifmmode \cos^2\theta_{\overline{MS}}
                     \else $\cos^2\theta_{\overline{MS}}$\fi}}
\def\ai{\alpha_i}
\def\aii{\alpha_i^{-1}}
\def\rZ{{\rm Z}}
\def\rW{{\rm W}}
\def\rG{{\rm GUT}}
\def\rt{{\rm threshold}}
\def\rS{{\rm SUSY}}
\def\rH{{\rm Higgs}}
\def\rF{{\rm Fam}}
\def\MG{M_\rG}
\def\MS{M_\rS}
\def\MZ{M_\rZ}
\def\MW{M_\rW}
\def\Mt{M_\rt}
\def\MSbar{{\overline{MS}}}
\def\DRbar{{\overline{DR}}}                                            
\newcommand{\Z}{\mbox{$Z^{0}$}}
\newcommand{\WW}{\mbox{$W^{\pm}$}}
\newcommand{\EE}{\mbox{$e^{+}e^{-}$}}
\newcommand{\MM}{\mbox{$\mu^{+}\mu^{-}$}}
\newcommand{\TT}{\mbox{$\tau^{+}\tau^{-}$}}
\newcommand{\GSW}{\mbox{\sc GSW}}
\newcommand{\QCD}{\mbox{\sc QCD}}
\newcommand{\QED}{\mbox{\sc QED}}
\newcommand{\SLC}{\mbox{\sc SLC}}
\newcommand{\LEP}{\mbox{\sc LEP}}
\newcommand{\CERN}{\mbox{\sc CERN}}
\newcommand{\PETRA}{\mbox{\sc PETRA}}
\newcommand{\DESY}{\mbox{\sc DESY}}
\newcommand{\SLAC}{\mbox{\sc SLAC}}
\newcommand{\ALEPH}{\mbox{\sc ALEPH}}
\newcommand{\DELPHI}{\mbox{\sc DELPHI}}
\newcommand{\OPAL}{\mbox{\sc OPAL}}
\newcommand{\DEG}{\mbox{$^{\circ}$}}
\newcommand{\DELSIM}{\mbox{\tt DELSIM}}
\newcommand{\DELANA}{\mbox{\tt DELANA}}
\newcommand{\SUSY}{\mbox{\sc SUSY}}
\newcommand{\GUT}{\mbox{\sc GUT}}
\newcommand{\LL}{{\ifmmode {\cal L} \else ${\cal L}$\fi}}
\newcommand{\hz}{{\ifmmode {\rm Hz} \else ${\rm Hz}$\fi}}
\newcommand{\khz}{{\ifmmode {\rm kHz} \else ${\rm kHz}$\fi}}
\newcommand{\mhz}{{\ifmmode {\rm mHz} \else ${\rm mHz}$\fi}}
\newcommand{\as}{{\ifmmode \alpha_s  \else $\alpha_s$\fi}}
\newcommand{\asmz}{{\ifmmode \alpha_s(M_Z) \else $\alpha_s(M_Z)$\fi}}
\newcommand{\astau}{{\ifmmode \alpha_s(M_{\tau})
                       \else $\alpha_s(M_{\tau})$\fi}}
\newcommand{\ca}{{\ifmmode C_a  \else $C_a$\fi}}
\newcommand{\tf}{{\ifmmode T_{\mbox{\scriptsize Fermion}}
           \else $T_{\mbox{\scriptsize Fermion}}$\fi}}
\newcommand{\ts}{{\ifmmode T_{\mbox{\scriptsize Scalar}}
           \else $T_{\mbox{\scriptsize Scalar}}$ \fi}}
\newcommand{\mhiggs}{{\ifmmode M_{\mbox{\scriptsize Higgs}}
           \else $M_{\mbox{\scriptsize Higgs}}$\fi}}
\newcommand{\mthres}{{\ifmmode M_{\mbox{\scriptsize threshold}}
           \else $M_{\mbox{\scriptsize threshold}}$ \fi}}
\newcommand{\msbar}{{\ifmmode \overline{MS} \else $\overline{MS}$\fi}}
\newcommand{\drbar}{{\ifmmode \overline{DR} \else $\overline{DR}$\fi}}
\newcommand{\lamms}{{\ifmmode \Lambda_{\overline{MS}}
                       \else $\Lambda_{\overline{MS}}$\fi}}
\newcommand{\PL}{Phys. Lett.}
\newcommand{\PRL}{Phys. Rev. Lett.}
\newcommand{\NP}{Nucl. Phys.}
\newcommand{\rr}{{{\ifmmode {\cal R}_2 }\else ${\cal R}_2 $\fi}}
\newcommand{\rrr}{{{\ifmmode {\cal R}_3 }\else ${\cal R}_3 $\fi}}
\newcommand{\rrrr}{{{\ifmmode {\cal R}_4 }\else ${\cal R}_4 $\fi}}
\newcommand{\jdd}{{{\ifmmode {\cal D}_2 }\else ${\cal D}_2 $\fi}}
\newcommand{\jddd}{{{\ifmmode {\cal D}_3 }\else ${\cal D}_3 $\fi}}
\newcommand{\jdddd}{{{\ifmmode {\cal D}_4 }\else ${\cal D}_4 $\fi}}
\newcommand{\rrre}{{{\ifmmode {\cal R}_3^{E0}}\else ${\cal R}_3^{E0}$\fi}}
\newcommand{\rrrp}{{{\ifmmode {\cal R}_3^P}\else ${\cal R}_3^P$\fi}}
\newcommand{\jdde}{{{\ifmmode {\cal D}_2^{E0}}\else ${\cal D}_2^{E0} $\fi}}
\newcommand{\jddp}{{{\ifmmode {\cal D}_2^P}\else ${\cal D}_2^P$\fi}}
\newcommand{\ycut}{{{\ifmmode y_{cut} }\else $y_{cut}$\fi}}
\newcommand{\ymin}{{{\ifmmode y_{min} }\else $y_{min}$\fi}}
\newcommand{\sph}{{{\ifmmode {\cal S} }\else ${\cal S} $\fi}}
\newcommand{\apl}{{{\ifmmode {\cal A} }\else ${\cal A} $\fi}}
\newcommand{\thr}{{{\ifmmode {\cal T} }\else ${\cal T} $\fi}}
\newcommand{\obl}{{{\ifmmode {\cal O} }\else ${\cal O} $\fi}}
\newcommand{\cpa}{{{\ifmmode {\cal C} }\else ${\cal C} $\fi}}
\newcommand{\eec}{{{\ifmmode {\cal E}{\cal E}{\cal C} }\else
${\cal E}{\cal E}{\cal C}$\fi}}
\newcommand{\aeec}{{{\ifmmode {\cal A}{\cal E}{\cal E}{\cal C} }
\else ${\cal A}{\cal E}{\cal E}{\cal C}$\fi}}
\newcommand{\hjm}{{\ifmmode {\bf M^2_{high}}
                   \else   ${\bf M^2_{high}}$\fi}}
\newcommand{\ljm}{{\ifmmode {\bf M^2_{low}}
                   \else   ${\bf M^2_{low}}$\fi}}
\newcommand{\djm}{{\ifmmode {\bf M^2_{diff}}
                   \else   ${\bf M^2_{diff}}$\fi}}
\newcommand{\hjmt}{{\ifmmode {\bf M({\cal T})^2_{high}}
                    \else   ${\bf M({\cal T})^2_{high}}$\fi}}
\newcommand{\ljmt}{{\ifmmode {\bf M({\cal T})^2_{low}}
                    \else   ${\bf M({\cal T})^2_{low}}$\fi}}
\newcommand{\djmt}{{\ifmmode {\bf M({\cal T})^2_{diff}}
                    \else   ${\bf M({\cal T})^2_{diff}}$\fi}}
\newcommand{\djr}{{{\ifmmode {\bf {\cal D}_2}\else ${\bf {cal D}_2}$\fi}}}
\newcommand{\ma}{{{\ifmmode {\bf {\cal M}_{Major}}
\else ${\bf {\cal M}_{Major}}$\fi}}}
\newcommand{\mi}{{{\ifmmode {\bf {\cal M}_{Minor}}
\else ${\bf {\cal M}_{Minor}}$\fi}}}
\newcommand{\ps}{{\mbox{\bf PS}}}
\newcommand{\me}{\mbox{\bf ME}}
\newcommand{\ha}{{{\ifmmode {\frac{1}{2}}\else ${\frac{1}{2}}$\fi}}}

\pagestyle{empty}

\begin{flushright}
\vspace{-1.8cm}
        IEKP-KA/94-01    \\
hep-ph/9402266   \\
        March, 1994         \\
\end{flushright}

\vspace{0.7cm}

\begin{center}
{\bf \LARGE     Grand Unified Theories \\
and    Supersymmetry in \\%[0.1cm]
    Particle Physics  and  Cosmology\\}

\vspace{1.0cm}
{\bf    W.  de Boer\footnote{Email: Wim.de.Boer@cern.ch\\
Based on lectures at the Herbstschule Maria Laach, Maria Laach (1992) and 
the Heisenberg-Landau Summerschool, Dubna (1992).}\\
\baselineskip=13pt
{\it Inst.\ f\"ur Experimentelle Kernphysik, Universit\"at  Karlsruhe   \\}
\baselineskip=12pt
{\it Postfach 6980, D-76128 Karlsruhe , Germany                         \\}
}
\vspace{2.0cm}
{\bf    ABSTRACT}
\end{center}
\vspace{0.3cm}

\begin{center}\parbox{16cm}{\small
A review is given on the consistency checks of Grand Unified Theories (GUT),
 which unify the electroweak and strong nuclear forces into a single theory.
Such theories predict a new kind of force, which could provide answers
 to several open questions in cosmology.
The possible role of such a ``primeval'' force will be discussed 
in the framework of the Big Bang Theory.

Although such a force cannot be observed directly, there are several 
predictions of  GUT's, which can be verified at low energies.  
The Minimal Supersymmetric Standard Model (MSSM) distinguishes 
itself from other GUT's by a successful prediction of many 
unrelated phenomena with a minimum number of parameters.

Among them: a)   Unification of the couplings constants; 
b)  Unification of the masses; c)  Existence of dark matter; 
d) Proton decay;
e)  Electroweak symmetry breaking at a scale far below the unification scale.

A   fit  of the free parameters in the MSSM to these low energy constraints 
predicts the 
masses of the as yet unobserved superpartners of the SM particles,   constrains
  the     unknown top mass to  a range between 140 and  200 GeV, and requires
the second order QCD   coupling constant to be between 0.108 and 0.132.
               }
\end{center}
\vspace{0.5cm}
\begin{center}
(Published in
``Progress in Particle and Nuclear Physics, {\bf 33} (1994) 201''.)

\end{center}

\clearpage
\pagestyle{empty}
\begin{center}\parbox{10cm}{\small  {\it  ``The possibility that the 
universe was generated from nothing is very interesting and should be 
further studied. A most perplexing question relating to the singularity 
is this: what preceded the genesis of the universe? This question appears 
to be absolutely methaphysical, but our experience with metaphysics tells 
us that metaphysical questions are sometimes given answers by physics.''\\
 \hfill A. Linde (1982) }}
\end{center}

\pagestyle{plain}
\pagenumbering{roman}

\tableofcontents
%  \listoffigures
%  \listoftables
\clearpage
\pagenumbering{arabic}
\setcounter{page}{1}
\setcounter{chapter}{0}
\chapter{Introduction}
\label{ch1}
The questions concerning the origin of our universe have 
long been thought of as metaphysical and hence outside the realm of physics.

However, tremendous advances in experimental techniques 
to study both the very large scale structures of the 
universe with space telescopes as well as  the tiniest 
building blocks of matter -- the quarks and leptons -- 
with large accelerators, allow us ``to put things together'', 
so that the creation of our universe now has become an area 
of active research in physics.

The two corner stones in this field are:
\begin{itemize}
\item {\underline{\bf Cosmology}}, i.e. the study of 
the large scale structure and the evolution of the universe. 
Today the central questions are being explored in the 
framework of the {\it \underline{Big Bang Theory}} 
 (BBT)\cite{borner,kolb,linde,cosm,cosmI},
 which provides a
satisfactory explanation for the  three basic observations about our universe:
 the Hubble expansion, the 2.7 K microwave background 
radiation, and the density of elements (74\% hydrogen, 
24\% helium and the rest for the heavy elements).
\item{\underline{\bf
Elementary Particle Physics}}, i.e. the study of 
the building blocks of matter and the interactions 
between them. As far as we know, the building blocks 
of matter are pointlike particles, the quarks and 
leptons, which can be  grouped according to certain 
symmetry principles; their interactions have been 
codified in the so-called {\it \underline{Standard Model}} 
(SM)\cite{sm}. In this model all forces are described by 
{\it gauge field theories}\cite{book}, which form a 
marvelous synthesis of {\it Symmetry
Principles} and {\it Quantum Field Theories}. 
The latter combine the classical field theories 
with the principles of Quantum Mechanics and 
Einstein's Theory of Relativity.
\end{itemize}
\begin{figure}
%\begin{center}\vspace{-1.cm}
\vspace{-1.cm}
\epsffile[50 90 500 390]{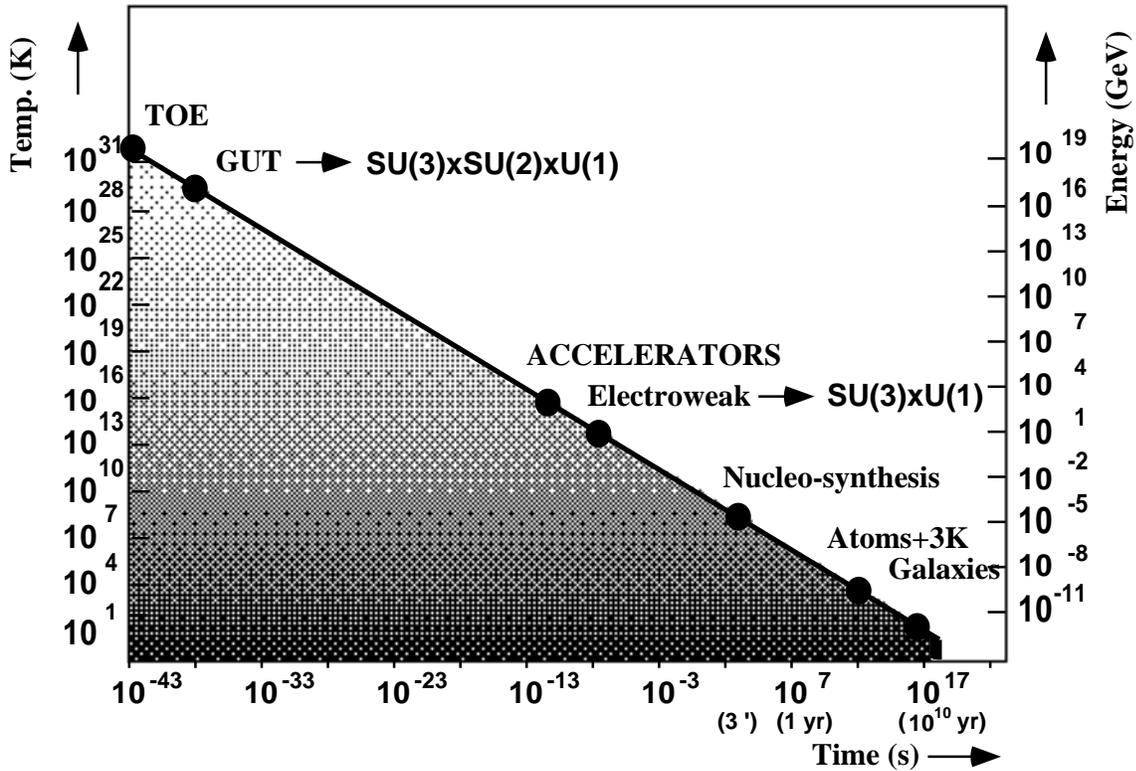}
\caption{The evolution of the universe
and the energy scale of some typical events: 
above the Planck scale of $10^{19}$
~ GeV gravity becomes so strong, that
one cannot neglect gravity implying the 
need for  a ``Theory Of Everything''
to describe all forces. Below that energy
 the well known strong  and electroweak
forces are assumed to be equally strong, implying
the possibility of a Grand Unified Theory (GUT) with
only a single coupling constant at the unification scale. 
After spontaneous symmetry breaking the gauge bosons of this
unified force become heavy and "freeze out". The remaining  forces
 correspond  to the well known $\su$~
symmetry at lower energies with their coupling constants
changing from the unified value at the GUT scale to the
low energy values; this evolution is attributed to calculable
radiative corrections. Future accelerators 
are expected to reach about 15 TeV	 
corresponding to a temperature of $10^{15}$
 K, which was reached about $10^{-12}$ s after the ``Bang''.
At about $10^2$~ GeV the gauge bosons
of the electroweak theory ``freeze out''
after getting mass through spontaneous
symmetry breaking and only the strong
and electromagnetic force play a role.
About three minutes later the temperature has 
dropped below the nuclear binding energy  and the 
strong force
 binds the quarks into nuclei (nucleosynthesis). 
Most of the particles annihilate with their antiparticles 
into a large number of photons after the photon energies become  
 too low to create new particles again. 
After about hundred thousand years the temperature 
is below the electromagnetic binding energies of atoms, 
so the   few remaining electrons and protons, 
which did not annihilate, form  the neutral atoms.
Then the universe becomes
transparent for  electromagnetic radiation
and the many photons stream away into the universe.  
The photons released at that time are now
observed as the 3 K microwave background
radiation. Then the neutral atoms start to cluster slowly 
into stars and galaxies 
under the influence of gravity.}
%\end{center}
\end{figure}

The basic observations, both in the {\it Microcosm} as well 
as in the {\it Macrocosm}, are well described by both models. 
Nevertheless, many questions remain unanswered. Among them:
\begin{itemize}
\item
What is the origin of mass?
\item What is the origin of matter?
\item What is the origin of the Matter-Antimatter Asymmetry in our universe?
\item Why is our universe so smooth and isotropic on a large scale?
\item Why is the ratio of photons to baryons in the universe 
so extremely large, on  the order of $10^{10}$?
\item What is the origin of  dark matter, which seems to
provide the majority of mass in our universe?
\item Why are the strong forces so strong and the electroweak forces so weak?
\end{itemize}
 Grand Unified Theories (GUT)\cite{gutbook,gutrev}, 
in which the known electromagnetic, weak, and strong 
nuclear forces are combined into a single theory,  
hold the promise of answering at least partially the 
questions raised above.  For example, they explain
 the different strengths of the known
 forces by  radiative
 corrections.  At high energies all  forces are equally
strong. The Spontaneous Symmetry Breaking (SSB) of a single 
unified force into  the electroweak  and strong forces 
occurs in such theories through
scalar fields, which  ``lock'' their phases
over macroscopic distances below the transition
temperature. A classical analogy is the build-up
of the magnetization in a ferromagnet below the
Curie-temperature: above the transition temperature
the phases of the magnetic dipoles are randomly
distributed and the magnetization is zero, but below
the transition temperature the phases are locked and the 
groundstate develops a nonzero magnetization.
 Translated in the jargon of particle
 physicists: the groundstate is called the vacuum and
the scalar fields develop  a nonzero ``vacuum expectation value''.  
Such a phase transition
might have   released   an enormous amount of energy, which would 
cause a rapid expansion (``inflation'') of the universe, thus   
explaining simultaneously the origin of matter,   its isotropic 
distribution and  the  flatness of our universe.

Given the importance of the questions at stake, GUT's 
have been under intense investigation during the last years.

The two directly testable predictions of the simplest GUT, namely
\begin{itemize}
\item
the finite lifetime of the proton
\item
and the unification of the three coupling constants of 
the electroweak and strong forces at high energies
\end{itemize}
 turned out to be a disaster for GUT's.
The proton was found to be much more stable
than predicted 
 and from the precisely measured coupling constants  at 
the new electron-positron collider LEP at the European 
Laboratory for Elementary Particle Physics -CERN- in Geneva one had
to conclude that the couplings did not unify, 
if extrapolated to high energies\cite{ekn,abf,lanluo}.

 However, it was shown later, that by introducing a
hitherto unobserved symmetry, called {\it \underline{Supersymmetry}} 
(SUSY)\cite{susybook,susyrev},
 into the Standard Model, both problems
disappeared:
unification  was obtained and the prediction of the proton life 
time could be pushed  above the present experimental lower limit!

The price to be paid for the introduction of SUSY is a doubling 
of the number of elementary particles, since it presupposes a 
symmetry between fermions and bosons, i.e. each  particle with 
even (odd)	 spin has a  partner with odd (even) spin. 
These supersymmetric partners have not been observed in nature, 
so the only way to save Supersymmetry is to assume that the 
predicted particles are too heavy to be produced by present 
accelerators. However, there are strong theoretical grounds to 
believe that they can
not be extremely heavy and in the minimal SUSY  model, the lightest	
	so-called Higgs particle will be relatively light, which implies
that it might even be detectable by upgrading the present LEP accelerator.
But SUSY particles, if they exist,  should    be observable	 
in the next generation of accelerators, since mass estimates  from 
the unification of the precisely measured coupling constants are
in the TeV region\cite{abf} and the lightest Higgs particle 
is expected to be of the order of $\mz$, as will be discussed in the last chapter.

It is the purpose of the present paper to discuss the experimental
tests of GUT's. The following experimental
constraints have been considered:
\begin{itemize}
\item Unification of the gauge coupling constants;
\item Unification of the Yukawa couplings;
\item Limits on proton decay;
\item Electroweak breaking scale;
\item Radiative $b\rightarrow s\gamma$ decays;
\item Relic abundance of dark matter.
\end{itemize}
%A flood of papers on these subjects have 
%emerged in the last years. Some recent 
%contributions of the groups involved are 
%given in refs. 
% \cite{susyrev,ekn,abf,lanluo,abfI,eknII,
%rrb,car,acpz,aczpt,nan,bek,ir 
%,arn,roskane,}
It is surprising that one can find  solutions within the
{\it minimal} SUSY model, which can describe
 all these
independent results simultaneously.
The constraints on the couplings, 
the unknown top-quark mass and the masses 
of the predicted SUSY particles will be discussed in detail.

The paper has been organized as follows: 
In chapters	\ref{ch2} to \ref{ch4}   the Standard Model,  
 Grand Unified Theories
(GUT)  and Supersymmetry are introduced. 
In chapter \ref{ch5}  the problems in cosmology will 
be discussed and why cosmology ``cries'' for 
Supersymmetry. Finally, in chapter \ref{ch6}  the consistency 
checks of GUT's through comparison with data are performed and in
 chapter \ref{ch7}    the results  are summarized.

\chapter{The Standard Model.}
\label{ch2} 
\section{Introduction.}
 The field of elementary particles has developed very rapidly
during the last two decades, after the success of QED as a gauge field
theory of the electromagnetic force could be extended to the
weak--  and strong forces.
The success largely started with the November Revolution in 1974,
when the charmed quark was discovered simultaneously
at SLAC and Brookhaven, for which B. Richter and S.S.C Ting
were awarded the Nobel prize in 1976. This discovery
 left little doubt that  the pointlike
 constituents inside the proton
and other hadrons are real, existing quarks and not some
mathematical objects to classify the hadrons, as they
were originally proposed by Gellman and independently by
Zweig\footnote{Zweig called the constituents ``aces'' and believed
they really existed inside the hadrons. This belief was
not shared by the referee of Physical Review, so his paper
was rejected and circulated only as a CERN preprint, albeit
well-known \cite{quark2}.}.

The existence of the charmed quark paved the way for
a symmetry between quarks and leptons, since with charm
one now had  four quarks ($u,d,c$ and $s$) and four leptons ($e,
 \mu, \nu_e$ and $\nu_\mu$), which fitted nicely into
the $SU(2)\otimes U(1) $ unified theory of the electroweak
interactions  proposed by Glashow, Salam and Weinberg (GSW) \cite{sm} for
the leptonic sector and extended to include quarks as well as
leptons by Glashow, Iliopoulis and Maiani (GIM) \cite{glas} as early as 1970.
Actually, from the absence of flavour changing neutral
currents, they predicted the charm quark with a mass
around 1-3 GeV and indeed the charmed quark was found
four years later  with a mass of about 1.5 GeV.  This
discovery became known as  the November
Revolution, mentioned above.

The unification of the electromagnetic  and weak interactions had
already been  forwarded by Schwinger  and Glashow in the sixties.
 Weinberg and Salam solved the problem
of the heavy gauge boson masses, required in order to explain the short 
range of the weak interactions, by introducing spontaneous symmetry breaking via
the Higgs-mechanism. This introduced gauge boson masses without
explicitly breaking the gauge symmetry.

The Glashow-Weinberg-Salam theory led to three important predictions:
\begin{itemize}
\item neutral currents, i.e. weak interactions without
changing the electric charge. In contrast to the charged currents
the neutral currents  could occur with leptons from different
generations in the initial state, e.g.
 $\nu_\mu e\rightarrow \nu_\mu e$ through the exchange of a new neutral 
gauge boson.
\item the prediction of the heavy gauge boson masses around 90 GeV.
\item a scalar neutral particle, the Higgs boson.
\end{itemize}

The first prediction  was confirmed in 1973 by the observation of
$\nu_\mu$ scattering without muon in the final state in the Gargamelle 
bubble chamber at CERN.
Furthermore, the predicted parity violation for the neutral currents
was observed in polarized electron-deuteron scattering and
in optical effects in atoms.
These successful experimental 
verifications\cite{book} led to the
award of the Nobel prize in 1979 to Glashow, Salam and Weinberg.
In 1983 the second prediction was confirmed by the
discovery of  the $W$ and $Z$ bosons  at CERN in $p\bar{p}$
collisions, for which C. Rubbia and S. van der Meer were
awarded the Nobel prize in 1985.

The last prediction has not been confirmed:
the Higgs boson is still at large  despite intensive searches.
It might  just be too heavy to be produced with the present accelerators.
No predictions for its mass exist within the Standard Model.
In the supersymmetric extension of the SM the mass is predicted
to be on the order of 100 GeV, which might be in reach
after an upgrading of LEP to 210 GeV. These predictions will be
discussed in detail in the last chapter, where a comparison with
available data will be made.

In between the gauge theory of the strong interactions,
as proposed by Fritzsch and Gell-Mann \cite{frit}, had established itself
firmly after the discovery of its gauge field, the gluon,
in 3-jet production in $\epem$ annihilation at the DESY
laboratory in Hamburg.
The colour charge of these gluons, which causes the
gluon self-interaction, has been established firmly
at CERN's Large Electron Positron storage ring, called  LEP.
This gluon self-interaction leads to asymptotic freedom,
as shown by Gross and Wilcek \cite{gros} and independently by Politzer
 \cite{poli},
thus explaining why the quarks can be observed as almost
free pointlike particles inside hadrons, and why they are
not observed as free particles, i.e. they are confined
inside these hadrons. This simultaneously explained
the success of the Quark Parton Model, which assumes quasi-free
 partons inside the hadrons. In this case the cross sections, if 
expressed in dimensionless
scaling variables,
are independent of energy. The
observation of scaling in deep inelastic
lepton-nucleon scattering led to the award of the
Nobel Prize to Freedman, Kendall and Taylor in 1990.
Even the observation of logarithmic
scaling violations, both in DIS and $\epem$ annihilation,
 as predicted by QCD, were observed and could be used
for precise determinations of the strong coupling constant of
 QCD\cite{asdis,assca}.

The discovery of the beauty quark at Fermilab in Batavia(USA)
in 1976 and the $\tau$-lepton at SLAC, both in 1976,
led to the discovery of the third generation of quarks and
leptons, of which the expected top quark is still missing.
Recent LEP data indicate that its mass is around 166 GeV\cite{lep},
thus explaining why it has not yet been discovered at
 the present accelerators. The third generation had been introduced
into the Standard Model long before by Kobayashi and Maskawa
in order to be able to explain the observed CP violation in
the kaon system within the Standard Model.

From the total decay width of the $\zz$ bosons, as measured at LEP,
one concludes that it couples to three different neutrinos
with a mass below $ \mz/2\approx  45$ GeV.
This strongly suggests that the number of generations of
elementary particles is not infinite, but indeed three,
since the  neutrinos are massless in the Standard Model.
The three generations have been summarized in table \ref{t21}
together with the gauge fields, which are responsible
for the low energy interactions.

The gluons are believed to be massless, since there
is no reason to assume that the $SU(3)$ symmetry is broken,
so one does not need Higgs fields associated with the
low energy strong interactions.
The apparent short range behaviour of the strong interactions
is not due to the mass of the gauge bosons, but to the
gluon self-interaction leading to confinement, as will be
discussed in more detail afterwards.

 This chapter has been organized as
follows: after a short description of the SM, we
discuss it shortcomings and unanswered questions.
They form the motivation for extending the SM
towards a  Grand Unified Theory, in which
the electroweak-- and strong forces are unified
into a new force with only a single coupling constant.
The Grand Unified Theories will be discussed in the next chapter.
Although such unification can only happen  at
extremely high energies -- far above the range
of present accelerators --  it still has
strong implications on low energy physics, which
can be tested at present accelerators.

\section{The \SM}

Constructing a gauge theory  requires the following steps to be taken:
\begin{itemize}
\item Choice of a symmetry group on the basis of the symmetry
of the observed interactions.
\item  Requirement of local gauge invariance under transformations
of the symmetry group.
\item Choice of the Higgs sector to introduce spontaneous
symmetry breaking, which allows the generation of masses
without breaking explicitly  gauge invariance. Massive gauge bosons 
are needed to obtain the short-range behaviour of the weak interactions. 
Adding ad-hoc  mass terms,  which are {\it not} gauge-invariant, leads 
to non-renormalizable field theories. In this
case the infinities of the theory cannot be absorbed in the parameters 
and fields of the theory. With the Higgs mechanism
the theory is indeed renormalizable, as was shown by G. 't Hooft\cite{hoof}.

\item
Renormalization of the couplings and masses in the theory in order
to relate the bare charges of the theory to known data.
The Renormalization Group analysis leads to the concept of ``running'', 
i.e. energy dependent
coupling constants, which allows the absorption of
infinities in the theory into the coupling constants.
\end{itemize}

\begin{table}[thb]
\begin{center}
\begin{tabular}{|c||c|c|c|c|}\cline{2-5}
\multicolumn{1}{ c }{ } &
\multicolumn{4}{|c|}{Interactions}  \\ \cline{2-5}
\multicolumn{1}{ c }{     }  &
\multicolumn{1}{|c }{strong} &
\multicolumn{1}{|c }{electro-weak} &
\multicolumn{1}{|c|}{gravitational} &
\multicolumn{1}{|c|}{unified ?}\\ \hline
\multicolumn{1}{|c||}{Theory}  &
\multicolumn{1}{|c }{\QCD} &
\multicolumn{1}{|c }{GSW} &
\multicolumn{1}{|c|}{quantum gravity ?} &
\multicolumn{1}{|c|}{SUGRA ?}\\ \hline
 Symmetry &$SU(3)$   &$SU(2)\times U(1)$ & ? & SU(5)?\\ \hline
Gauge &$g_1 \cdots g_8$&photon    &   G & X,Y  ?\\
 bosons & gluons &\WW~ ,\Z~  bosons &graviton& GUT bosons?\\ \hline
 charge&colour          &weak isospin       &mass&  ?  \\
       &                &weak hypercharge   &    &  \\   \hline
%Strength    &     0.2        &    0.03      &$<10^{-30}$&  ?  \\ \hline
\end{tabular}
\end{center}
\caption[The fundamental forces]{The fundamental forces. The question marks
indicate areas of intensive research.}
\label{t21}
\end{table}

\subsection{Choice of the Group Structure.}
Groups of particles observed in nature show very similar properties, 
thus suggesting the existence of symmetries. For example, the quarks 
come in three colours, while the weak interactions suggest the grouping 
of fermions into doublets. This leads naturally to the $SU(3)$ 
and $SU(2)$ group structure for the strong and weak interactions, respectively.
The electromagnetic interactions don't change the quantum numbers 
of the interacting particles, so the simple $U(1)$
group is sufficient.

Consequently, the  \SM\ of the strong and 
electroweak interactions 
is based on the symmetry of the following unitary\footnote{
Unitary transformations rotate vectors, but leave their length
constant.
%The determinant of unitary rotation %matrices  is $\pm$1,
SU(N) symmetry groups are Special Unitary groups
with   determinant +1.} groups:
\bq \label{egr}  \su.\eq
The need
for three colours
arose in connection with the
existence of
hadrons consisting of three quarks with identical quantum numbers.
  According to the Pauli principle fermions are
not allowed to be in the same state, so labeling them with different
colours solved the problem\cite{book}.
More direct experimental evidence for colour
 came  from the
decay width of the $\pi^0$  and
  the total hadronic cross section in $\epem$ annihilation\cite{book}. Both are
 proportional to the number of quark species and both require
the number of colours to be three.

Although colour was introduced first as an ad-hoc quantum number for the
reasons given above, it became later evident, that its role was
much more fundamental, namely that it acted as the source of the
field for the strong interactions (the ``colour'' field), just like
the electric charge is the source of the electric field.

 The ``charge'' of the
weak interactions is the third component of the
``weak'' isospin $T_3$.  The charged weak interactions  only operate
on left-handed particles, i.e. particles with the spin
aligned opposite to their momentum  (negative  helicity), so only
left-handed particles are given  weak isospin $\pm$ 1/2 and right-handed
particles  are put into singlets (see table \ref{t22}). Right-handed
neutrinos do not exist in nature, so within each generation one has
15 matter fields: 2(1) left(right)-handed leptons and
2x3 (2x3) left(right)-handed quarks (factor 3 for colour).

 The electromagnetic
interactions  originate  both from the exchange of the
neutral gauge boson of the $SU(2)$ group as well as the one from
the $U(1)$ group. Consequently the ``charge'' of the $U(1)$ group
 cannot  be identical with the electric charge, but it is the
  so-called weak hypercharge -$Y_W$-,
which is related to the electric charge via the Gell-Mann-Nishijima
relation:
\bq Q=T_3+\frac{1}{2}Y_W .\label{gmn}\eq
The quantum number $Y_W$ is $(B-L)$ for left handed doublets
and $2Q$ for righthanded singlets, where the baryon
 number $B$=1/3 for quarks
and 0 for leptons, while the lepton number $L$ =1 for leptons and
0 for quarks. Since $T_3$ and $Q$ are conserved, $Y_W$ is also a
conserved quantum number.
The electro-weak quantum numbers for the elementary particle
spectrum are summarized in table \ref{t22}.

\begin{table}[thb]
\begin{center}
\begin{tabular}{|c|c c c | c c c|}\cline{2-7}
\multicolumn{1}{ c }{ } &
\multicolumn{3}{|c|}{Generations}&
\multicolumn{3}{|c|}{Quantum Numbers}          \\ \cline{1-1}
helicity  & 1.         &      2.    &      3.    &  Q  &$T_3$ & $Y_W $\\
\hline\hline
          &            &            &            &     &      &    \\
          &
    $  \left(\begin{array}{c} \nu_e     \\  e   \end{array}\right)_L$  &
    $  \left(\begin{array}{c} \nu_{\mu} \\\mu   \end{array}\right)_L$  &
    $  \left(\begin{array}{c} \nu_{\tau}\\\tau  \end{array}\right)_L$  &
    $        \begin{array}{r}     0     \\ -1   \end{array}         $  &
    $        \begin{array}{r}     1/2   \\ -1/2 \end{array}         $  &
    $        \begin{array}{r}     -1    \\ -1   \end{array}         $  \\
  L       &            &            &            &     &      &    \\
          &
    $  \left(\begin{array}{c} u \\ d^{\prime}  \end{array}\right)_L$  &
    $  \left(\begin{array}{c} c \\ s^{\prime}  \end{array}\right)_L$  &
    $  \left(\begin{array}{c} t \\ b^{\prime}  \end{array}\right)_L$  &
    $        \begin{array}{r}    2/3      \\-1/3  \end{array}      $  &
    $        \begin{array}{r}    1/2      \\ -1/2 \end{array}      $  &
    $        \begin{array}{r}    1/3      \\  1/3 \end{array}      $  \\
          &            &            &            &     &      &    \\
\hline
          &            &            &            &     &      &    \\
          &$  e  _R$   &$ \mu  _R$  &$ \tau  _R$& -1  & 0    & -2 \\
          &            &            &            &     &      &    \\

    $        \begin{array}{c}  R  \\         \end{array}      $  &
    $        \begin{array}{c} u_R   \\ d_R    \end{array}     $  &
    $        \begin{array}{c} c_R   \\ s_R    \end{array}     $  &
    $        \begin{array}{c} t_R   \\ b_R    \end{array}     $  &
    $        \begin{array}{r}    2/3      \\ -1/3 \end{array}      $  &
    $        \begin{array}{r}     0       \\  0   \end{array}      $  &
    $        \begin{array}{r}    4/3      \\ -2/3 \end{array}      $  \\
          &            &            &            &     &      &    \\
\hline
\end{tabular}
\end{center}
\caption[The electro-weak quantum numbers]
{The electro-weak quantum numbers (electric charge $Q$, third component 
of weak isospin $T_3$ and weak hypercharge $Y_W$) of the particle spectrum.
The neutrinos $\nu_e$, $\nu_{\mu}$ and $\nu_{\tau}$ are the weak isospin 
partners of
the electron($e$), muon($\mu$)
and tau($\tau$) leptons, respectively. The up($u$),
down($d$), strange($s$), charm($c$), bottom($b$) and top($t$) quarks
come in three colours, which have not been indicated. The primes for the 
left handed quarks
$d^{\prime}$, $s^{\prime}$ and $b^{\prime}$ indicate    the  interaction
eigenstates of the electro-weak theory, which are mixtures of the  mass 
eigenstates, i.e. the real particles. The mixing matrix is the 
Cabibbo-Kobayshi-Maskawa matrix.
 The weak hypercharge $Y_W$ equals $B-L$
for the left-handed doublets and $2Q$ for the right-handed singlets.}
\label{t22}
\end{table}
\begin{figure}
\begin{center}
\vspace{-0.9cm}
\mbox{\epsfysize=8.0cm\epsfxsize=14.cm\epsfbox{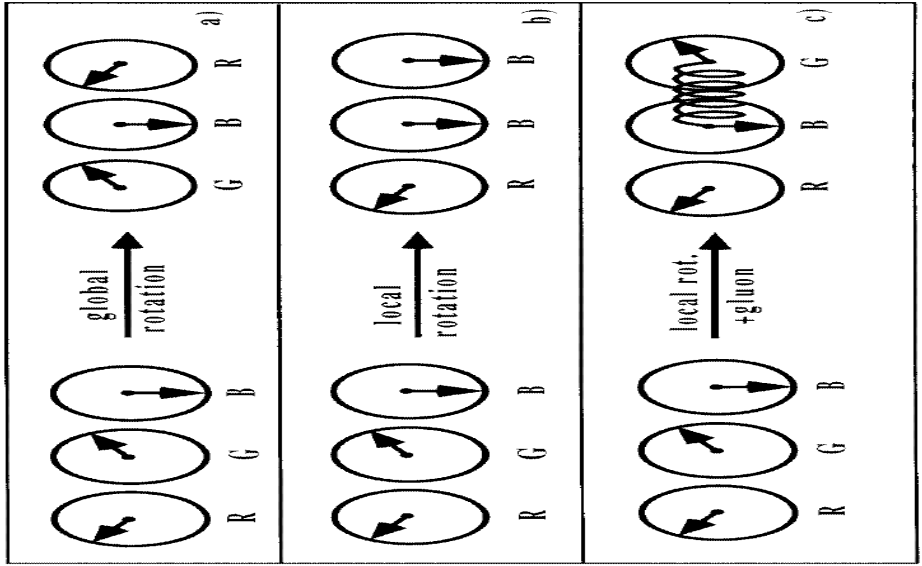}}
% H.S. E
\caption{   Global rotations leave  the
baryon colourless (a). Local rotations
change the colour locally, thus changing
  the colour of
the baryon (b), unless the colour
 is restored by the exchange of a gluon   (c).}
\label{f21}
\end{center}
\end{figure}

\begin{figure}
\begin{center}\vspace{-1.5cm}
\mbox{\epsfysize=10.cm\epsfxsize=8.cm\epsfbox{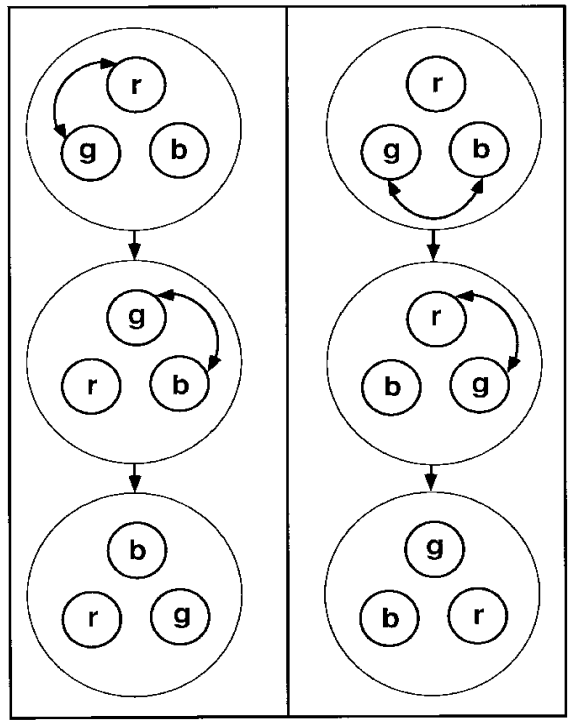}}
% H.S. E
\caption{Demonstration of the non-abelian character of the SU(3)
rotations inside a colourless baryon: on the left-hand side one first
exchanges a red-green gluon, which exchanges the colours of the quarks, 
and then a green-blue gluon; on  the right-hand
side the order is reversed. The final
result is not the same, so these operations do not commute.}
\label{f22}
\end{center}
\end{figure}

\section{Requirement of local gauge invariance.}
The Lagrangian $\LL$ of a free fermion can  be written as:
\bq \LL=i\overline{\Psi}\gamma^\mu\partial_\mu\Psi-m\overline{\Psi}\Psi, 
\label{lag} \eq
where the first term  represents the kinetic energy of the matter field
$\Psi$ with mass $m$  and the second term is the energy corresponding
to the mass $m$.  The Euler-Lagrange equations  for this $\LL$
yield the Dirac equation for a free fermion.

The unitary groups $SU(N)$ introduced above represent
rotations in $N$ dimensional\footnote{The $SU(N)$ groups
can be represented by $N\times N$ complex matrices $A$ or $2N^2$
real numbers. The unitarity requirement ($A^{\dag}=A^{-1}$) imposes $N^2$ 
conditions, while
    requiring the
 determinant to be one
 imposes one more constraint, so in total the matrix is
represented by $N^2-1$ real numbers.}
 space. The bases for the space are
provided by the eigenstates of the matter fields, which
 are the colour triplets in case of $SU(3)$, weak isospin doublets in case 
of $SU(2)$ and singlets for $U(1)$.

Arbitrary rotations of the states can be represented by
\bq U=exp(-i\vec{\alpha}\cdot\vec{F})=
exp(-i\sum_{k=1}^{N^2-1} \alpha_k \cdot F_k ) \eq
where $\alpha_k$ are the rotation parameters and $F_k$ the
rotation matrices.   $F_k$ are the eight  3x3 Gell-Mann
matrices for $SU(3)$,   denoted by $\lambda$ hereafter, and  the well 
known Pauli matrices  for  $SU(2)$  denoted by $\tau$.

The Lagrangian is invariant
 under the  $SU(N)$ rotation, if  $\LL(\Psi^\prime)$
=$\LL(\Psi)$, where $\Psi^\prime=U\Psi$.
The mass term is clearly invariant: $m\overline{\Psi}^\prime\Psi^\prime=
m\overline{\Psi} U^{\dag}  U\Psi=
m\overline{\Psi}\Psi$, since $U^{\dag} U=1$ for
unitary matrices. The kinetic term is only invariant under global
transformations, i.e. transformations where $\alpha_k$ is
everywhere the same in space-time. In this case $U$ is
independent of $x$ and can be treated as a constant multiplying $\Psi$,
which leads to:
$\overline{\Psi}U^{\dag} \gamma^\mu\partial_\mu U\Psi=
\overline{\Psi}U^{\dag} U\gamma^\mu\partial_\mu \Psi=
\overline{\Psi}\gamma^\mu\partial_\mu \Psi$.

However, one could also require $\it local$
instead of  global gauge invariance, implying that the interactions 
should be invariant under rotations of the symmetry group for each 
particle  separately. 
 The motivation is
simply  that the interactions
should be the same for particles belonging to the same multiplet of a 
symmetry group. For example,
the interaction between a green and a blue quark should
be the same as the interaction between a green and a red
quark; therefore it should be allowed to perform
a local colour transformation of a single quark.
The consequence of requiring local gauge invariance is
dramatic: it requires the introduction of intermediate
gauge bosons whose quantum numbers
 completely determine the possible interactions
between the matter fields, as was first shown by Yang and Mills in 1957
for the isopin symmetry of the strong interactions.

 Intuitively this is quite clear. Consider a hadron consisting
of a colour triplet of quarks in a colourless
groundstate. A global rotation of all
quark fields will leave the groundstate invariant, as shown
schematically in fig. \ref{f21}. However, if a quark field
is rotated locally, the groundstate is not colourless anymore,
unless  a ``message'' is mediated to the other quarks to
change their colours as well. The ``mediators'' in $SU(3)$
are the gluons, which carry a colour charge themselves and
the local colour variation of the quark field is restored by
the gluons.

The colour charge of the gluons is a consequence of
the non-abelian character of SU(3), which implies that rotations in colour space
do not commute, i.e. $\lambda_a \lambda_b \neq \lambda_b \lambda_a$,
as demonstrated in fig. \ref{f22}. If the gluons would all be colourless,
they would not change the colour of the quarks and their exchange
would be commuting.

Mathematically, local gauge invariance is introduced  by replacing
the derivative $\partial_\mu$ with the covariant\footnote{The
term originates from Weyl, who tried to introduce local gauge invariance
for gravity, thus introducing the derivative in curved space-time, which
varies  with the curvature, thus being covariant.}
 derivative $D_\mu$,
which is required to have the following property:
\bq D^\prime\Psi^\prime=UD\Psi \label{cov}, \eq
i.e. the covariant derivative of the field
has the {\it same} transformation
properties as the field in contrast to the normal derivative.
Clearly with this requirement $\LL$ is manifestly gauge invariant,
since in each term of eq. \ref{lag} the transformation leads to
the product $U^{\dag} U=1$ after substituting  $\partial_\mu\rightarrow D_\mu.$

%If the rotation parameters are chosen to be
%$x$-dependent and proportional to the %coupling constants, i.e.
%\bq \alpha=g\vec{\beta}(x) ,\eq
% one can write the rotations  as:
%\bq %U~=~exp\left[\frac{ig^\prime}{2}\beta_1(x)
%Y_W\right]
%~exp\left[\frac{ig}{2}\vec{\beta}_3(x)\cdot %\vec{\tau}\right]
%~exp\left[\frac{ig_s}{2}\vec{\beta}_8(x)
%\cdot \vec{\lambda}\right] \label{u} ,\eq
%where $\beta_1$ is the rotation angle of
% the $U(1)$
%group in the weak hypercharge space,
% $\vec{\beta}_3$ are the three rotation %angles
%of the $SU(2)$ group in weak  isospin space %and $\vec{\beta}_8$ the 
%eight rotation 
%angles of the $SU(3)$ group in colour space
%and $g^\prime ,~ g$ and $g_s$ are the %coupling constants of
%the corresponding groups.
%The weak hypercharge space is %one-dimensional, so
%$\beta_1$ is just  a  number.
%transitions between mass eigenstates with different quantum
%numbers. For each group there exists a conserved
%quantum number:  the weak hypercharge for the $U(1)$ group,
%the weak isotopic spin of the $SU(2)$ group and the
%colour charge of the $SU(3)$ group. The constants
%$g^\prime, g$ and $g_s$ define the effective\footnote{
%As we will see, the charges are energy dependent due
%to vacuum polarization effects.} strength of these charges.

For infinitesimal transformations
% the exponentials in eq. \ref{u} can be %expanded, in which case
the covariant derivative can be written as\cite{book}:
\bq D_\mu=\partial_\mu+\frac{ig^\prime}{2} B_\mu {Y_W}
+  \frac{ig}{2} {\bf \vec{W}_\mu}\cdot{\bf\vec{\tau}}
+ \frac{ig_s}{2} {\bf \vec{G}_\mu}\cdot{\bf\vec{\lambda}}
, \label{cov1} \eq
where $ B_\mu,\bf \vec{W}_\mu $ and $\bf \vec{G}_\mu $
are the field quanta (``mediators'') of the
$U(1)$, $SU(2)$ and $SU(3)$ groups 
and $g^\prime , ~g$ and $g_s$   the corresponding  coupling constants.
% which behave under the infinitesimal
%transformations $U$ as follows:
%\bq %B^\prime_\mu=B_\mu+\frac{1}{g^\prime}
%\partial_\mu\beta_1(x) \eq
%\bq \vec{W^\prime_\mu}=\vec{W_\mu}+
%\frac{1}{g}\partial_\mu\vec{\beta}_3(x)-\v
%ec{W_\mu}\times\vec{\beta}_3(x)\label{wpmu%}
%\eq
%\bq \vec{G}^\prime_\mu=\vec{G}_\mu+
%\frac{1}{g_s}\partial_\mu\vec{\beta}_8(x)-
%\vec{G}_\mu\times\vec{\beta}_8(x)\eq

%\begin{figure}[tb]
%   HNS  A
%\hspace{4.cm}
%\epsfysize=8.cm
%\epsfxsize=7.cm
%\epsffile[70 62 385 410]{f23.eps}
%   HNS  E
%\caption{Feynman  diagrams of couplings 
%between gauge bosons
%and fermions for
%  the first generation.}
%\label{f23}
%\end{figure}
The term ${\bf \vec{W}_\mu}\cdot{\bf\vec{\tau}}$ can be explicitly written as:
\begin{displaymath} W^1_\mu\tau_1+W^2_\mu\tau_2+W^3_\mu\tau_3=
W^1_\mu\left(\ba{cc}0&1\\ 1&0\ea\right)
+W^2_\mu\left(\ba{cc}0&-i\\ i&0\ea\right)+
W^3_\mu\left(\ba{cc}1&0\\ 0&-1\ea\right)=   \end{displaymath} \bq
\left(\ba{cc}W_\mu^3&W_\mu^1-iW_\mu^2 \\
W_\mu^1+iW_\mu^2&-W_\mu^3\ea\right)    \equiv
\left(\ba{cc}W_\mu^3&\sqrt{2}W^+  \\
\sqrt{2}W^-&-W_\mu^3\ea\right)
\label{wtau}\eq
The operators $W^\pm$ in the off-diagonal elements act as
lowering- and raising operators for the weak isospin.
For example, they transform  an electron into a neutrino
and vice-versa, while the operator $W_\mu^3$ represents the
neutral current interactions between a fermion and antifermion.
% This can be seen, if we take the field %$\Psi$ to be the
%left-handed lepton doublet $(\nu_e,e)$. %The term
% $\overline{\Psi}\;W_\mu\cdot\tau\;\Psi$ %yields
%four terms:
%\bq %\LL_{int}=-\frac{g}{2}
%\left(\overline{\nu_e}\gamma^\mu
%W_\mu^3\nu_e
%-\overline{e}\gamma^\mu W_\mu^3e\right)
%-\frac{g}{\sqrt{2}}
%\left(\overline{\nu}\gamma^\mu W_\mu^+e
%+\overline{e}\gamma^\mu W_\mu^- \nu\right)
%\label{int}\eq
%These interactions can be represented by %the  Feynman
%graphs shown in fig. \ref{f23}.
%However,
%the interactions between electrons cannot %be considered
%to represent the electromagnetic %interactions, since the
%same operator acts between the neutrinos, %which don't possess
%an electric charge.
% The $W_\mu$ gauge %fields cannot
%represent the mediators of the weak %interactions, since
%the latter have to be massive.  Mass terms %for $W_\mu$, such as
%$M^2W_\mu W^\mu$, are not gauge invariant, %as can be checked
%  from
% the transformation laws for the fields. %(eq. \ref{wpmu}).
%The real fields $\gamma, Z^0,$ and $W^\pm$
%can be obtained from the gauge fields
%after spontaneous symmetry breaking, as %will be discussed in the
%next section.

After substituting the Gell-Mann matrices $\lambda$
 the term ${\bf G_\mu\cdot \lambda}=
{\textstyle \sum_{k=1}^8 }G_k\lambda_k$
can be written similarly as:\\
\bq \left(\ba{ccc}G_\mu^3+\frac{1}{\sqrt{3}}G^8
&G^1-iG^2&G^4-iG^5 \nn\\
G_\mu^1+iG^2&-G^3+\frac{1}{\sqrt{3}}G^8&G^6-iG^7\nn\\
G_\mu^4+iG^5&G^6+iG^7&-\frac{2}{\sqrt{3}}G^8
\ea\right)\equiv \left(\ba{ccc}G_\mu^3+\frac{1}{\sqrt{3}}G^8
&\sqrt{2}G_{rg}& \sqrt{2}G_{rb}\nn\\
\sqrt{2}G_{gr}&-G^3+\frac{1}{\sqrt{3}}G^8&\sqrt{2}G_{gb}\nn\\
\sqrt{2}G_{br}&\sqrt{2}G_{bg}&-\frac{2}{\sqrt{3}}G^8
\ea\right)
\label{glam}\eq

%Similarly, if $\Psi$ represents a colour triplet $\Psi_r=(1,0,0)$
%$\Psi_g=(0,1,0)$ and $\Psi_b=(0,0,1)$
This term induces transitions between the colours.
For example, the off-diagonal element $G_{rg}$ acts like a raising
operator between a green $\Phi_g=(0,1,0)$ and red $\Phi_r=(1,0,0)$ field.
The terms on the diagonal don't change the colour. Since the trace of
the matrix has to be zero, there are only two independent
gluons, which don't change the colour. They are linear combinations
of the diagonal matrices $\lambda_3$ and $\lambda_8$.

%The left- and right-handed fields
% transform differently under $SU(2)\otimes %U(1)$ rotations:
%\bq {\bf %L^\prime}=exp(\frac{ig^\prime}{2}
%\beta_1(x)Y_L)~
%~exp(\frac{ig}{2}\vec{\beta_3}(x)\cdot %\vec{\tau})~
%~{\bf L} ,\eq
%\bq {\bf %e_R^\prime}=exp(\frac{ig^\prime}{2}
%\beta(x)Y_R)~
%~{\bf e_R}
%\eq
%Consequently, fermion mass terms are not %invariant under $SU(2)$
%rotations, as can be seen by writing the %mass terms
%explicitly as:
%\bq -m\overline{\bf \Psi}{\bf %\Psi}=-m\overline{\bf \Psi}
%(P_L^2+P_R^2){\bf \Psi}=
%-m\overline{\bf \Psi}_{\bf L}{\bf \Psi_R}
%-m\overline{\bf \Psi}_{\bf R}{\bf %\Psi_L}\eq
%where
%$P_{R,L}=(1\pm \gamma_5)/2$ are the usual
%projection operators, which select
%the left or right handed components of the %fields and  satisfy
%the following relations:
%\bq
%1=P_L+P_R=P_L^2+P_R^2 ~~~{\rm and} %~~~\overline{\bf \Psi}
%P_R =\overline{\bf \Psi}_{\bf L}. \eq

%Thus Dirac  mass terms can always be %written as combinations
%of left- and right-handed components and %since these components
%transform differently under $SU(2)$ %rotations, they
%break the $SU(2)$ symmetry. In the next %section we will
%see how effective fermion mass terms can %be introduced
%by the interaction with a scalar field %without
%breaking  the gauge invariance  %explicitly.
  The $W_\mu$ gauge  fields cannot
 represent the mediators of the weak  interactions, since
 the latter have to be massive.  Mass terms  for $W_\mu$, such as
  $M^2W_\mu W^\mu$, are not gauge invariant,  as can be checked
   from
  the transformation laws for the fields. %(eq. \ref{wpsemu}).
 The real fields $\gamma, Z^0,$ and $W^\pm$
 can be obtained from the gauge fields
 after spontaneous symmetry breaking via the 
Higgs mechanism, as  will be discussed in the
 next section.

\section{The Higgs mechanism.}
\subsection{Introduction.}
The problem of mass for the fermions and weak gauge bosons
can be solved by assuming that  masses are generated
dynamically through the interaction with a scalar field, which
is assumed to be present everywhere  in  the vacuum,
i.e. the space-time in which interactions take place.

 The vacuum or equivalently the groundstate,
i.e. the state with the lowest potential energy,
 may have a non-zero  (scalar) field value  represented by $\Phi=v~exp(i\phi)$; 
$v$ is called the vacuum expectation value (vev).  
The same minimum is reached
for an arbitrary value of the phase $\phi$, 
so there exists an infinity
of different, but equivalent groundstates.
 This degeneracy of the ground state
takes on a special significance in a quantum field theory,
because the vacuum is required to be unique, so the phase
cannot be arbitrarily at each point in space-time.
Once a particular value of the phase is chosen, it has to remain
the same everywhere, i.e. it cannot change
locally.
A scalar field with a nonzero vev
therefore breaks local gauge invariance\footnote{
An amusing analogy was proposed by A. Salam:
A number of guests sitting around a round dinner table
all have a serviette on the same side of the plate
with complete symmetry. As soon as one guest picks
up a serviette, say on the lefthand side,
the symmetry is broken and all guests have to follow suit
and take the serviette on the same side., i.e. the phases
are locked together everywhere in the ``vacuum'' due to the
``spontaneously broken symmetry''.}. More
details can be found in the nice introduction by Moriyasu\cite{book}.

 Nature has many examples of broken symmetries.
Superconductivity is a well known example.
Below the critical temperature the electrons bind into
Cooper pairs\footnote{
The interaction of the conduction electrons with the
 lattice  produces an attractive force. When
the electron energies are sufficiently small, i.e.
below the critical temperature, this attractive force overcomes
the Coulomb repulsion and binds the
electrons into Cooper pairs, in which the
momenta and spins of the electrons are in opposite directions, so the 
Cooper pair  forms a scalar
field and its quanta have a charge two times the electron charge.}.
 The density of Cooper pairs corresponds to the vev.
Owing to the weak binding, the effective size of a Cooper pair is
large, about $10^{-4} $ cm, so every Cooper  pair
overlaps with about  $10^6$ other Cooper pairs and this
overlap ``locks'' the phases of the wave function over
macroscopic distances: ``Superconductivity is a remarkable
manifestation of Quantum Mechanics on a truly macroscopic
scale'' \cite{lon}.

In the superconducting phase the photon gets an effective mass
through the interaction with the Cooper pairs in the ``vacuum'', 
which is apparent in  the Meissner effect:
 the magnetic field
 has a very short penetration depth into the superconductor or
equivalently the photon is very massive.
Before the phase transition the vacuum would have zero Cooper
pairs, i.e. a zero vev, and the magnetic field can penetrate
the superconductor without attenuation as expected for
 massless photons.

This example of Quantum Mechanics and spontaneous symmetry breaking in
superconductivity  has been transferred
almost literally to elementary particle physics by Higgs
and others\cite{hig}.
\begin{figure}[tb]
%\begin{center}
%\vspace{-1.1cm}
%\mbox{\epsfysize=8.cm\epsfxsize=15.cm\epsfbox{f23.eps}}
%\hspace*{2cm}
\epsfysize=8.cm\epsfxsize=15.cm
\epsffile[50 25 1970 990]{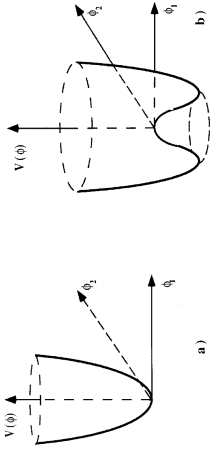}
\caption{Shape of the Higgs potential for $\mu^2>0$ (a) and $\mu^2<0$ (b);
$\phi_1$ and $\phi_2$ are the real and imaginary parts of the Higgs field.}
\label{f23}
%\end{center}
\end{figure}
For the self-interaction of the Higgs field one considers
a potential analogous to the one proposed by
Ginzburg and Landau for superconductivity:

\bq V(\Phi)=\mu^2\; \Phi^{\dagger} \Phi + \lambda
(\Phi^{\dagger}\Phi)^2 \eq
where $\mu^2$ and  $\lambda$ are constants.
The potential has a parabolic shape, if $\mu^2 > 0$,
but takes the shape of a Mexican hat for $\mu^2<0$,
 as pictured in fig. \ref{f23}. In the latter case the field free vacuum,
i.e. $\Phi =0 $, corresponds to a local maximum, thus forming
an unstable equilibrium. The groundstate corresponds to
a minimum with a nonzero value for the field:
\bq |\Phi|=\sqrt{\frac{-\mu^2}{2\lambda}}.\eq

In superconductivity $\mu^2$ acts like the critical temperature $T_c$:
above $T_c$ the electrons are free particles, so  their
phases can be rotated arbitrarily at all points in space,
but below $T_c$ the individual rotational freedom is lost,
because the electrons form a coherent system, in which all
phases are locked to a certain value. This corresponds to
a single point in the minimum of the Mexican hat,
 which represents a vacuum with a nonzero vev
 and a well defined phase,
thus defining a unique vacuum.
The coherent system can still be rotated as a whole so it is
invariant under global but not under local rotations.

\subsection{Gauge Boson Masses and the Top Quark Mass.}
After this general introduction about the Higgs mechanism,
one has to consider the number of Higgs fields needed to
break the $SU(2)_L\otimes U(1)_Y$ symmetry to the $U(1)_{em}$
symmetry. The latter must have one massless gauge boson, while
the $W$ and $Z$ bosons must be massive.
This can be achieved by choosing $\Phi$ to be a complex
 $SU(2)$ doublet with definite hypercharge ($Y_W=1$):
\bq \Phi(x)=\left(\ba{cc}\phi_1^+(x)&+i\phi^+_2(x)\\
\phi^0_1(x)&+i\phi^0_2(x)\ea \right) \label{phix} \eq
In order to understand the interactions of the Higgs field
with other particles, one considers the following
Lagrangian for a scalar field:
\bq \LL_H=(D_\mu \Phi)^{\dag} (D^\mu \Phi) -V(\Phi). \label{lhig}\eq
The first term is the usual kinetic energy term for a scalar
particle, for which the Euler-Lagrange equations
lead to the Klein-Gordon equation of motion.
Instead of the normal derivative, the covariant derivative
%\bq D_\mu=\partial_\mu+
%\frac{ig^\prime}{2} B_\mu {Y_W}
%-  \frac{ig}{2} {\bf %\vec{W}_\mu}\cdot{\bf\vec{\tau}}
% \label{cov2} \eq
is used in eq. \ref{lhig} in order to ensure local gauge
invariance under $SU(2)\otimes U(1)$ rotations.

The vacuum is known to be neutral. Therefore the groundstate
of $\Phi$ has to be of the form (0,v).
Furthermore $\Phi(x) $ has to be constant everywhere in order to have
zero kinetic energy, i.e. the derivative term in $\LL_H$
disappears.
%\footnote{Solutions with kinetic energy %nonzero are
%the so-called instanton solutions, which %we will not consider here.}

The quantum fluctuations of the field around the ground state
  can be parametrised as follows, if we include an arbitrary
$SU(2)$ phase factor:
\bq \Phi=e^{i\vec{\zeta}(x)\cdot\vec{\tau}}
\left(\ba{c}0\\v+h(x)\ea\right).\eq
The (real) fields $\zeta(x)$ are  excitations of the field
{\it along} the potential minimum. They correspond to
the massless Goldstone bosons of a global symmetry, in this
case three for the three rotations of the $SU(2)$ group.
However, in a {\it local} gauge theory these massless
bosons can be eliminated by a suitable rotation:
\bq \Phi^\prime=e^{-i\vec{\zeta}(x)\cdot\vec{\tau}}\Phi(x)=
\left(\ba{c}0\\v+h(x)\ea\right).
\label{hmin}\eq
Consequently the field $\zeta$ has no physical significance.
Only the real field $h(x)$ can be interpreted as a real
(Higgs) particle.
The original field $\bf \Phi$ with four degrees of freedom
has lost three degrees of freedom; these are recovered
as the longitudinal polarizations of the three heavy gauge
bosons.
%\footnote{As Coleman   put it: the gauge %bosons
%have eaten the Goldstone bosons and grown %fat.}.

%The mass of the Higgs boson can be expressed in the $\mu$
%parameter of the potential by expanding the potential in terms
%of  $h$:
%\bq V(h) =\frac{1}{2}\mu^2v^2+2\mu^2h^2+4\lambda v h^3+\lambda h^4 .\eq
%The quadratic term of $h$ represents the mass term, so
%\bq m_h =\sqrt{-4\mu^2} \eq
%In terms of $v$ and $h$ the Lagrangian can be written as:
%\bq \LL_H=\frac{1}{2}\partial^\mu \;h \partial_\mu\; h + \frac{(v+h)^2}
%{8}\chi^*\left[(g^\prime B_\mu+2gW_\mu) (g^\prime B^\mu+2gW^\mu)\right]
%\chi ,\label{lhig1}\eq
%where $\chi\equiv (\ba{c}0\\1\ea)$. The neutral Higgs field $h$
%has the mass
%\bq m_h^2=-2\mu^2 .\eq
The kinetic part of eq. \ref{lhig}  gives rise to mass terms for
the vector bosons, which can be written  as ($Y_W=1)$:
\bq \LL_H=\frac{1}{4}\left[\left({g} (W_\mu^1\tau_1+W_\mu^2\tau_2+W_\mu^3\tau_3)+
g^\prime B_\mu\right)\Phi\right]^{\dag}
\left[\left({g} (W^{\mu 1}\tau_1+W^{\mu 2}\tau_2+W^{\mu 3}\tau_3)+
g^\prime B^\mu\right)\Phi\right] \eq
or substituting for $\Phi$ its vacuum expectation value $v$ one obtains
from the off-diagonal terms (by writing the $\tau$ matrices explicitly, see eq.
\ref{wtau})

\bq
 \left(\frac{gv}{2}\right)^2\left( (W_\mu^1)^2
 +(W_\mu^2)^2\right) \eq
and from the diagonal terms:
\bq
\frac{1}{2}\left(\frac{v^2}{2}\right)
\left( -g W_\mu^{3\dag}+g^\prime B_\mu^{\dag}\right)
\left( -g W^{3\mu }+g^\prime B^\mu\right)=\nn
\frac{1}{2}\left(\frac{v^2}{2}\right)
\left(  B_\mu^{\dag}¸ W_\mu^{3\dag}\right) \left(\ba{cc}
+g^{\prime 2}& -gg^\prime\\-gg^\prime&g^2\ea \right)
\left(\ba{c} B^\mu\\W^{3\mu}\ea\right).\label{mas}\eq
Since mass terms of physical fields  have to be diagonal, one
obtains the ``physical'' gauge fields of the
broken symmetry by diagonalizing the mass term: 
\bq \left(  B_\mu^{\dag}¸ W_\mu^{3\dag}\right) ~U^{-1}U ~M ~U^{-1}U~
\left(\ba{c} B^\mu\\W^{3\mu}\ea\right)\eq
where $U$ represents a unitary matrix
\bq U=\frac{1}{\sqrt{g^{\prime 2}+g^2}}
\left(\ba{cc}g & g^\prime\\-g^\prime&g\ea\right)
\equiv
\left(\ba{cc}\cw & \sw\\-\sw&\cw\ea\right)
\eq
Consequently the real fields become a mixture of the gauge fields:
\bq \left(
\ba{c} A^\mu \\ Z^\mu  \ea\right)~=~U~\left(
\ba{c} B^\mu \\ W^{3\mu}  \ea\right)
 \eq
and the matrix $UMU^{-1}$ becomes a diagonal matrix
for a suitable mixing angle $\theta_W$.

In these fields the mass terms 
  have the
form \bq M_W^2W_\mu^+ W^{-\mu}+\frac{1}{2}(A_\mu,Z_\mu)
\left(\ba{cc}0&0\\ 0& M_Z^2\ea \right)
\left(\ba{c}A^\mu \\Z^\mu \ea \right)\label{mas1} \eq
with
\bq M_W^2=\frac{1}{2} g^2v^2  \label{emw} \eq
\bq M_Z^2=\frac{g^{\prime\; 2}+g^2}{2} v^2.
\label{emzz} \eq
  Here $v$ is the vacuum expectation value of the Higgs potential, which
for the known gauge boson masses and couplings  can be calculated to 
be\footnote{Sometimes
the Higgs field is normalized by $1/\sqrt{2}$, 
in which case $v\approx 256$ GeV.}:
\bq v \approx 174~ {\rm GeV}. \label{v174}\eq

%The complete Lagrangian  
%for the leptons of one generation can be
%written as:
%\bqa  \LL &=&
%\overline{\bf L}\gamma^\mu\left(\partial_\mu+\frac{ig^\prime}{2}
%Y_L{\bf B_\mu}+\frac{ig}{2}\vec{\tau}\cdot\vec{W_\mu}\right){\bf \L}\nn\\
%&&
%+\overline{\bf e}_R\gamma^\mu
%\left(\partial_\mu+\frac{ig^\prime}{2}
%Y_R{\bf  B_\mu}\right){\bf e_R}\\
%&& -\frac{1}{4}{\bf B_{\mu\nu} B^{\mu\nu}}
%-\frac{1}{4}{\bf \vec{W}_{\mu\nu} \vec{W}^{\mu\nu}}
%\label{lag1}\eqa
%Here ${\bf L}$ stands for the left-handed doublet:
%\bq {\bf L}\equiv\left(\ba{c}\nu_L\\e_L\ea\right){\rm with}
%\ba{ccc} \nu_L &=&\frac{1}{2}(1-\gamma_5)\Psi_\nu\\
%e_L &=& \frac{1}{2}(1-\gamma_5)\Psi_e\ea \eq
%and
%\bq \bf e_R=\frac{1}{2}(1+\gamma_5)\Psi_R,
%\eq
%where $\Psi$ are the lepton fields and $1\pm \gamma_5$ the usual
%projection operators, which select
%the left- or right handed components of the fields.

The neutral part of the Lagrangian, if expressed in terms
of the physical fields, can be written as:
\bqa \LL_{int}^{neutr}&=& -A_\mu[g^\prime \;\cw (\overline{e}_R
\gamma^\mu e_R +\frac{1}{2}\overline{\nu}_L\gamma^\mu \nu_L+
\frac{1}{2} \overline{e}_L\gamma^\mu e_L)                 %\nn \\&&
 -\frac{1}{2}g \;\sw ( \overline{\nu}_L\gamma^\mu \nu_L-
\overline{e}_L\gamma^\mu e_L)]\nonumber \\
&&+Z_\mu[g^\prime  \sw (\overline{e}_R
\gamma^\mu e_R +\frac{1}{2}\overline{\nu}_L\gamma^\mu \nu_L+
\frac{1}{2} \overline{e}_L\gamma^\mu e_L)              %%%%%%%\nn\\&&
 +\frac{1}{2}g  \cw  ( \overline{\nu}_L\gamma^\mu \nu_L-
\overline{e}_L\gamma^\mu e_L)]
\nn
\label{nc} \eqa
 The photon field should only couple to the electron
fields and not to the neutrinos,
so the terms proportional to $g'\cw$ and $g\sw$
should cancel and the coupling to the electrons
has to be the electric charge $e$.
 This can  be achieved
by  requiring:
\bq g^\prime\cw =g\;\sw =e~. \eq
Hence
\bq \tw =\frac{g^\prime}{g}~~;
\sws=\frac{g^{\prime 2}}{g^2+g^{\prime 2}}
~ ~{\rm and} ~~
e=\frac{gg^\prime}{\sqrt{g^2+g^{\prime 2}}}.
  \label{sws}\eq

\begin{figure}[tb]
%   HNS  A
\begin{center}
\epsfysize=4.cm
\epsfxsize=6.cm
\hspace{1.cm}
\epsffile[0 20 270 200]{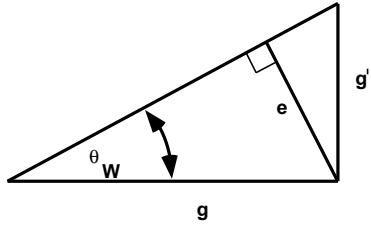}
\caption{Geometric picture of the relations between the
electroweak coupling constants.}
\label{f24}
\end{center}
\end{figure}

A geometric picture of these relations
is shown in  fig. \ref{f24}. From these
relations and the relations between masses and couplings (\ref{emw}
and \ref{emzz}) one finds the famous relation between
the electroweak mixing angle and the gauge boson masses:
\bq M_W= \cw \cdot M_Z \;\;\; {\rm or}~~~\sws =
1-\frac{M^2_W}{M^2_Z}\label{sin22}\;.\label{emwmzsin} \eq

The value of $M_W$ can also be related to the precisely measured
muon decay constant $G_\mu=1.16639(2)\cdot 10^{-5}$ GeV$^{-2}$.
If calculated in the SM, one finds:
\bq \frac{G_\mu}{\sqrt{2}}=\frac{e^2}{8\sws M_W^2}.\eq
This relation can be used to calculate the
gauge boson masses from  measured coupling constants
$\alpha$, $G_\mu$ and $\sw$:
\bqa M_W^2 &=& \frac{\pi\alpha}{\sqrt{2}G_\mu}\cdot \frac{1}{\sws}\\
M_Z^2 &=& \frac{\pi\alpha}{\sqrt{2}G_\mu}\cdot \frac{1}{\sws\cws}\\
\eqa
Inserting $\sws=0.23$ and $1/\alpha=137.036$  yields $\mz$=88 GeV.
However, these relations are only at tree level. Radiative
corrections  depend on the as yet unknown top mass.
Fitting the unknown top mass to the measured $\mz$ mass, the
electroweak asymmetries and the cross sections at LEP yields\cite{lep}:
\bq M_{top}=166^{+17~}_{-19~}(stat.)~ ^{+19~}_{-22~}  ~(unknown \; Higgs). 
\label{mtlep}\eq

Also the fermions can interact with the scalar field, albeit
not necessarily with the gauge coupling constant.
The Lagrangian for the interaction of the leptons with the Higgs field
can be written as:
\bq \LL_{H-L}=-g_Y^e\left[{  \overline{L}{\bf \Phi}
    e_R+\overline{e}_R{\bf \Phi}^{\dag}L}\right]. \eq
Substituting the vacuum expectation value for $\Phi$ yields
\bq \frac{-g_Y^e}{\sqrt{2}}\left[{  (\overline{\nu}_L,\overline{e}_L)
\left(\ba{c}0\\v\ea \right)e_R+
\overline{e}_R~(0,v)~
\left(\ba{c}\nu_L\\e_L\ea \right)}\right]\nn
=\frac{-g_Y^e v}{\sqrt{2}}\left[{  \overline{e}_L e_R
 +\overline{e}_R e_L}\right]=
\frac{-g_Y^e v}{\sqrt{2}}{  \overline{e}e }\label{mass}\eq
The Yukawa coupling constant $g_Y^e$ is a free parameter,
 which has to be adjusted
such that $m_e=g_Y^e v\sqrt{2}$.
Thus  the  coupling $g_Y$ is proportional to the
mass of the particle  and consequently the coupling of the
Higgs field to fermions is proportional to the mass of the fermion,
a prediction of utmost importance to the experimental search
for the Higgs boson.

Note that the neutrino stays massless with the choice of the Lagrangian, 
since no mass term
for the neutrino appears in eq. \ref{mass}.

\subsection{Summary on the Higgs mechanism.}
 In summary, the Higgs mechanism assumed the existence of a scalar field
$\Phi=\Phi_0 ~exp~(i\theta(x))$. After spontaneous symmetry breaking the phases
are ``locked'' over macroscopic distances, so the field averaged
over all phases is not zero anymore and $\Phi$ develops a
vacuum expectation value. The interaction of the fermions and
gauge bosons with this coherent system of scalar fields $\Phi$ gives
rise to effective particle masses, just like the interaction
of the electromagnetic field with the Cooper pairs inside a
superconductor can be described by an effective photon mass.

The vacuum corresponds to the groundstate with minimal potential energy
and zero kinetic energy. At high enough temperatures
the thermal fluctuations of the Higgs particles
about the groundstate become so strong that the coherence is lost,
i.e. $\Phi(x)=constant$ is not true anymore. In other words a phase transition
from the ground state with broken symmetry ($\Phi \neq 0$) to
the symmetric groundstate takes place. In the symmetric phase the groundstate
is invariant again under local $SU(2)$ rotations, since the phases
can be adjusted locally without changing the groundstate with $<\Phi>=0$.
 In the latter case all masses disappear, since they are
proportional to $<\Phi>=0$.

Both, the fermion  and  gauge boson masses are generated through the
interaction with the Higgs field. Since the interactions are proportional
to the coupling constants, one finds a relation between masses
and coupling constants.
For the fermions the Yukawa coupling constant is proportional to the fermion
mass and the mass ratio of the $W$ and $Z$ bosons is {\it only}
dependent on the electroweak mixing angle
 (see eq. \ref{sin22}).
This mass relation is in excellent agreement with experimental data after
including radiative corrections.
Hence, it is the first indirect evidence that the gauge bosons
masses are indeed generated by the interaction with a scalar
field, since otherwise there is no reason to expect
the masses of the charged and neutral gauge bosons to be
related in such a specific way via the couplings.
%   HNS  A
\begin{figure}
\epsfysize=6.cm
\hspace*{3.cm}
\epsffile[60 130 475 350]{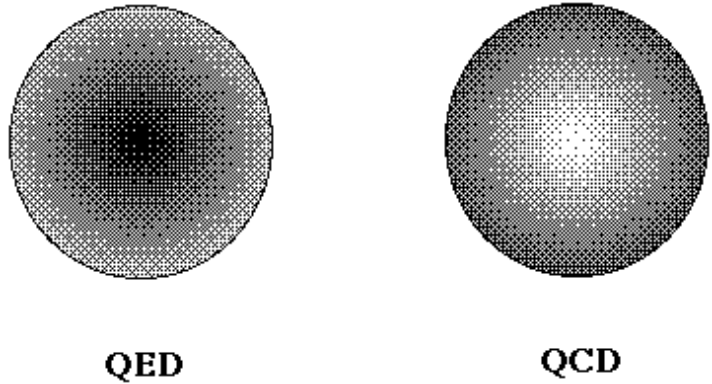}
\caption{The effective charge distribution around an electric charge (QED) and
 colour charge (QCD). At higher $Q^2$ one probes smaller distances, thus
 observing a larger (smaller) effective charge, i.e. a larger (smaller)
coupling constant in QED (QCD).}
\label{f25}
\end{figure}
\begin{figure}[t]
\begin{center}
\mbox{\epsfysize=5.cm\epsfbox{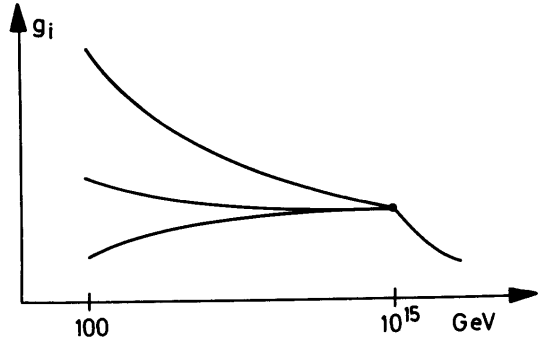}}
\caption{Running of the three coupling constants in the Standard Model owing 
to the different space charge
distributions (compare fig. 2.5.}
\label{f26}
\end{center}
\end{figure}
\section{ Running Coupling Constants}

In a Quantum Field Theory the coupling constants are only
effective constants at a certain energy. They are energy, or equivalently
distance dependent through virtual corrections,  both in QED and in QCD.

%   HNS  E
\begin{figure}[t]
\epsfysize=8.cm
\hspace*{1.cm}
\epsffile[65 495 525 785]{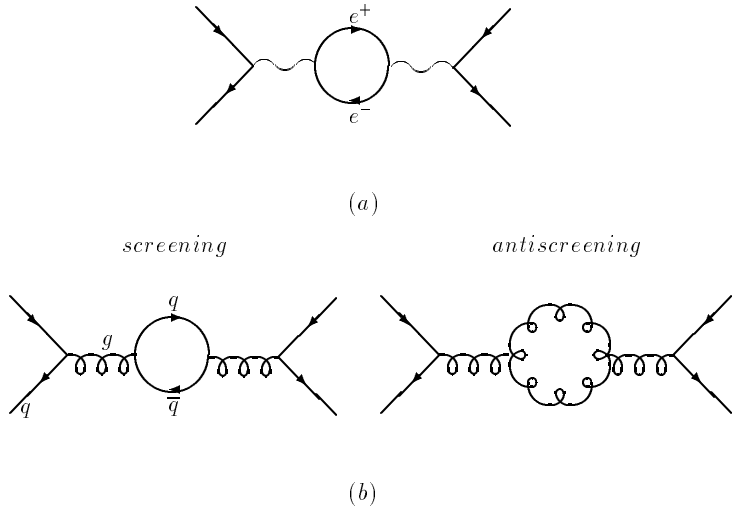}
\caption{Loop corrections in QED (a) and QCD (b). In QED only the fermions 
contribute in the loops, which causes a screening of the bare charge. 
In QCD also the bosons contribute through the gluon selfinteraction, 
which enhances the bare charge. This antiscreening dominates over  
the screening.}
\label{f27}
\end{figure}
\begin{figure}[tb]
%   HNS  A
\hspace*{2.5cm}
\epsffile[5 60 315 290]{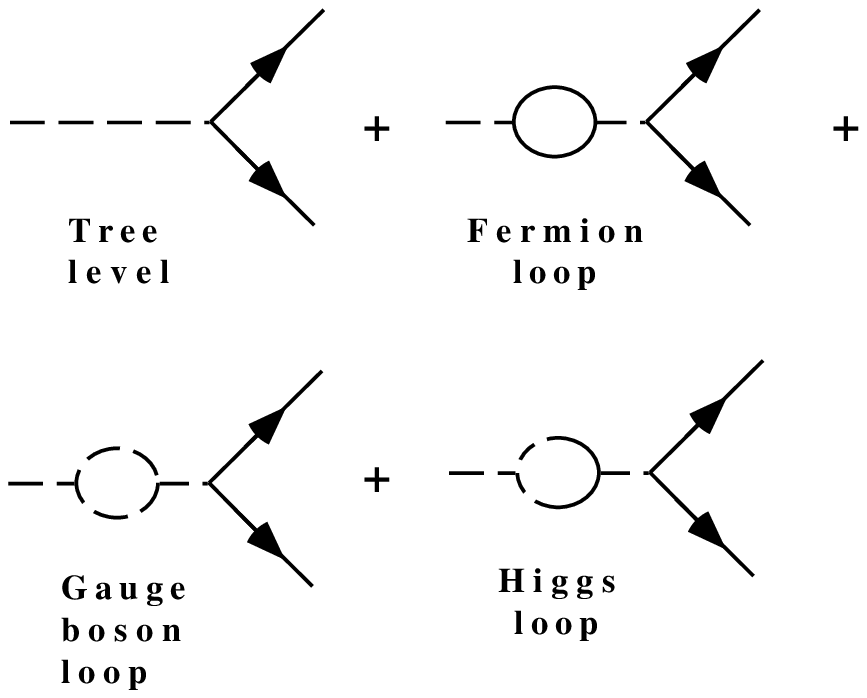}
\caption{First order vacuum polarization
diagrams.}
\label{f28}
\end{figure}
%   HNS  E

However, in               QED the coupling constant increases
as function of $Q^2$, while in QCD the coupling constant decreases.
A simple picture for this behaviour is the following:
\begin{itemize}
\item
The electric field around a pointlike electric charge diverges like $1/r$.
In such a strong field electron-positron pairs can be created
with a lifetime determined by Heisenberg's uncertainty relations.
These virtual $\epem$ pairs orient themselves in the electric field,
thus giving rise to vacuum polarization, just like the atoms
in a dielectric are polarized by an external electric field.
This vacuum polarization screens the ``bare'' charge, so at a large
distance one observes only an effective charge. This causes
deviations from Coulomb's law, as observed  in the well-known
Lamb shift of the energy levels of the hydrogen atom.     If  the  electric
    charge is probed at higher energies (or  shorter  distances),
    one penetrates the shielding  from  the  vacuum  polarization
    deeper and observes more of the bare charge, or  equivalently
    one observes a larger coupling constant.
\item
    In QCD the situation is more complicated: the  colour  charge
    is  surrounded by a cloud  of  gluons  {\it and}  virtual  \qq~
    pairs;   since  the  gluons  themselves  carry  a  colour
    charge, one has two contributions: a shielding  of  the  bare
    charge by the \qq~ pairs
%   and transversely polarized gluons
%   (in the so-called
% Coulomb gauge)
    and an  increase  of the
    colour charge by the gluon
    cloud. The net  effect  of  the  vacuum  polarization  is  an
    increase of the total colour charge, provided not too many \qq~pairs
    contribute,  which is the case if the number of generations is     
    below 16, see hereafter.
    If one probes this charge at smaller distances,
    one penetrates part of the ``antishielding'', thus observing
    a smaller colour charge at higher energies.
    So it is the fact that gluons carry colour themselves
    which makes the coupling decrease at small distances
    (or high energies).
   This property is called
   asymptotic freedom and it   explains
           why  in  deep  inelastic  lepton-nucleus  scattering
   experiments the quarks inside a  nucleus  appear   quasi
  free  in spite of the fact that they are tightly  bound
   inside a nucleus.
   The increase of \as~ at large distances explains  qualitatively
    why it is
     so difficult
    to separate the quarks inside a hadron: the larger the
   distance  the more       energy one needs to separate them
even further.
    If 
    the energy of the colour field
    is too high, it is transformed into mass, thus
generating new quarks, which then recombine
    with the old ones to form new hadrons, so one always
    ends up with a system of hadrons instead of free quarks.
\end{itemize}

The space charges from the virtual pairs surrounding an
 electric charge and colour charge are shown schematically in fig. \ref{f25}.
The different vacuum polarizations
lead to the energy dependence of the coupling constants sketched in fig. 
\ref{f26}.
The colour field becomes infinitely dense at the QCD scale 
$\Lambda\approx 200$ MeV
(see hereafter).
So the confinement radius of typical
hadrons is $\cal{O}$(200 MeV) or one Fermi ($10^{-13}$ cm).

The vacuum polarization effects
 can be calculated from the loop diagrams
to the gauge bosons.
The main difference between the charge distribution
in QED and QCD originates from the diagrams shown  in fig.
\ref{f27}. In addition one has
to consider diagrams of the type shown
in fig. \ref{f28}.
The  ultraviolet divergences ($Q^2\rightarrow\infty$)
in these diagrams can be absorbed in the
coupling constants in a renormalizable theory. All other divergences are 
canceled
at the amplitude level by summing the appropriate amplitudes.
The first step in such calculations is the regularization of
the divergences, i.e. separating the divergent parts in the
mathematical expressions. The second step is the renormalization
of physical quantities, like charge and mass, to absorb
the divergent parts of the amplitudes, i.e. replace
the ``bare'' quantities of the theory with measured  quantities.
%For example,
% for $\epem\rightarrow\mu^+\mu^-$ the matrix element
%at the Born level is:
%\begin{equation}
%M^0=e^2\overline{u}_4
%\gamma_{\mu}u_3\left(\frac{-g^{\mu\nu}}{g^2}
%\right)\overline{u}_2\gamma_{\nu}u_1
%\end{equation}
%%the loop corrections to the photon propagator yield:
%\begin{equation}
%\Pi^{\mu\nu}(Q^2)=-\frac{1}{Q^4}\int \frac{d^4k}{(2\pi)^4}
%%%Tr\left[ie\gamma^{\mu}
%\frac{1}{k-m}ie\gamma^{\nu}\frac{1}{k-q-m}\right]
%\end{equation}
%which diverges if $k\rightarrow\infty$.
For example, the loop corrections to the photon
propagator diverge, if the
momentum transfer $k$ in the loop is integrated to infinity.
If one introduces a cutoff $\mu_0$ for large values of $k$,
one finds for the regularized amplitude of the sum of the Born term  $M_0$
and the loop corrections\cite{bjor}:
\begin{equation}
M^1=e^2\left(1-\frac{\alpha}{3\pi}\ln\frac{\mu_0^2}{m^2}\right)
    \left(1+\frac{\alpha}{3\pi}\ln\frac{Q^2}{m^2}\right)\;M^0
   ~ {\rm for}~ Q^2>>m^2
\end{equation}
%and
%\bq
%%M^1=\left(1-\frac{\alpha}{3\pi}
%%\ln\frac{\mu_0^2}{m^2}\right)
%    \left(1-\frac{\alpha}{15\pi}
%%\ln\frac{Q^2}{m^2}\right)\;M^0
 %   ~{\rm for}~ Q^2<<m^2
%\eq
The divergent part depending on the cutoff parameter
$\mu_0$ disappears, if one replaces the ``bare'' charge $e$ 
by the renormalized charge  $e_R$:
\begin{equation}
e_R^2\equiv e^2\left(1-
\frac{\alpha}{3\pi}
\ln\frac{\mu_0^2}{m^2}\right)
\end{equation}
i.e.  the ``bare'' charge, occurring  in the
Dirac equation,  is renormalized to a measurable quantity $e_R$.
For $e_R$ one usually takes the Thomson limit for Compton scattering,
i.e. $ \gamma e\rightarrow \gamma e$ for $k\rightarrow 0 $:
\bq
\sigma_T=\frac{8\pi}{3} \frac{\alpha^2}{m_e^2}
\eq
with $\alpha={e_R^2}/{4\pi}=1/137.036$ and $m_e=0.00051 $ GeV.

After regularization and renormalization to a measured quantity
(in this case using the so called ``on shell'' scheme, i.e.
one uses the mass and charge of a free
 electron as measured at low energy), one is left with a
$Q^2$ dependent but finite part of the vacuum polarization.
This  can be absorbed in a $Q^2$ dependent coupling constant,
which in case  of QED becomes for $Q^2>>m^2$:
\bq
\alpha(Q^2)=\alpha\left(1+\frac{\alpha}{3\pi}
\ln\frac{Q^2}{m^2_e}\right)
\eq
 %The regularization     is usually done
%with the dimensional regularization scheme of 't Hooft and
%Veltman\cite{hoof}. In $n=4-2\epsilon$ dimensions the bare
%coupling constant has the dimension of a mass. In order to
%make it dimensionless,
%one introduces an arbitrary parameter $\mu$
%with the dimension of a mass and defines the coupling as
%$g(\mu^2)=\mu^\epsilon g$ and $\as=g^2/4\pi=\as(\mu^2)$
%>The diagram in Fig. 9a  contributes a term
%$\approx \frac{\alpha(\mu^2)}{3\pi} \ln \frac{Q^2}{\mu^2}$
% to the cross section, if $Q^2 >>  \mu^2$.
%In QED it is customary
% to choose for $\mu$    the electron mass $m_e$.
%%In this case one can absorb
%the  divergent vacuum polarization in
%    an effective coupling constant  by modifying the
%fine structure constant $\alpha=e^2/4\pi$ as follows:
%\begin{equation}
%\alpha(Q^2) = \alpha (1+
%  \frac{\alpha}{3\pi} \ln \frac{Q^2}{m_e^2})
%\end{equation}
If one sums more loops, this yields terms
$(\frac{\alpha}{3\pi})^n(\ln\frac{Q^2}{m_e^2})^m$
and retaining  only the leading logarithms (i.e. n=m),
 these terms can be summed to:
\begin{equation}
\alpha(Q^2) = \frac{\alpha}{ (1-
  \frac{\alpha}{3\pi} \ln \frac{Q^2}{m_e^2})}
\label{alphaq}
\end{equation}
since
\begin{equation}
\sum_{n=0}^{\infty} x^n = \frac{1}{1-x}.
\end{equation}
Of course, the total $Q^2$ dependence is obtained
 by summing over  all possible fermion loops
in the photon propagator.

These vacuum polarization effects are non-negligible.
For example, at LEP accelerator energies $\alpha$ has increased from
its low energy value $1/137$ to 1/128 or about 6\%.

%Whenever the virtual photon momentum $Q$ is timelike
%and $Q^2$ above threshold
%for pair production, one gets an additional
%imaginary contribution
%corresponding to real particles in the loop
%of Fig. 9a\cite{bjo}.
%%\medskip\\  %\indent

The diagrams of fig. \ref{f27}b        yield similarly to eq. \ref{alphaq}:
\begin{equation}
\as(Q^2)=\as(\mu^2)\left[1+\frac{\as(\mu^2)}{4\pi}
\left(11-\frac{2N_f}{3}\right)
\ln \frac{Q^2}{\mu^2}\right]^{-1}
\label{alphasq}
\end{equation}
Note that \as~ decreases  with increasing $Q^2$
if $11-      2N_f/3>0$ or $N_f<16$,
thus leading  to asymptotic freedom at high energy.
This is in  contrast to  the $Q^2$ dependence of
$\alpha(Q^2)$ in eq. \ref{alphaq}, which increases
with increasing $Q^2$.
Since \as~ becomes infinite at small $Q^2$, one cannot take this  scale
as a reference scale.
Instead one could choose as renormalization point the
``confinement scale'' $\Lambda$, i.e.
$\as\rightarrow \infty$, if $Q^2\rightarrow \Lambda$.
In this case  eq. \ref{alphasq} becomes independent of $\mu$, since 
the 1 in brackets becomes negligible, so one obtains:
\begin{equation}
\as(Q^2)=\frac{4\pi}
{(11-\frac{2n_f}{3})\ ln \frac{Q^2}{\Lambda^2}}
\end{equation}

%A physical quantity should not depend on the spurious
%parameter $\mu$, at least if one calculates it to all orders.
%If one calculates only up to a finite order, one can minimize
%the higher order terms by a suitable choice of $\mu$.
%In lowest  order  $\mu$ is arbitrary, but in higher orders the loop
%calculations  contain terms $ln \frac{Q^2}{\mu^2}$ and to keep
%these terms small, it is best to choose $\mu^2$ to be of the same
%order as $Q^2$, where $Q^2$ is the relevant physical scale
%of the process.
The definition of $\Lambda$ depends on the renormalization scheme.
The most widely used  scheme
is the $\overline{MS}$ scheme\cite{msb}, which we will use
 here.
Other schemes can be used as well and simple relations between
the definitions of $\Lambda$ exist\cite{robe}.
%\medskip\\ \indent

 %   In summary one can absorb the divergent vacuum polarization
%diagrams in the coupling constant, which then becomes dependent %on
%$Q^2$.

The higher order  corrections are usually calculated
with the renormalization group technique, which yields for
the $\mu$ dependence of a coupling constant $\alpha$~:
\bq %\label{rge}
\mu\frac{\partial\alpha}{\partial\mu}=\beta_0 \alpha^2+\beta_1\alpha^3+
       \beta_2\alpha^4+...
\label{rgeI}\eq
The first two terms in this perturbative expansion
are renormalization-scheme independent.
%\bq
%\beta_0=-\frac{1}{2\pi}
%\left[\frac{11}{3}C_A
%-\frac{4}{3}T_F~N_f\right]
%\end{equation}
%\begin{equation}
%\beta_1=-\frac{1}{4\pi^2}\left[\frac{34}{3}C_A^2
%-\frac{20}{3}C_G~T_F~N_f-4C_F~T_F~N_f\right]
%\eq
%where the group factors $C$ and $T$ only %dependent
%on the group structure.
 Their specific values
are given in the appendix.
The first order solution of eq. \ref{rgeI} is simple:
\bq \frac{1}{\alpha(Q^2)}=\frac{1}{\alpha(Q^2_0)}-
\beta_0 \ln(\frac{Q^2}{Q_0^2} )     \label{srgeI} \eq
where $Q^2_0$ is a reference energy.
One observes a linear relation between the
change in the inverse of the coupling constant and
the logarithm of the energy.
The slope depends on the sign of $\beta_0$, which
is positive for QED, but negative for QCD, thus
leading to asymptotic freedom in the latter case.
The second order corrections are so small, that they
do not change this conclusion.
Higher order terms depend on the
renormalization prescription.
In higher orders there are also  corrections
from Higgs particles and gauge bosons
in the loops. Therefore the running of a given coupling
constant depends slightly on the value of the
other coupling constants and the Yukawa  couplings.
These higher order corrections cause the RGE equations
to be coupled, so one has to solve a large number
of coupled differential equations.
All these equations are summarized in the
appendix.

\chapter{Grand Unified Theories.}
\label{ch3}
\section{Motivation }
The Standard Model describes all observed interactions
between elementary particles with astonishing precision.
Nevertheless, it cannot be considered to be the ultimate
theory because the many unanswered questions
remain a problem.
Among them:
\begin{itemize}
\item
\underline{\bf The Gauge Problem }\\
Why are there three independent  symmetry groups?
\item
\underline{\bf The Parameter Problem }\\
How can one reduce the number of free parameters?
(At least 18 from the couplings, the mixing parameters,
the Yukawa couplings and the Higgs potential.)
\item
\underline{\bf The Fermion Problem }\\
Why are there three generations of quarks and leptons?
What is the origin of the the symmetry between quarks
and leptons? Are they composite particles of more
fundamental objects?
\item
\underline{\bf The Charge Quantization Problem }\\
Why do protons and electrons have exactly opposite electric
charges?
\item
\underline{\bf The Hierarchy Problem }\\
Why is the weak scale so small compared with the GUT scale, 
i.e. why is $M_W\approx 10^{-17}~M_{Planck}?  $
\item
\underline{\bf The Fine-tuning Problem }\\Radiative 
corrections to the Higgs masses and gauge boson
masses have quadratic divergences. For example, $\Delta M_H^2
\approx
{\cal{O}}(M_{Planck}^2)$.  In other words,
 the corrections to the Higgs  masses
  are many orders of magnitude larger than the
masses themselves, since they are expected
to be of the order of the electroweak gauge boson masses.
 This requires extremely unnatural fine-tuning in the
parameters of the Higgs potential.
This ``fine-tuning'' problem is solved in the supersymmetric
extension of the SM, as will be discussed afterwards.
\end{itemize}
\section{Grand Unification }
The problems mentioned above can be partly solved by assuming the
symmetry groups \su~ are part of a larger group $G$, i.e.
\bq G\supset \su. \eq
The smallest group $G$ is the SU(5) group\footnote{$G$
 cannot be the direct product of the $SU(3),~SU(2) $ and $U(1)$ groups, since
this would not represent a new unified force with a single
coupling constant, but still require three independent coupling
constants.}\cite{su5}, so the minimal extension of the SM
towards a GUT is based on the $SU(5)$ group. Throughout this
paper we will only consider this minimal extension.
The group $G$  has a single coupling constant for all interactions
and the observed differences in the couplings at low energy
are caused by radiative corrections.
As discussed before, the strong coupling constant decreases
with increasing energy, while the electromagnetic
one increases with energy, so that at some high energy
they will become equal. Since the changes with energy
are only logarithmic  (eq. \ref{srgeI}), 
the unification scale is high,
namely of the order of $10^{15}-10^{16}$ GeV, depending
on the assumed particle content in the loop diagrams.

%The basic representation of the $SU(N)$ group is a  multiplet
%of dimension $N$ and the next higher one is a multiplet
%of the dimension $N\times N$.
In the $SU(5)$ group\cite{su5} the 15 particles  and antiparticles
of the first
generation can be fit into the $\overline{5}$-plet\footnote{The
bar indicates the complementary representation of the fundamental
representation.} and 10-plet:

\bq\overline{5}=\left(\ba{c}d_g^C\\d_r^C\\d_b^C\\e^-\\-\nu_e
\ea\right)~~10=\frac{1}{\sqrt{2}}\left(\ba{ccccc}
0&+u_b^C&-u_r^C&-u_g&-d_g\\
 -u_b^C&0&+u_g^C&-u_r&-d_r\\
+u_r^C &-u_g^C &0 &-u_b&-d_b\\
 +u_g&+u_r&+u_b&0&-e^+\\
+d_g&+d_r&+d_b&+e^+&0\ea\right)_L \eq
The superscript $C$ indicates the charge conjugated particle,
i.e. the antiparticle and all particles are chosen to be left-handed, 
since a left-handed antiparticle transforms
like a right-handed particle. Thus the superscript $C$
implies a right-handed singlet with weak isospin equal zero.

With this multiplet structure the  sum of the
quantum numbers $Q$, $T_3$ and $Y$ is zero within one
multiplet, as required, since the corresponding operators
are represented by traceless matrices.

Note  that there is no space for the antineutrino in these
multiplets, so within the minimal $SU(5)$
the neutrino must be massless, since for a
massive particle the right-handed helicity state
is also present. Of course, it is possible to put a right-handed
neutrino into a singlet representation.

$SU(5)$ rotations can be represented by $5\times 5$ matrices.
Local gauge invariance requires the introduction of $N^2-1=24$
gauge fields (the ``mediators''), which cause the interactions
between the matter fields.
The gauge fields transform under
 the adjoint representation of the $SU(5)$ group, which
can be written in matrix form as (compare eqns. \ref{wtau}
and \ref{glam}):
\bq 24=\left(\ba{ccc|cc}
G_{11}-\frac{2B}{\sqrt{30}} & G_{12} & G_{13} & X_1^C & Y_1^C\\
G_{21} & G_{22}-\frac{2B}{\sqrt{30}} & G_{23} & X_2^C & Y_2^C\\
G_{31} & G_{32} & G_{33}-\frac{2B}{\sqrt{30}} & X_3^C & Y_3^C\\
\hline
X_1&X_2&X_3&\frac{W^3}{\sqrt{2}}+\frac{3B}{\sqrt{30}} & W^+\\
Y_1&Y_2&Y_3&W^-&-\frac{W^3}{\sqrt{2}}+\frac{3B}{\sqrt{30}} \\
\ea
\right) \label{gsu}\eq
The $G$'s represent the gluon fields of equation \ref{glam},
while the $W$'s and $B$'s are the gauge fields of the $SU(2)$
symmetry groups. The $X$ and $Y$'s are new gauge bosons, which
 represent interactions, in which  quarks are
transformed into leptons and vice-versa,
as should be  apparent if one operates with this matrix on
the $\overline{5}$-plet. Consequently, the $X$ ($Y$) bosons, which
couple to the  electron (neutrino)
 and $d$-quark must have electric charge $4/3$ (1/3).
\begin{figure}
%   HNS  A
% \vspace*{1cm}
  \hspace*{0.8cm}
\epsffile[16 42 546 152]{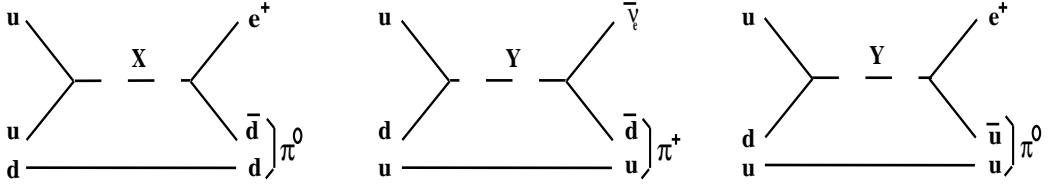}
%\vspace{1cm}
%   HNS  E
\caption{GUT proton decays through the exchange of $X$ and $Y$ gauge bosons.}
\label{f31}
\end{figure}
\section{$SU(5)$ predictions}

\subsection{Proton decay }

The $X$ and $Y$ gauge bosons can introduce
 transitions between quarks and leptons, thus violating
lepton and baryon number\footnote{The difference between
lepton and baryon number B-L is   conserved
in these transitions.}.
This can lead to the following proton and neutron decays
(see fig. \ref{f31}):
\bq
\ba{cc} p\rightarrow e^+\pi^0 & n\rightarrow e^+\pi^- \\
 p\rightarrow e^+\rho^0 & n\rightarrow e^+\rho^- \\
p\rightarrow e^+\omega^0 & n\rightarrow {\nu}\omega^0\\
p\rightarrow e^+\eta & n\rightarrow \overline{\nu}\pi^0 \\
p\rightarrow \overline{\nu}\pi^+
& n\rightarrow \overline{\nu}_\mu K^0 \\
p\rightarrow \overline{\nu}\rho^+
 & \\
p\rightarrow \overline{\nu}_\mu K^+
  \ea \eq

The decays with kaons in the final state are allowed
through flavour mixing, i.e. the interaction eigenstates
are not necessarily the mass eigenstates.

For the lifetime of the nucleon one writes in analogy
to  muon decay:
\bq \tau_p\approx \frac{M^4_X}{\alpha_5^2 m_p^5} \label{taup}\eq
The proton mass $m_p$ to the fifth power originates from
the phase space in case the final states are much lighter
than the proton, which is the case for the dominant decay
mode: $p\rightarrow e^+\pi_0$.
After this prediction of an unstable proton in grand unified
theories, a great deal of activity developed and the
lower limit on the proton life time increased to\cite{pdb}
\bq \tau_p > 5\cdot 10^{32} ~{\rm yrs} \eq
for the dominant decay mode
$p\rightarrow e^+\pi^0$.
From equation \ref{taup} this implies
 (for $\alpha_5=1/24$, see chapter \ref{ch6})
\bq M_X \geq 10^{15}~ {\rm GeV.} \eq

From the extrapolation of the couplings
in the $SU(5)$ model to high energies
one expects the unification scale to be reached well below
$10^{15}$ GeV, so the proton lifetime measurements
exclude the minimal $SU(5)$ model as a viable
GUT.  As will be discussed later, the supersymmetric
extension of the $SU(5)$ model has the unification point
well above $10^{15}$ GeV.
\subsection{Baryon Asymmetry }
The heavy gauge bosons responsible for the unified force
 cannot   be produced with conventional
accelerators, but  energies above $10^{15}$
were easily accessible
during the birth of our universe.
This could have led to an excess of matter over antimatter
right at the beginning, since the $X$ and $Y$ bosons can
decay into pure matter, e.g. $X\rightarrow uu$, which
is allowed because the charge of the $X$ boson is 4/3.
As pointed out by Sakharov\cite{sak} such an excess is possible
 if both C and CP are violated, if the baryon number B is
violated, and if the process goes through a phase of
non-equilibrium. All three conditions are possible within the
$SU(5)$ model. The non-equilibrium phase  happens
if the hot universe cools down and arrives at a temperature,
too low to generate $X$ and $Y$ bosons anymore, so only
the decays are possible. Since the  CP violation is
expected to be small, the excess of matter over antimatter
will be small, so most of the matter annihilated with
antimatter into enormous number of photons.
This would explain why the number of
photons over baryons is so large:
\bq \frac{N_\gamma}{N_b}\approx 10^{10} \eq
However, later it was realized that the
electroweak phase transition may wash out
any (B+L) excess generated by GUT's.
One then has to explain the observed
baryon asymmetry by the electroweak baryogenesis, which 
is actively studied\cite{baryo}.

\subsection{Charge Quantization }
From the fact that quarks and leptons are assigned to
the same multiplet the charges must be related,
since the trace of any generator  has to be zero.
For example, the charge operator Q on the fundamental
representation yields:
\bq
Tr Q=Tr (q_{\overline{d}},q_{\overline{d}},
q_{\overline{d}},e,0) =0\eq
or in other words, in $SU(5)$ the electric charge  of the
$d$-quark has to be 1/3 of the charge of an electron!
Similarly, one finds the charge of the $u$-quark is 2/3 of the
positron charge, so the total charge of the proton (=$uud$)
has to be exactly opposite to the charge of an electron.
\begin{figure}[tb]
%   HNS  A
 \vspace*{2cm}
 \hspace*{2.9cm}
\epsffile[25 75 500 200]{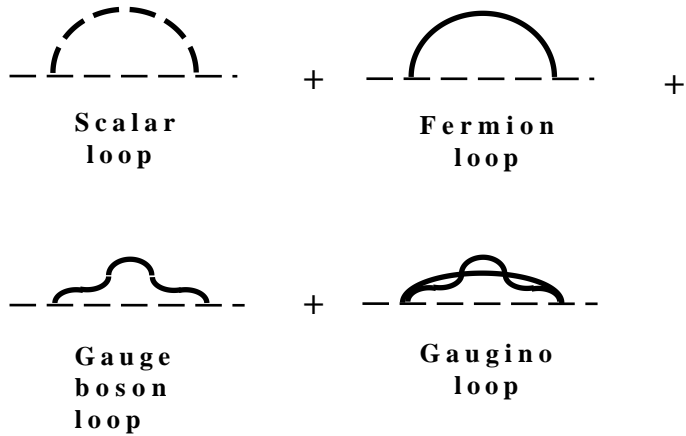}
\caption{Radiative corrections to particle masses.}
\label{f32}
\end{figure}
\subsection{Prediction of \sws }
If the  $SU(2)$ and $U(1)$ groups have equal coupling constants,
 the electroweak mixing angle can be calculated
easily, since it is given by the ratio $g^{\prime\; 2}
/(g^2+g^{\prime\; 2})$ (see eq. \ref{sws}), which would be
 1/2 for equal
coupling constants. However, the argument is slightly more
subtle, since for unitary transformations the rotation matrices have to be
 normalized such that
\bq Tr\;{F_k F_l}=\delta_{kl}. \label{trf}\eq
This normalization is not critical in case one has
independent coupling constants for the subgroups, since
a ``wrong'' normalization for a rotation matrix can
always be corrected by a redefinition of the
corresponding coupling constants,
as is apparent from  equation \ref{cov1}. This freedom is lost,
if one has a single coupling constant, so one has to be careful
about the relative normalization. It turns out, that the
Gell-Mann and Pauli rotation matrices of the $SU(3)$ and
$SU(2)$ groups have the correct normalization, but the
normalization of the weak hypercharge operator needs to be
changed. Defining $1/2Y_W=CT_0$ and substituting this
into the Gell-Mann-Nishijima relation \ref{gmn} yields:
\bq Q=T_3+CT_0 \eq
Requiring the same normalization for $T_3$ and $T_0$ implies
from equation \ref{trf}:
\bq Tr\;{Q^2}=(1+C^2)~ Tr\; {T_3^2} \eq
or inserting numbers from the $\overline{5}$-plet of $SU(5)$
yields:
\bq   1+C^2=\frac{Tr\;Q^2}{Tr\;T_3^2}=
\frac{3\cdot 1/9+1}{2\cdot 1/4} =\frac{8}{3}.\label{c} \eq

Replacing in the covariant derivative (eq. \ref{cov1})
$1/2Y_W$ with $CT_0$ implies $g^\prime CT_0
\equiv g_5 T_0$ or:
\bq g_5=C g^\prime,\label{c53}\eq
where   $C^2= {5}/{3}$
from eq. \ref{c}.
With this normalization the electroweak mixing angle after unification
becomes:
\bq \sws=\frac{g^{\prime 2}}{(g^2+g^{\prime\; 2})}=
\frac{g_5^2/C^2}{(g_5^2+g^2_5/C^2)}=\frac{1}{1+C^2}
=\frac{3}{8}.
\eq
The manifest disagreement with the experimental value of 0.23
at low energies brought the $SU(5)$ model originally into
discredit, until it was noticed that the running
of the couplings between the unification scale and low
energies could reduce the value of $\sws$ considerably. %\cite{glashow}.
 As we will show in the last chapter, with the very precise
measurement of $\sws$ at LEP, unification
of the three coupling constants within the $SU(5)$  model
is excluded, and just as in the case of the proton life time,
supersymmetry comes to the rescue and unification is perfectly
possible within the supersymmetric extension of $SU(5)$.

Note that the prediction of $\sws=3/8$ is not specific
to the $SU(5)$ model, but is true for any group
with $\su$ as subgroups, implying that $Q$, $T_3$ and $Y_W$
are generators with  traces equal zero and
thus leading to the predictions given above.

\section{Spontaneous Symmetry Breaking in $SU(5)$ }
The $SU(5)$ symmetry is certainly broken, since the
new force corresponding to the exchange of the
$X$ and $Y$ bosons would lead to very rapid proton
decay, if these new gauge bosons were massless.
As mentioned above, from the limit on the proton life
time these $SU(5)$ gauge boson have to be very heavy, i.e.
masses above $10^{15}$ GeV.
The generation of masses can be obtained again in a gauge
invariant way via the Higgs mechanism.
The Higgs field is chosen in the adjoint
 representation $\underline{24}$ and
the minimum $<\Phi_{24}>$ can be chosen in the following
way:
\bq <\Phi_{24}>=v_{24}~\left(\ba{ccc|cc}
1&&&&\\
&1&&&\\
&&1&&\\
\hline
&&&-\frac{3}{2}&\\
&&&&-\frac{3}{2}\ea\right)
\eq

The 12 X,Y gauge bosons of the $SU(5)$ group require
a mass:
\bq M_X^2=M_Y^2=\frac{25}{8}g_5^2v_{24}^2 \label{mxy} \eq
after `eating' 12 of the 24 scalar fields in the
adjoint representation, thus providing the
longitudinal degrees of freedom.
The field $\Phi_{24}$ is   invariant under the
rotations of the $\su$ group, so this symmetry is
not broken and the corresponding gauge bosons,
including the $W$ and $Z$ bosons, remain massless.
after the first stage of $SU(5)$ symmetry breaking.

The usual breakdown of the electroweak symmetry
to $SU(3)_C\otimes U(1)_{em} $ is achieved by a 5-plet
$\Phi_5$ of Higgs fields, for which the minimum
of the effective potential can be chosen at:
\bq <\Phi_5>={v_5}\left(\ba{c}0\\0\\0\\0\\1\ea\right)
\eq
The fourth and fifth component of  $\Phi_5$
 correspond to the $SU(2)$ doublet $(\Phi^+,\Phi^0)$
of the SM (see eq. \ref{phix}).
Since the total charge in a representation has to be
zero  again, the first triplet of complex fields
in $\Phi_5$, which transforms as (3,1)$_{-2/3}$ and
 $(3^*,1)_{2/3}$, must have  charge $|1/3|$.
Since they couple to all fermions with mass,
 they can induce  proton decay:
\bq u+d\rightarrow  H^{1/3}\rightarrow
\ba{c}e^+ +\overline{u}\\\overline{\nu}_e +\overline{d}
\ea \label{h3}\eq
Such decays can be suppressed only by
sufficiently high masses of the coloured
Higgs triplet. These can obtain high masses through
interaction terms between  $\Phi_5$ and $\Phi_{24}$.

Note that from eq. \ref{mxy} $<\Phi_{24}>$ has to be of the
order of  $M_X$, while  $\Phi_5$ has to be of
 the order of $M_W$, since
\bq M_W^2=\frac{1}{2}(g_5 v_5)^2 \eq
and
\bq M_Z=\frac{M_W}{\cos \theta_W} \eq
or more precisely $v_5= 1/\sqrt{ G_F}$= 174 GeV.
The minimum of the Higgs potential involves both
 $\Phi_5$ and $\Phi_{24}$. Despite this mixing, 
the ratio $v_5/v_{24}\approx 10^{-13}$  has to be preserved
(hierarchy problem).
Radiative corrections spoil usually such a fine-tuning, 
so $SU(5)$ is in trouble.
As will be discussed later, also here supersymmetry  
offers solutions
for both this fine-tuning and the hierarchy problem.

\section{Relations between Quark and Lepton Masses }
The Higgs 5-plet $\Phi_5$ can be used to generate
fermion masses.
Since the $\overline{5}$-plet of the matter fields contains
both leptons and down-type quarks, their masses are
related, while the up-type quark masses are free parameters.
At the GUT scale one expects:
\bqa
m_d&=&m_e\\
m_s&=&m_\mu\\
m_b&=&m_\tau  \label{mbt}
\eqa
Unfortunately the masses of the light quarks have large 
uncertainties from
the binding energies in the hadrons,
but the b-quark mass can be correctly predicted from the
$\tau$-mass after including radiative corrections (see
fig. \ref{f32} for typical graphs).

Since the corrections from graphs involving  the strong coupling
constant $\alpha_s$ are dominant,
one expects in first order\cite{bmas}
\bq \frac{m_b}{m_\tau}
={\cal{O}}\left(\frac{\alpha_s(m_b)}
{\alpha_s(M_X)}\right)={\cal{O}} (3) \eq
More precise formulae are given in the appendix
and will be used in the last chapter in a quantitative
analysis, since the b-quark mass   gives a rather
strong constraint on the evolution of the couplings
and through  the radiative corrections involving
the Yukawa couplings on the top quark mass.

\chapter{Supersymmetry}\label{ch4}
\section{Motivation }\label{s41}
Supersymmetry\cite{susybook,susyrev} presupposes a symmetry
between fermions and bosons,
which can be realized in nature only if one assumes each particle
with spin j has a supersymmetric partner with
spin j-1/2. This leads to a doubling of the particle
spectrum (see table \ref{t41}), which are
assigned to two supermultiplets: the vector multiplet 
for the gauge bosons and the chiral multiplet for the matter fields.
 Unfortunately
the supersymmetric particles or ``sparticles'' have
not been observed so far, so either supersymmetry
is an elegant idea, which has nothing to do
with reality, or
supersymmetry is not an exact symmetry, in which case the
sparticles can be heavier than the particles.
\begin{table} 
\begin{center} \begin{displaymath}
\begin{array}{|c|c||c|c|}  \hline
  \multicolumn{2}{|c||}{\mbox{VECTOR MULTIPLET }} 
&\multicolumn{2}{|c|}{\mbox{CHIRAL MULTIPLET}} \\
\hline\hline
\rule{0cm}{0.6cm}
  J=1 & J=1/2&J=1/2&J=0\\
\hline
\rule{0cm}{0.6cm}
%g& \~{g} &Q_L,U_L^C,D_L^C&
g& \tilde{g} &Q_L,U_L^C,D_L^C&
\tilde{Q}_L,\tilde{U}_L^C,\tilde{D}_L^C\\
\rule{0cm}{0.6cm}
 W^\pm,W^0&\tilde{W}^\pm,\tilde{W}^0&
L_L,E_L^C&\tilde{L}_L,\tilde{E}_L^C\\
\rule{0cm}{0.6cm}
B&\tilde{B}&\tilde{H}_1,\tilde{H}_2&H_1,H_2\\
\hline
\end{array} \end{displaymath}
\end{center}
\caption{Assignment of gauge fields to the vector superfield
and the matter fields to the chiral superfield.}
\label{t41}
  \end{table}
Many people opt for the latter way out, since there are many
good reasons to believe in  supersymmetry:
\begin{itemize}
\item
\underline{\bf SUSY solves the fine-tuning problem }\\
As mentioned before, the radiative corrections in the $SU(5)$
model have quadratic divergences from the diagrams in
fig. \ref{f28}, which lead to $
\Delta M_H^2 \approx {\cal{O}}( M_X^2)$,
where $M_X$ is a cutoff scale, typically the unification 
scale if no other scales introduce new physics beforehand.

However, in SUSY the loop corrections
contain both fermions (F) and bosons (B) in the loops,
which according to the Feynman rules contribute
with an opposite sign, i.e.
\bq \Delta M_H^2\approx {\cal{O}}(\alpha)~|M_B^2-M_F^2|\approx
{\cal{O}}(10^{-2})M_{SUSY}^2 \label{appr}\eq
where $M_{SUSY}$ is a typical SUSY mass scale.
In other words, the fine-tuning problem disappears, if
the SUSY partners are not too heavy compared with the known 
fermions. An estimate of the required SUSY breaking scale 
can be obtained by considering that
the masses of the weak gauge bosons and Higgs masses are 
both obtained by multiplying   the vacuum expectation 
value of the Higgs field (see previous chapter) with a 
coupling constant, so one expects $M_W\approx M_H$.
Requiring that the radiative corrections
are not much larger than the masses themselves, i.e. 
$\Delta M_W<M_W$, or replacing $M_W$ by $M_H$, 
$\Delta M_H<{\cal{O}} (10^2)$,   yields after substitution 
into eq. \ref{appr}:
\bq
M_{SUSY}\leq 10^3~ {\rm GeV.} \label{susest}\eq
\item
\underline{\bf SUSY offers a solution for the hierarchy problem }\\
The possible explanation for the small ratio 
$M_W^2/M_X^2\approx 10^{-28}$
 is simple in SUSY models:  large
  radiative corrections from the top-quark
Yukawa coupling to the Higgs sector drive one of the Higgs
masses squared negative, thus changing the shape of
the effective potential from the parabolic shape to
the Mexican hat (see fig. \ref{f24}) and triggering electroweak
symmetry breaking\cite{ewbr}.
 Since radiative corrections are logarithmic in energy,
this automatically leads to a large hierarchy between
the scales. In the SM one could invoke
a similar mechanism for the triggering of electroweak
symmetry breaking, but in that case the quadratic
divergences in the radiative corrections would upset
the argument.

In the MSSM the electroweak scale is governed
by the starting values of the parameters  at the 
GUT scale and the top-quark mass.
This   strongly constrains   the SUSY mass spectrum,
as will be discussed in the last chapter.

\begin{figure}
%   HNS  A
\vspace{-3.cm}
\begin{center}
\mbox{\epsfysize=22.cm\epsfxsize=12.cm\epsfbox{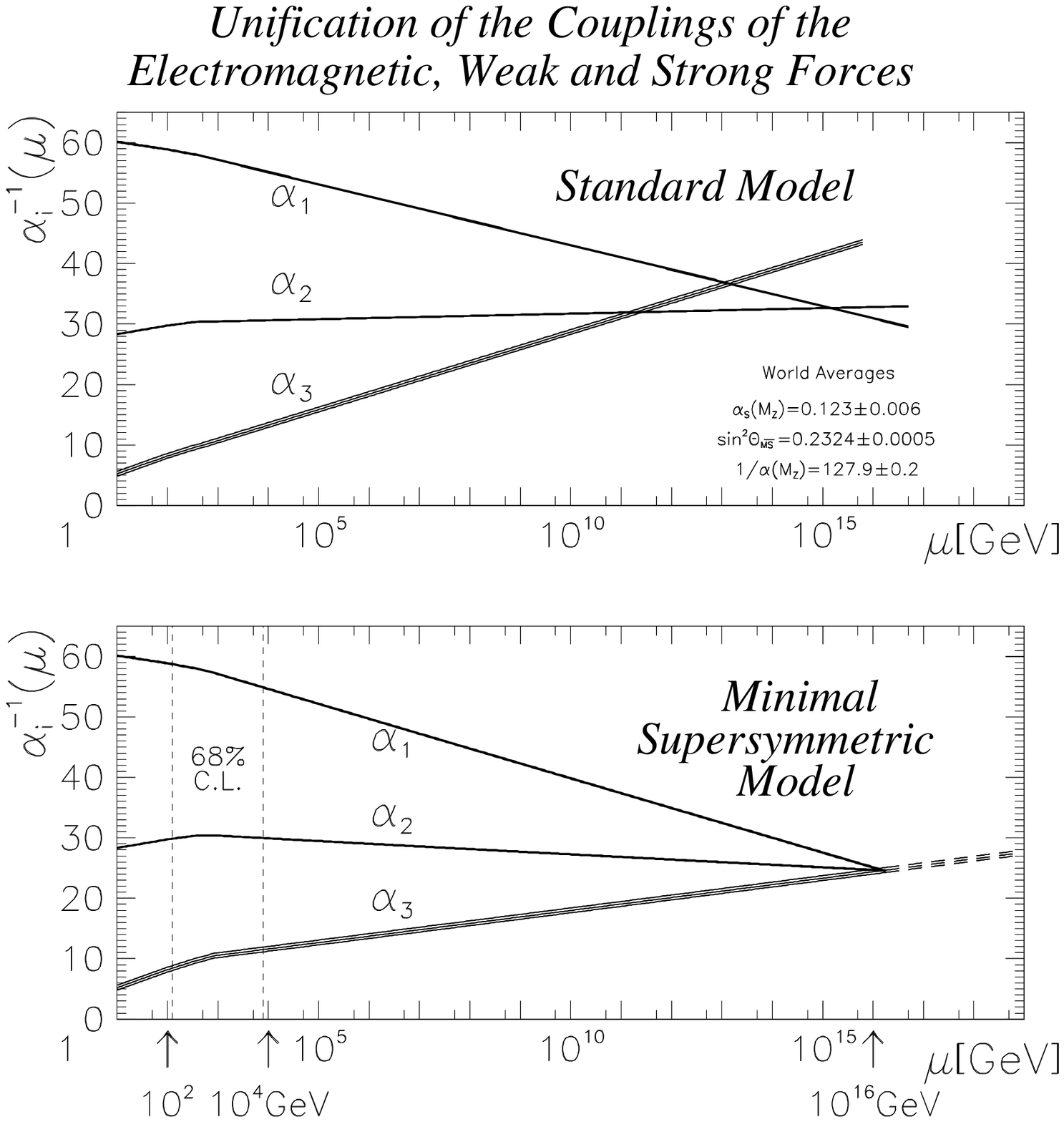}}
%   HNS  E
\caption{Evolution of the inverse of the three coupling
constants in the Standard Model (SM) (top) and in the
supersymmetric extension of the SM (MSSM) (bottom). 
Only in the latter case unification is obtained. 
The SUSY particles are assumed to contribute only 
above the effective SUSY scale $\msusy$ of about  
one TeV, which causes the change in slope in the 
evolution of the couplings. The 68\% C.L. for this 
scale is indicated by the vertical lines (dashed). 
The evolution of the couplings was calculated  in 
second order (see section A.2 of  the appendix with 
the constants $\beta_i$ and $\beta_{ij}$ calculated for the
 MSSM   above $\msusy$
in the bottom part and for the SM elsewhere). The thickness of the lines
represents the error in the coupling           constants. }
\label{f41}
\end{center}
\end{figure}
\begin{figure}[tb]
%   HNS  A
\vspace*{0.2cm}
\begin{center}
%\epsffile[70 480 560 750]{f42.eps}
\mbox{\epsfysize=12.cm\epsfxsize=16.cm\epsfbox{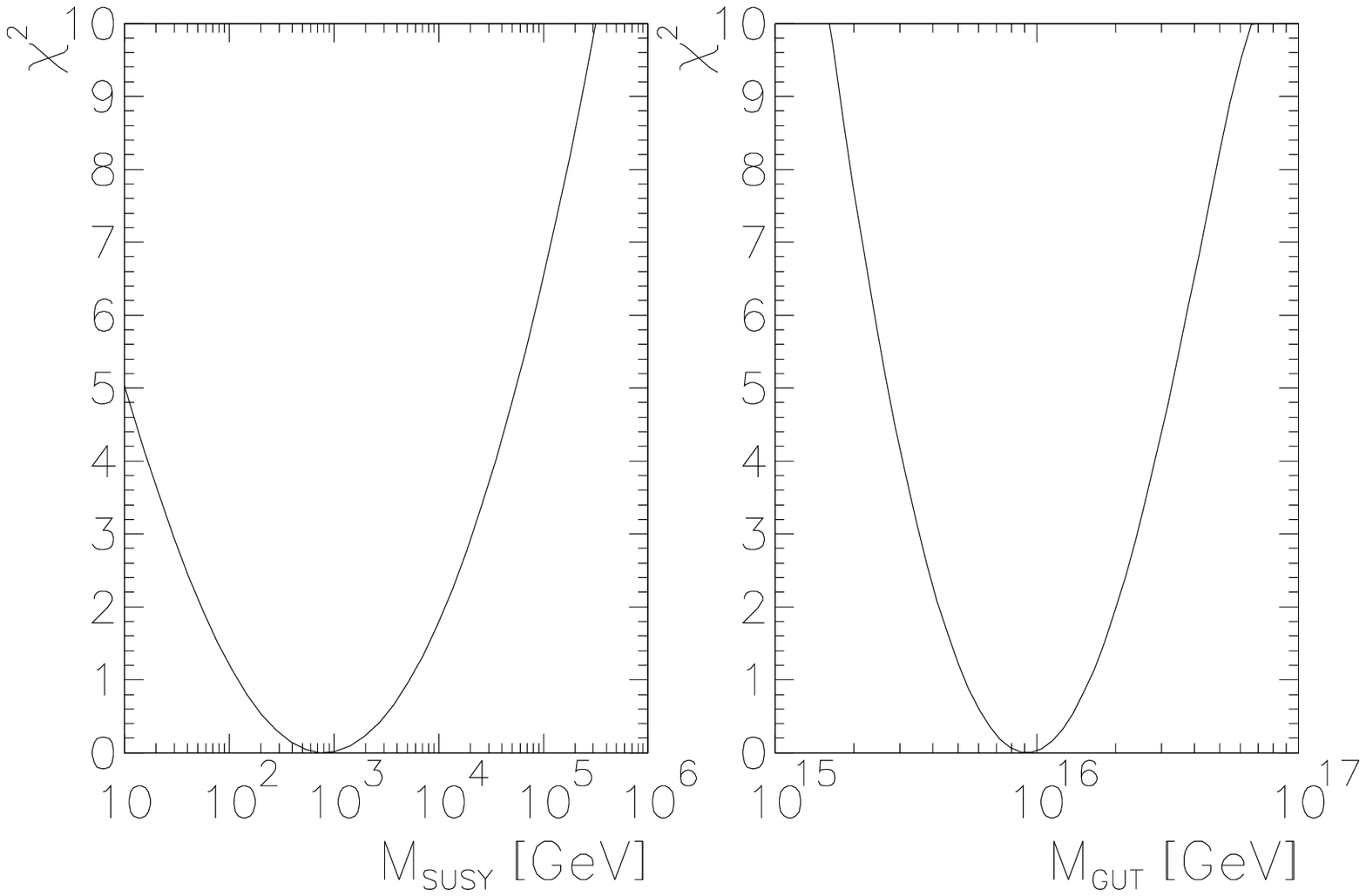}}
\vspace*{-6.cm}
%   HNS  E
\caption{ $\chi^2$ distribution for $\msusy$ and$\mgut$.}
\label{f42}
\end{center}
\end{figure}
\item
\underline{\bf SUSY yields unification of the coupling constants }\\
After the precise measurements of the $\su$ coupling
constants, the possibility of coupling constant unification
within the SM could be excluded, since after extrapolation
to high energies the three coupling constants would
not meet in a single point.
This is demonstrated in the upper part of 
fig. \ref{f41}, which
shows the evolution of the inverse of the couplings
as function of the logarithm of the energy. In this
presentation the evolution becomes a straight line in
first order, as is apparent from the solution
of the RGE (eqn. \ref{srgeI}).
The second order corrections, which have been included
in fig. \ref{f41}  by using eqs.  \ref{itera} from the 
appendix, are so small, that they cause no
visible deviation from a straight line.

A single unification point is excluded by more than
8 standard deviations.
The curve $1/\alpha_3$ meets the crossing point of the
other two coupling constants only for a starting
value at $\alpha_s(M_Z)$ = 0.07, while the measured value
is $0.12\pm 0.006$\cite{lep}.
This is an exciting result, since it means unification
can only be obtained, if new physics enters
between the electroweak and the Planck scale!

It turns out that within the SUSY model perfect
unification can be  obtained  if the SUSY masses are of the
order of one TeV. This is shown in the bottom part of 
fig. \ref{f41}; the SUSY particles are assumed to 
contribute effectively to the running of the coupling 
constants only for energies above the typical SUSY mass 
scale, which causes the change in the slope of the lines 
near one TeV.  From a fit requiring unification
one finds for the breakpoint $\msusy$ and the unification 
point $\mgut$\cite{abfI,fur}:
\bqa \msusy&=&10^{3.4~\pm~ 0.9~
 ~\pm ~0.4 } ~\gev \label{msus}\\
\mgut&=&10^{15.8~\pm ~0.3~\pm ~0.1}~ \gev \\
\agut^{-1}&=&26.3~\pm~1.9~\pm ~1.0,  \eqa
where $\agut\equiv g_5^2/4\pi$.
The first error originates from the uncertainty in the coupling constant,
while the second error is due to the uncertainty in the
mass splittings between the SUSY particles. 
The $\chi^2$ distributions of $\msusy$ and $\mgut$ 
for the fit in the bottom part 
of fig. \ref{f41}  are shown in fig. \ref{f42}. These figures are an update 
of the published figures using the 
newest values of the coupling constants,
as shown in the figure\cite{fur}.

Note that
the   parametrisation of the SUSY
mass spectrum with a
single mass scale is not adequate and leads
to uncertainties.
However, the errors
in the coupling constants (mainly in $\as$) are large
and the uncertainties from mass splittings between
the sparticles are more than a factor two smaller (see eq. \ref{msus}).
In the last chapter the unification including a more
detailed treatment of the mass splittings will be studied.

One can ask: {\it 'What is the significance of this observation?}
For many people it was the first ``evidence'' for
supersymmetry,
especially  since $\msusy$ was found in the
range where the fine-tuning problem does
not reappear (see eq. \ref{susest}).
Consequently    the results triggered a revival of 
the interest in SUSY, as was apparent from the fact that
ref. \cite{abf} with the $\chi^2$ fit of the unification of the
 coupling constants, as exemplified in figs. \ref{f41} and \ref{f42},
 reached  the Top-Ten of the citation list, thus leading to 
discussions in practically all
popular journals\cite{popular}. 

Non-SUSY enthusiasts 
were considering unification obvious:
  with a total of three free parameters ($\mgut,
~\agut$ and $\msusy$)
and three equations one can naively always find a solution.
The latter statement is certainly not true:
searching for other types of new physics with
the masses as free particles yields only rarely
unification, especially if one requires in addition
that the unification scale is above $10^{15}$ GeV in order
to be consistent with the proton lifetime limits and below
the Planck scale in order to be in the regime where
gravity can be neglected. From the 1600 models tried,
only a handful  yielded unification\cite{abfI}. 
The reason is simple:
introducing new particles usually alters all three
couplings simultaneously, thus giving rise to strong
correlations between the slopes of the three lines.
For example, adding a fourth family of particles with
an arbitrary mass will never yield unification, since
it changes the slopes of all three coupling by the same
amount, so if with three families unification cannot
be obtained, it will not work with four families either,
even if one has an additional free parameter!
Nevertheless,
unification does not prove supersymmetry, it only
gives an interesting hint. The real proof would be the observation
of the sparticles.
\item
\underline{\bf Unification with gravity }\\
The space-time symmetry group is the Poincar\'e group.
Requiring local gauge invariance under the transformations
of this group leads to the Einstein theory of gravitation.
Localizing  both the internal and the space-time symmetry groups
yields   the Yang-Mills gauge fields and the gravitational fields.
This paves the way for the unification of gravity with the strong
and electroweak interactions. The only non-trivial unification
of an internal symmetry and the space-time symmetry group
is the supersymmetry group, so supersymmetric theories
automatically include gravity\cite{supergrav}.
Unfortunately supergravity models
are inherently non-renormalizible,
which prevents up to now clear
predictions.
Nevertheless, the spontaneous symmetry
breaking of supergravity is important
for the low energy spectrum of supersymmetry\cite{supergrav1}.
The most common scenario is the {\it 	hidden sector} 
scenario\cite{weinberg},
in which one postulates two sectors of fields: the visible 
sector containing all the particles of the GUT's described 
before and the hidden sector, which contains fields which 
lead to symmetry breaking of supersymmetry at some large 
scale $\Lambda_{SUSY}$. One assumes that none of the 
fields in the hidden sector contains quantum numbers 
of the visible sector,
so the two sectors only communicate via gravitational 
interactions. Consequently,
the effective scale of supersymmetry breaking in the 
visible sector is suppressed by a power of the Planck scale, i.e.
\bq \msusy\approx  \frac{\Lambda^n_{SUSY}}{M^{n-1}_{Planck}},\eq
where $n$ is model-dependent (e.g. $n=2$ in the Polonyi model). 
Thus the SUSY breaking scale can be  large, above $10^{10}$ GeV,
while still producing a small breaking
scale in the visible sector.
 In this case the fine-tuning problem
can be avoided in a natural way and it is
gratifying to see that the first experimental hints for $\msusy$ are
indeed in the mass range consistent with
eq. \ref{susest}.

The hidden sector scenario  leads to an effective low-energy 
theory with explicit soft breaking terms, where soft implies
that no new quadratic divergences are generated\cite{soft}. 
 The soft-breaking terms in string-inspired supergravity models 
have been studied recently in refs. \cite{superstring}.
 A final theory, which simultaneously solves the cosmological 
constant problem\cite{coscon}
and explains the origin of supersymmetry breaking, needs 
certainly a better understanding of  superstring theory.
\item
\underline{\bf The unification scale in SUSY is large }\\
As discussed in chapter \ref{ch3}, the limits on the 
proton lifetime require the
unification scale to be
above $10^{15}$ GeV,  which is the case for the MSSM.
In addition, one has to consider proton decay via graphs
of the type shown in fig. \ref{f43}.
These yield a strong constraint on the mixing in
the Higgs sector\cite{arn}, as will be discussed in detail
in the last chapter.
\item
\underline{\bf Prediction of dark matter }\\
The lightest supersymmetric particle (LSP) cannot decay
into normal matter, because of R-parity conservation
(see the next section for a definition of R-parity).
In addition R-parity forbids a coupling
between the LSP and normal matter. 

Consequently,
the LSP is an ideal candidate for dark matter\cite{lspdark}, which
is believed to account for a large fraction of  all  mass
in the universe (see next chapter).
The mass of the dark matter particles  is expected to be below
one TeV\cite{dimo}.
\end{itemize}

\begin{figure}[tb]
%   HNS  A
\hspace*{1.7cm}
\epsffile[5 65 350 145]{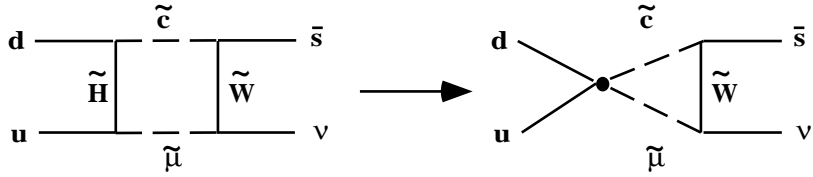}
%   HNS  E
\caption{ Examples of  proton decay in the minimal supersymmetric model
via wino and Higgsino exchange.}
\label{f43}
\end{figure}

\section{SUSY interactions }
\label{s42}
The quantum numbers and the gauge couplings
 of the particles and sparticles have to
be the same, since they belong to the same multiplet
structure.

The interaction of the sparticles with normal matter
is governed by a new {\it multiplicative}
quantum number called R-parity, which is needed
in order to prevent baryon- and lepton number
violation. Remember that quarks, leptons and Higgses are
all contained in the same
chiral supermultiplet, which allows couplings between
quarks and leptons.
Such transitions, which could lead to rapid proton decay,
are not observed in nature.
Therefore, the SM particles are assigned a positive
R-parity and the supersymmetric partners are R-odd.
Requiring R-parity conservation implies that:
\begin{itemize}
\item
sparticles can be produced only  in pairs
\item
the lightest supersymmetric particle
is stable, since its decay into normal matter would change
R-parity.
\item
the interactions of particles and sparticles
can be different.  For example, the photon couples to
electron-positron pairs, but the photino does
not couple to
selectron-spositron pairs, since in the latter case
the R-parity would change from -1 to +1.
\end{itemize}

\section{The SUSY Mass Spectrum }
\label{s43}
Obviously SUSY cannot be an exact symmetry of nature; or
else the supersymmetric partners would have the
same mass as the normal particles.
As mentioned above, the supersymmetric partners should
be not too heavy, since otherwise the hierarchy problem
reappears.

Furthermore, if one requires that the breaking terms 
do not introduce quadratic divergences, only the so-called soft
breaking terms are allowed\cite{soft}.

Using the supergravity inspired
breaking terms, which assume a common mass
$m_{1/2}$ for the gauginos and another common mass $m_0$
for the scalars, leads to the following  breaking term in the Lagrangian 
(in the notation of ref. \cite{barb}):
\begin{eqnarray} {\cal L}_{Breaking} & = &
-m_0^2\sum_{i}^{}|\varphi_i|^2-m_{1/2}\sum_{\alpha}^{}\lambda_\alpha
 \lambda_\alpha \label{2} \\ & -& 
   Am_0\left[h^u_{ab}Q_aU^c_bH_2+h^d_{ab}Q_aD^c_bH_1+
 h^e_{ab}L_aE^c_bH_1\right] - Bm_0\left[\mu H_1H_2\right].
    \end{eqnarray}

Here

\begin{tabular}{ll}
$h^{u,d,e}_{ab}$ & are the Yukawa couplings, \ $a,b =1,2,3$ run over the
generations \\
 $Q_a$ & are the SU(2) doublet quark fields \\
$U_a^c$ & are the SU(2) singlet charge-conjugated up-quark fields \\
$D_b^c$ & are the SU(2) singlet charge-conjugated down-quark fields \\
$L_a$ & are the SU(2) doublet lepton fields \\
$E_a^c$ & are the SU(2) singlet charge-conjugated lepton fields \\
$H_{1,2}$ & are the SU(2) doublet Higgs fields \\ $\varphi_i$ & are all
scalar fields \\ $\lambda_\alpha $ & are the gaugino fields \end{tabular}

The last two terms in ${\cal{L}}_{Breaking}$ originate from the
cubic and quadratic terms in the superpotential with
A, B and $\mu$ as free parameters.
In total we  now have three couplings $\alpha_i$ and five mass parameters 
$$m_0,~m_{1/2},~ \mu(t),~A(t),~B(t) $$
with the following boundary conditions at $\mgut$ $(t=0)$:
\bqa
 {\rm scalars:}&&
 \tilde{m}^2_Q=\tilde{m}^2_U=
\tilde{m}^2_D=\tilde{m}^2_L=\tilde{m}^2_E=
m_0^2;\\
 {\rm gauginos:}&&
 M_i=m_{1/2}, \ \ \ i=1,2,3;\\
 {\rm couplings:}&&
 \tilde{\alpha}_i(0)=\tilde{\alpha}_{GUT},\ \ \ i=1,2,3 .\eqa
 Here $M_1$, $M_2$, and $ M_3$ are the  gauginos masses of the $ U(1)$, 
$SU(2)$ and $SU(3)$ groups.  In $N=1$ supergravity one expects at the 
Planck scale $B=A-1$.
With these parameters and the initial conditions
at the GUT scale the masses of all SUSY
particles can be calculated via the renormalization
group equations.
\section{Squarks and Sleptons }
\label{s44}
 The squark and slepton masses all have the same value
at the GUT scale. However, in contrast to the
leptons, the squarks get additional
radiative corrections from virtual gluons
(like the ones in fig. \ref{f32} for quarks), which makes them heavier than the
sleptons at low energies.
These radiative corrections can be calculated from the
corresponding RGE, which have been assembled in the
appendix. The solutions are:
\begin{eqnarray}
\tilde{m}^2_{E_{L}}(t=66)&=&m^2_0+0.52m^2_{1/2}
-0.27\cos(2\beta)M_Z^2\label{mslep}\\
\tilde{m}^2_{\nu_{L}}(t=66)  &=&m^2_0+0.52m^2_{1/2}
+0.5\cos(2\beta) M_Z^2\\
   \tilde{m}^2_{E_{R}}(t=66) 
&=&m^2_0+0.15 m^2_{1/2} -0.23\cos(2\beta)M_Z^2 \\
\tilde{m}^2_{U_{L}}(t=66) 
&=&m^2_0+6.6m^2_{1/2}
+0.35\cos(2\beta)M_Z^2 \label{msq}\\
\tilde{m}^2_{D_{L}}(t=66) 
&=&m^2_0+6.6m^2_{1/2}
-0.42\cos(2\beta)M_Z^2\\
\tilde{m}^2_{U_{R}}(t=66) 
&=&m^2_0+6.2m^2_{1/2}+0.15\cos(2\beta)M_Z^2\\
\tilde{m}^2_{D_{R}}(t=66) &=&
m^2_0+6.1m^2_{1/2}-0.07\cos(2\beta)M_Z^2,
\label{sqsl}
\end{eqnarray}
where $\beta$ is the mixing angle between
the two Higgs doublets, which will be defined more precisely in 
section \ref{s46}. 
The coefficients depend on the couplings 
as shown explicitly in the appendix.
They were calculated for the parameters
from the typical fit shown in table \ref{t61} 
($\alpha_{GUT}=1/24.3$, $\mgut=2.0\cdot10^{16}$ GeV and $\sws=0.2324$).
For the third generation the Yukawa coupling is
not necessarily negligible.
If one includes only the correction from the
top Yukawa coupling $Y_t$\footnote{For large values
of the mixing angle $\tan\beta$ in the Higgs sector, the b-quark Yukawa coupling
can become large too. However, since the
limits on the proton lifetime limit
$\tan\beta$ to rather small values (see last chapter),
this option is not further considered here.}, one finds:
\begin{eqnarray}
\tilde{m}^2_{b_{R}}(t=66) &=&\tilde{m}^2_{D_{R}} \\
\tilde{m}^2_{b_{L}}(t=66) &=&\tilde{m}^2_{D_{L}}-
0.48m^2_0-1.21m^2_{1/2} \\
\tilde{m}^2_{t_{R}}(t=66) &=&\tilde{m}^2_{U_{R}}
+m_t^2 -0.96m^2_0-2.42m^2_{1/2}  \\
\tilde{m}^2_{t_{L}}(t=66) &=&\tilde{m}^2_{U_{L}} 
+m_t^2 -0.48m^2_0-1.21m^2_{1/2}
\label{blr} 
\end{eqnarray}
The numerical factors have been calculated
for $A_t(0)=0$ and the explicit dependence
on the couplings can be found in the appendix.
Note that only the left-handed b-quark gets corrections from
the top-quark Yukawa coupling through
a loop with a charged Higgsino and a top-quark.
The subscripts $L$ or $R$ do not
indicate the helicity, since the squarks
and sleptons have no spin.
The labels just indicate in analogy
 to the non-SUSY particles, if they
are $SU(2)$ doublets or singlets.
The mass eigenstates are mixtures
of the
$L$ and $R$ weak interaction states. Since the mixing is
proportional to the
Yukawa coupling, we will only consider the mixing for the top quarks.
After mixing the mass eigenstates are: 
 (using the same numerical input as for the light  quarks):
\begin{eqnarray}
\tilde{m}^2_{t_{1,2}}(t=66) &=&
\frac{1}{2}\left[\tilde{m}^2_{t_{L}}+\tilde{m}^2_{t_{R}} \pm
\sqrt{(\tilde{m}^2_{t_{L}}-\tilde{m}^2_{t_{R}}
)^2+4m^2_t(A_tm_0 + \mu/\tan\beta )^2}
\right]
 \nn
 &\approx& 
\frac{1}{2}\left[0.6m_0^2+9.2m_{1/2}^2+2m_t^2-0.19\cos(2\beta)M_Z^2 \right]
\nn
 & &\pm
\frac{1}{2}\sqrt{\left[1.6m_{1/2}^2+0.5m_0^2-0.5\cos(2\beta)M_Z^2\right]^2+
 4m_t^2(A_tm_0 + \mu/\tan\beta )^2}\nn. \label{t12}
\end{eqnarray}
 where   the values of $A_t$ and $\mu $ at the  weak 
 scale can be calculated as:
 \bq A_t(M_Z)=4.6A_t(0)+1.7\frac{m_{1/2}}{m_0} 	
   \eq
\bq \mu(M_Z)=0.63\mu(0) \eq
%with $A_t(0)=0$.

Note that for large values of $A_t(0)$ or $\mu$ combined with a  small 
$\tan\beta$ the splitting becomes 
large and  
one of the stop masses can become very
small; since the stop mass lower limit is
about 46 GeV\cite{reino}, this yields
a constraint on the possible values
of $m_t,~\mha,~\tb$ and $\mu$.
\begin{figure}
\begin{center}
%  HNS   A
\mbox{\epsfysize=18.cm\epsfxsize=15.cm\epsfbox{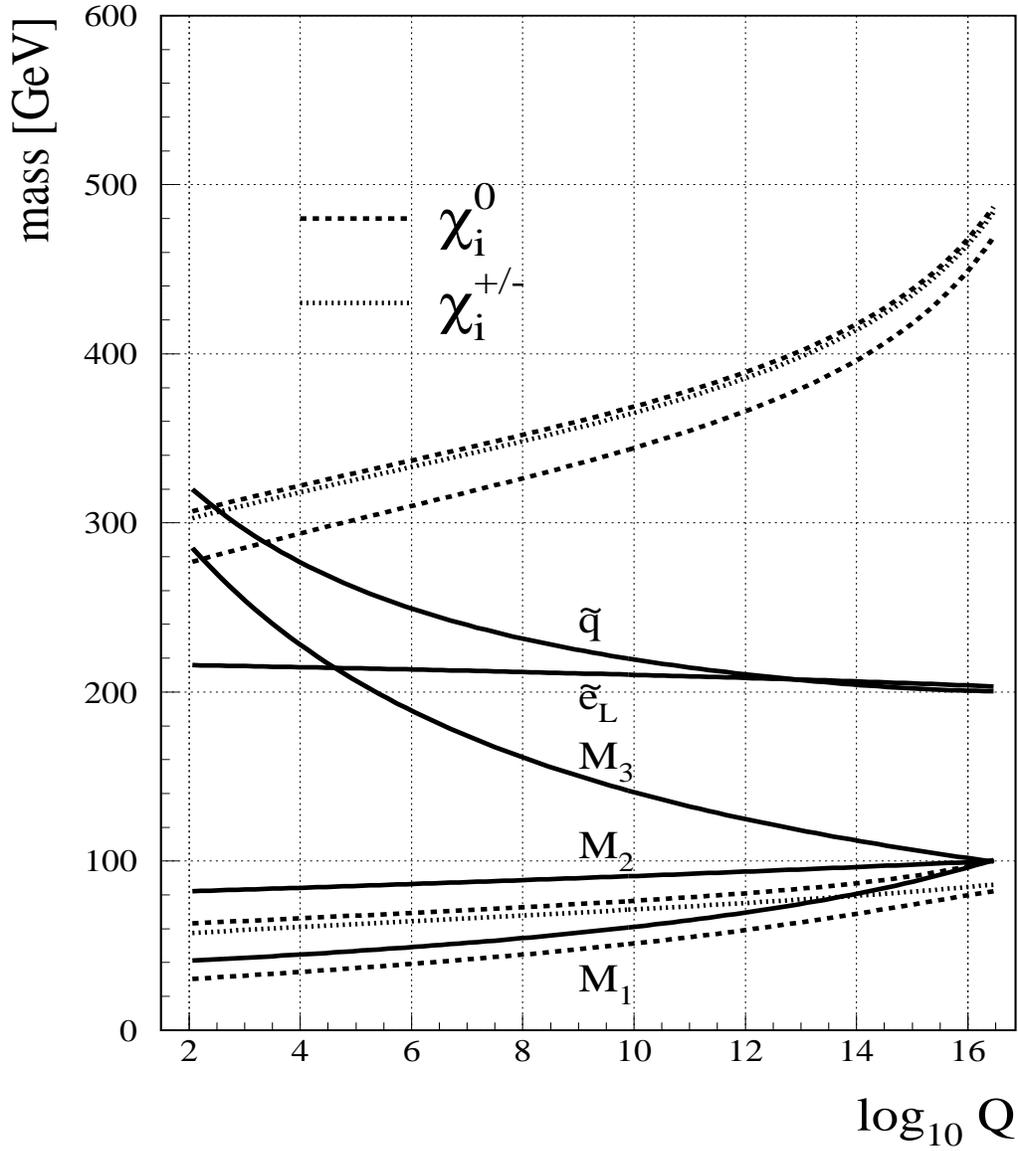}}
%  HNS   E
\caption{Typical running of the squark ($\tilde{q}$), 
slepton ($\tilde{e}_L$), and  gaugino ($M_1,~M_2,~ M_3$) 
masses (solid lines). The dashed lines indicate the
running of the four neutralinos and two charginos.
}
\label{f44}
\end{center}
\end{figure}

%\begin{figure}
%\begin{center}
%  HNS   A
%\mbox{\epsfysize=14.cm\epsfxsize=14.cm\epsfbox{f46.eps}}
%  HNS   E
%\caption{Evolution of the mass parameters 
%in the Higgs potential for small values
%of $\mu$ and $\mha$ (left-hand side, $\mha 
%(\mu)\approx 200(600)$ GeV) and large
%values (right-hand side¸ $\mha 
%(\mu)\approx 4000(3000)$ GeV). In the 
%latter case the corrections from the top 
%Yukawa couplings are so large
%that $m_2$  becomes negative. However, in 
%both cases the determinant %$m_1^2~m_2^2-m_3^4$ becomes negative (top
%curves),
%thus causing spontaneous symmetry %breaking.
%For large values of $\mha$ and $(\mu)$
%the one-loop correction to $\mz$ become
%very large (top right-hand corner).}
%
%\label{f46}
%\end{center}
%\end{figure}

\section{Charginos and Neutralinos }
\label{s45}
The solutions of the RGE group equations for the gaugino masses
are simple:
\bq
M_i(t)=\frac{\tilde{\alpha}_i(t)}{\tilde{\alpha}_i(0)}m_{1/2}.
\eq
Numerically at the weak scale $(t=2\ln(\mgut/\mz)=66)$ one finds 
(see fig. \ref{f44}):
\bqa
M_3(\tilde{g})&\approx& 2.7m_{1/2},\\ 
 M_2(M_Z)&\approx& 0.8m_{1/2},\\  
 M_1(M_Z)&\approx& 0.4m_{1/2}.\label{gaugino}\eqa
Since the gluinos obtain corrections from
the strong coupling constant $\alpha_3$, they grow
heavier than the gauginos of the $\su$ group.

The  calculation of the
mass eigenstates is more complicated, since both Higgsinos and gauginos
are spin 1/2 particles, so the mass eigenstates
are in general mixtures of the weak interaction eigenstates.
  The mixing of
the Higgsinos and gauginos, whose mass eigenstates
are called charginos and neutralinos for the charged
and neutral fields, respectively,
  can be parametrised by  the following Lagrangian:
 $$ {\cal L}_{Gaugino-Higgsino}=
 -\frac{1}{2}M_3\bar{\lambda}_a\lambda_a
 -\frac{1}{2}\bar{\chi}M^{(0)}\chi -(\bar{\psi}M^{(c)}\psi + h.c.)  $$
where $\lambda_a , a=1,2,\ldots ,8,$ are the Majorana gluino fields and
$$ \chi = \left(\begin{array}{c}\tilde{B} \\ \tilde{W}^3 \\
\tilde{H}^0_1 \\ \tilde{H}^0_2
\end{array}\right), \ \ \ \psi = \left( \begin{array}{c}
\tilde{W}^{+} \\ \tilde{H}^{+}
\end{array}\right),$$
are  the Majorana neutralino and Dirac chargino fields,
 respectively.
Here all the terms in the Lagrangian were assembled into
matrix notation (similarly to the mass matrix
for the mixing between $B$ and $W^0$ in the SM, eq. \ref{mas}).
The mass matrices can be written as\cite{susyrev}:
\bq M^{(0)}=\left(
\begin{array}{cccc}
M_1 & 0 & -M_Z\cos\beta \sw & M_Z\sin\beta \sw \\
0 & M_2 & M_Z\cos\beta \cw   & -M_Z\sin\beta \cw  \\
-M_Z\cos\beta \sw & M_Z\cos\beta \cw  & 0 & -\mu \\
M_Z\sin\beta \sw & -M_Z\sin\beta \cw  & -\mu & 0
\end{array} \right)\label{neutmat}\eq
\bq M^{(c)}=\left(
\begin{array}{cc}
M_2 & \sqrt{2}M_W\sin\beta \\ \sqrt{2}M_W\cos\beta & \mu
\end{array} \right)\label{charmat} \eq
%Note that the  off-diagonal terms, like
%$M_Z\cos\beta\sin\theta_W\propto gv_1$, %are the same as in the SM 
%(eq. \ref{mas}).
The last matrix leads to two chargino eigenstates $\tilde{\chi}_{1,2}^{\pm}$
with mass eigenvalues
\bq M^2_{1,2}=\frac{1}{2}\left[M^2_2+\mu^2+2M^2_W \mp
\sqrt{(M^2_2-\mu^2)^2+4M^4_W\cos^22\beta +4M^2_W(M^2_2+\mu^2+2M_2\mu
\sin 2\beta )}\right].\eq
The dependence on the parameters at the GUT
scale can be estimated by substituting for $M_2$ and $\mu$  
their values at the weak scale:   
$M_2(M_Z)\approx 0.8m_{1/2}$ and $\mu(M_Z) \approx 0.63\mu(0)$.
In the case favoured by the fit discussed
in chapter \ref{ch6} one finds $\mu >>M_2\approx M_Z$, in which case the
charginos eigenstates are approximately
$M_2$ and $\mu $.

The four neutralino mass eigenstates   are denoted by
$\tilde{\chi}_i^0(i=1,2,3,4)$ with masses
 $M_{\tilde{\chi}_1^0}\leq \cdot\cdot\cdot\leq
 M_{\tilde{\chi}_4^0}$.
 The sign of the  mass eigenvalue corresponds
to the CP quantum number of the Majorana neutralino state.

In the limiting case $M_1,M_2,\mu >>M_Z$ one can neglect the
off-diagonal elements and the mass eigenstates become:
\bq \tilde{\chi}_i^0=[\tilde{B},\tilde{W}_3,
\frac{1}{\sqrt{2}}(\tilde{H}_1-\tilde{H}_2),
\frac{1}{\sqrt{2}}(\tilde{H}_1+\tilde{H}_2)] \eq
with eigenvalues $|M_1|,|M_2|, |\mu|,$ and $|\mu|$,
respectively.
In other words, the bino and neutral wino do not
mix with each other nor with the Higgsino eigenstates
in this limiting case.
As we will see in a quantitative analysis, the data
indeed prefer  $M_1,M_2,\mu > M_Z$, so the LSP is bino-like,
which has consequences for dark matter searches.

\section{Higgs Sector }\label{s46}

The Higgs sector of the SUSY model has to be
extended with respect to the one of the SM for two
reasons:
\begin{itemize}
\item
the Higgsinos have spin 1/2, which implies they contribute
to the gauge anomaly, unless one has pairs of Higgsinos
with opposite hypercharge, so in addition to the
 Higgs doublet with $Y_W$=1 one needs a second one with
$Y_W$=-1:

\bq H_1(1,2,-1)= \left(\begin{array}{c}H^0_1 \\
H^-_1\end{array}\right), \ \ \
 H_2(1,2,1)= \left(\begin{array}{c}H^+_2 \\
H^0_2\end{array}\right)\eq
\item
The introduction of the second Higgs doublet solves
simultaneously the problem that a single doublet
can give mass to only either the up- or down-type
quarks, as is apparent from the fact that only
the neutral components have a non zero vev,
since else the vacuum would not be neutral. So one
can write:
\bq <H_1>= \left(\begin{array}{c}v_1 \\
0\end{array}\right), \ \ \
 <H_2>= \left(\begin{array}{c}0 \\
v_2\end{array}\right).\label{vev}\eq
% (see Eq. \ref{hmin}).
 In the SM the conjugate field can give mass to the other
type. However, supersymmetry is a spin-symmetry,
in which the matter - and Higgs fields are contained
in the same chiral supermultiplet. This forbids
couplings between matter fields and conjugate Higgs fields.
With the two Higgs fields introduced above,
$H_1$ generates mass to the down-type matter fields, while
$H_2$ generates mass for the up-type matter fields.
\end{itemize}
The supersymmetric model with two Higgs doublets is
called the Minimal Supersymmetric Standard Model (MSSM).
The mass spectrum can be analyzed by
considering again the expansion around the vacuum
expectation value, given by eq. \ref{vev}:
\bq H_1= \left(\ba{c}v_1 +\frac{1}{\sqrt{2}}
\left(H^0\cos\alpha-h^0\sin\alpha
+iA^0\sin\beta-iG^0\sin\beta\right)\\
H^-\sin\beta-G^-\cos\beta\ea\right)\eq
\bq H_2= \left(\ba{c}
H^+\cos\beta+G^+\sin\beta\\
v_2 +\frac{1}{\sqrt{2}}
\left(H^0\sin\alpha+h^0\cos\alpha
+iA^0\cos\beta+iG^0\sin\beta\right)
\ea\right)\eq
Here $H ,h $ and $A $ represent
the fluctuations around the vacuum
corresponding to the real Higgs fields, while
the $G$'s represent the Goldstone fields, which
disappear in exchange for the longitudinal
polarization components of the heavy gauge bosons. 
The imaginary  and real sectors do not mix, since
they have different CP-eigenvalues;
$\alpha$ and $\beta$
are the mixing angles in these different sectors.
The mass eigenvalues of the imaginary components
are CP-odd, so one is left with 2 neutral CP-even
Higgs bosons $H^0$ and $h^0$,
 1 CP-odd neutral Higgs bosons $A^0$, and 2 CP-even
charged Higgs bosons.

The
complete tree level potential for the neutral Higgs sector, 
assuming colour and charge conservation, reads:
\bq V(H_1^0,H_2^0)= %\frac{g^2}{2}|H_1^0^*H_2^0|^2+
\frac{g^2+g^{'2}}{8}(|H_1^0|^2-|H_2^0|^2)^2+
m^2_1|H_1^0|^2+m^2_2|H_2^0|^2-m^2_3(H_1^0H_2^0+h.c.)\label{vhiggs}\eq
Note that in comparison with the potential in the SM,
the first terms
do  not have   arbitrary coefficients anymore, but these
are restricted to be the gauge coupling constants
in supersymmetry, again because the Higgses belong
to the same chiral multiplet as the matter
fields\footnote{In principle one should consider
the running of the gauge couplings between the
electroweak scale and the mass scale of the Higgs bosons.
However, since the Higgs bosons are expected to be
below the TeV mass scale, this running
is small and can be neglected, if one considers
all other sources of uncertainty in the MSSM.}.
The last three terms in the potential arise from the
soft breaking terms  with the following boundary
conditions at the GUT scale:
\bq m_1^2(0)=m^2_2(0)=\mu(0)^2 +m_0^2, \ m^2_3(0)=-B\mu(0)m_0,
\label{m0} \eq
where $\mu(0)$ is the value of $\mu$ at the GUT scale.
Since $\mu$ generates mass for the Higgsinos,
one expects $\mu$ to be small compared with the GUT scale.
A low $\mu$ ~value   can be obtained dynamically,
if one adds a singlet scalar field to the MSSM.
 This will not
be considered further. Instead $\mu$ is considered to be
a free parameter to be determined from data
(see  chapter \ref{ch6}).

From the potential one can derive easily the    
five Higgs masses  in terms of these parameters by diagonalization
of the mass matrices:
\bq
M_{ij}^2=\frac{1}{2}\frac{\partial^2 V_H}{\partial\phi_j\partial\phi_j}
\label{mij}\eq
 where $\phi_i$ is a generic notation
for the real or imaginary part of the Higgs
field. Since the Higgs particles
are quantum field oscillations around the minimum, 
eq. \ref{mij} has to be evaluated
at the minimum.
One finds zero masses for the Goldstone bosons.  
These would-be Goldstone bosons
$G^\pm$ and $G^0$ are ``eaten'' by
the $SU(2)$ gauge bosons.
For the masses of the five remaining Higgs particles one finds\cite{susyrev}:\\
CP-odd neutral Higgs $ A$:
 \bq m^2_A = m^2_1+m^2_2.\eq
 Charged Higgses $H^{\pm}$:
\bq m^2_{H^{\pm}}=m^2_A+M^2_W .\eq
CP-even neutral Higgses $H,h$:
\bq m^2_{H,h}=
\frac{1}{2}\left[m^2_A+M^2_Z \pm
\sqrt{(m^2_A+M_Z^2)^2-4m^2_AM_Z^2\cos^22\beta}\right].\eq
By convention $m_H>m_h$.
The mixing angles $\alpha$ and $\beta$
are related by
\bq \tan~2\alpha=-
\frac{m^2_A+M_Z^2}{m^2_A-M_Z^2}\tan~2\beta \eq
  $v_1$ and $v_2$ have been chosen
real and positive, which implies $0\leq\beta\leq\pi/2$.
Furthermore,
the electroweak breaking conditions
 require $\tan\beta >1$, so
\bq \pi/4 < \beta < \pi/2.\label{beta}\eq
From the mass formulae at tree level
one obtains the once celebrated
SUSY mass relations:
\bq m_{H^\pm}\geq M_W \eq
\bq m_h\leq m_A\leq M_H   \eq
\bq m_h\leq M_Z\cos 2\beta\leq M_Z \eq
\bq m_h^2+m_H^2=m_A^2+M_Z^2.\eq
After including radiative corrections
the lightest neutral Higgs
$m_h$ becomes considerably heavier
and these relations are not valid anymore.
The mass formulae including the radiative
corrections are given in the appendix.
\begin{figure}[tb]
%  HNS   A
\epsfysize=5.5cm
\epsfxsize=11.cm
\hspace{2.0cm}
\epsffile[75 120 455 290]{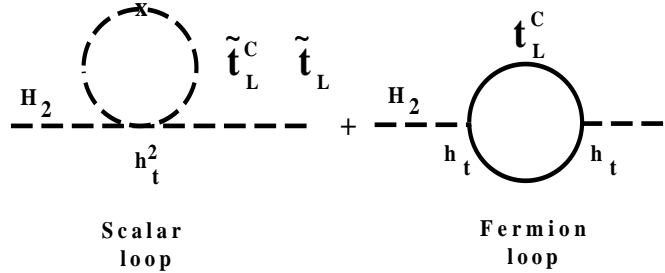}
%  HNS   E
\caption{Corrections to the Higgs self-energy from
Yukawa type interactions.}
\label{f45}
\end{figure}

\section{Electroweak Symmetry Breaking }
\label{s47}
The coupling $\mu$ plays   an important role in the
shape of the potential and consequently
in the pattern of electroweak symmetry breaking, which
occurs if the minimum of the potential is not obtained
for $<H_1>$=$<H_2>$=0. In the SM this condition
could be introduced ad-hoc by requiring the
coefficient of the quadratic term to be negative.
In supersymmetry this term is restricted by the
gauge couplings\cite{ewbr}.  A non-trivial minimum can
only be obtained by the soft breaking terms, 
if the mass matrix for the Higgs sector, given 
by $M_{ij}^2=\frac{\partial^2 V}{\partial H_i\partial H_j}$, 
has a negative eigenvalue. This is obtained if the 
determinant is negative, i.e.
\bq  |m_{3}^2(t)|^2> m_1^2(t)~ m_2^2(t). \label{m1} \eq
In order that the new minimum is below the trivial
minimum with $<H_1>$=$<H_2>$=0, one has to require
in addition $V_H(v_1,v_2)<V_H(0,0)<V_H(\infty,\infty)$, which is
fulfilled if
\bq m_1^2(t)+m_2^2(t)\geq 2|m_{3}^2(t)|.\label{m2} \eq
If one compares eqns. \ref{m1} and \ref{m2}
and notices from eq. \ref{m0} that $m_1=m_2$,
one realizes that these conditions cannot be fulfilled
simultaneously, at least not at the GUT scale.

However,  at lower energies there are
substantial radiative corrections, which can cause
differences between $m_2$ and $m_1$, since the first one
involves mass corrections proportional to  the top 
Yukawa coupling $Y_t(0)$,
while for the latter these corrections are proportional to the
bottom Yukawa coupling. Typical diagrams are shown
in fig. \ref{f45}. From the RGE for the mass parameters   
in the Higgs potential  one finds at the weak scale:
\bqa
\mu^2(t=66)&=&0.40 \mu^2(0)\\
m_1^2(t=66)&=&m_0^2+0.40\mu^2(0)+0.52
m^2_{1/2}\\
m^2_2(t=66)&=&-0.44m_0^2+0.40\mu^2(0) -3.11m^2_{1/2}
\nn&&
-0.09A_t(0)m_0m_{1/2}-0.02A_t(0)^2m_0^2 \label{mh12} \\
m^2_3(t=66)&=&0.63m^2_3(0)+0.04\mu(0)m_{1/2}+0.19A_t(0)m_0\mu(0).
\eqa
The coefficients were evaluated for the
parameters of the   fit to the experimental
data (central column of table 
\ref{t61} in chapter \ref{ch6}).
The explicit dependence of the coefficients on the coupling constants 
is given in the appendix. 
The   coefficients of  the last three terms in $m_2$ depend  on
the top Yukawa coupling. This dependence   disappears
if the  masses of the stop and top quarks
in the diagrams of fig. \ref{f45} are equal. However,   
if the stop mass is heavier, the negative 
contribution of the diagram
with the top quarks dominates; in this 
case $m_2$ decreases much faster than $m_1$
with decreasing energy and the potential
takes the form of a mexican hat, as soon as
conditions \ref{m1} and \ref{m2} are satisfied. 
Since $A(t)$ is expected to be small, the dominant  
negative contribution   is proportional to $m_{1/2}$ (see eq. \ref{mh12}), 
so the electroweak breaking scale
is a sensitive function of both the initial conditions,
the top Yukawa coupling and the gaugino masses.

The minimum  of the potential can be found by
requiring:
\begin{eqnarray*}
\frac{\partial   V}{\partial   |H_1^0|} & = & 2m_1^2v_1-2m_3^2v_2 +
\frac{g^2+ g^{'2}}{2}(v_1^2-v_2^2)v_1 =0\\
%\\
%& & +\frac{3}{8\pi^2}h_t^2\mu(A_t m_0v_2+\mu v_1)
%\frac{f(\tilde{m}^2_{t1})-f(\tilde{m}^2_{t2})}{\tilde{m}^2_{%t1}-
%\tilde{m}^2_{t2}}= 0\\
%
\frac{\partial   V}{\partial   |H_2^0|} & = & 2m_2^2v_2-2m_3^2v_1 -
\frac{g^2+ g^{'2}}{2}(v_1^2-v_2^2)v_2 =0
%\\
%& & +\frac{3}{8\pi^2}\left\{ h_t^2 A_t m_0(A_t m_0v_2+\mu %v_1)
%%\frac{f(\tilde{m}^2_{t1})-f(\tilde{m}^2_{t2})}
%{\tilde{m}^2_{t1}-
%\tilde{m}^2_{t2}}\right.  \\
%& &\left.
%[(f(\tilde{m}^2_{t1})+f(\tilde{m}^2_{t2})-2f(m^2_t)]h_t^2v_2\right\}=0  )
\end{eqnarray*}
Here we substituted
$$<H_1>\equiv v_1=      v           \cos\beta , \ \
<H_2>\equiv v_2=      v           \sin\beta,                $$
where
$$ v^2=      v_1^2+v_2^2~~(v \approx 174~{\rm GeV}),
  \ \  \tan\beta \equiv \frac{v_2}{v_1}.$$

From the  minimization conditions  given above one  can derive   easily:
\begin{eqnarray}
v^2&=&\frac{\displaystyle 4}
{\displaystyle (g^2+g^{'2})(\tan^2\beta-1)}
\left\{ m_1^2-m_2^2\tan^2\beta\right\}\\
2m_3^2&=&(m_1^2+m_2^2)
\sin 2\beta\\
M^2_Z\equiv \frac{g^2+g^{' 2}}{2} v^2
&=&2\frac{\displaystyle m^2_1-m^2_2 \tan^2\beta }
{\tan^2\beta -1} \\
M^2_W\equiv\frac{g^2}{2} v^2&=&M_Z^2\cws
\label{break}\end{eqnarray}

The derivation of these formulae including  the one-loop radiative corrections
is given in the appendix.
\chapter{The Big Bang Theory }\label{ch5}
\section{Introduction}
In the 1920's Hubble discovered that most galaxies showed a
 redshift in the visible spectra, implying that they were moving
 away from each other.
This observation is one of the basic building
blocks of the Big Bang Theory\cite{cosm},
which assumes  the universe is expanding, thus solving the
problem that a static universe cannot be stable according the
 Einstein's equations of general relativity.
\begin{figure}[tb]
%  HNS  A
\begin{center}
\mbox{\epsfysize=8.cm\epsfbox{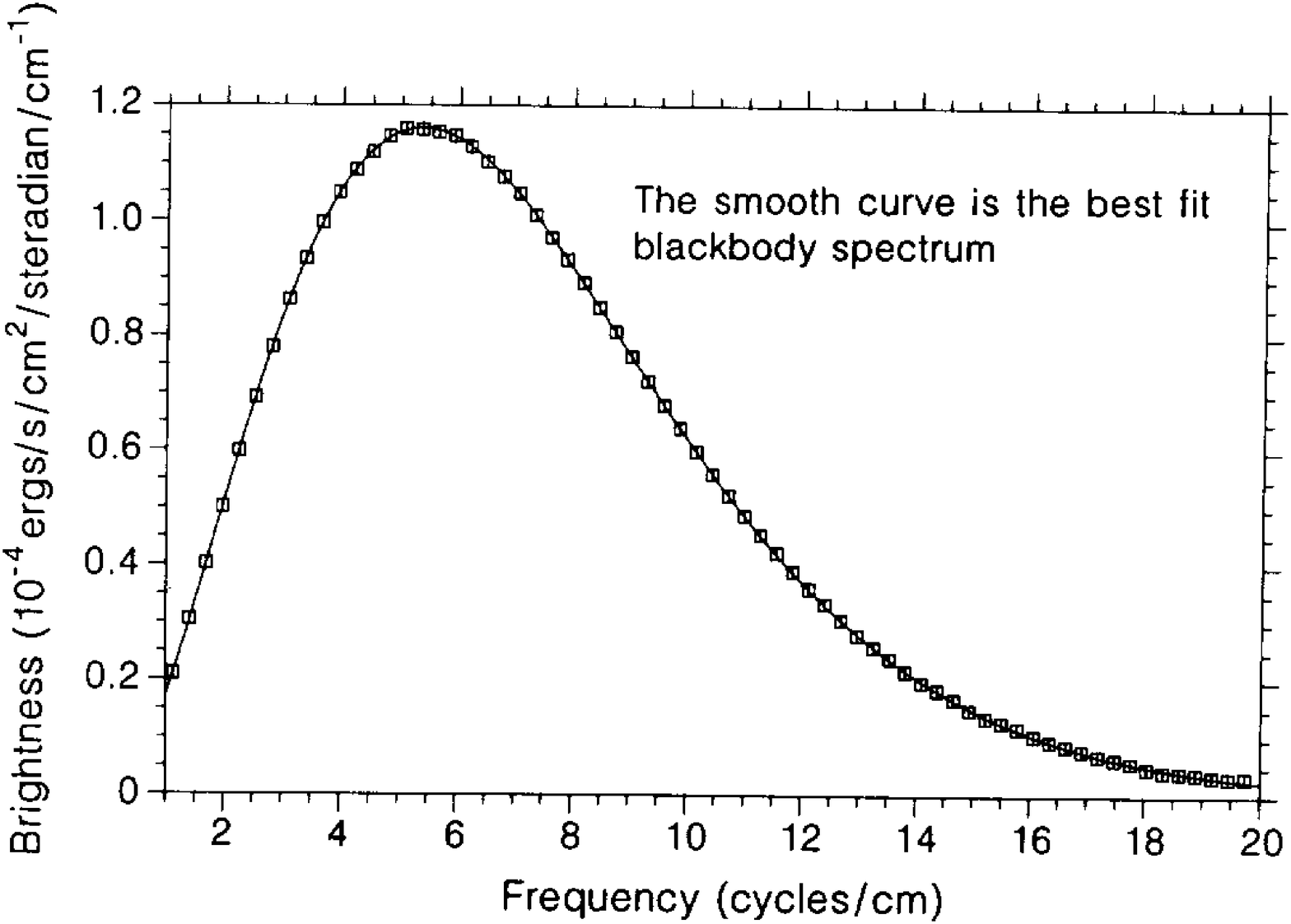}}
%   HNS   E
\caption{Spectrum of the microwave background radiation as measured by 
the COBE satellite. The curve is the black body radiation
corresponding to a temperature of $2.726$ K.}% From ref. \cite{borner}.}
\label{f51}
%   HNS   A
\end{center}
%   HNS  E
\end{figure}
\begin{figure}[tb]
% HNS  A
\begin{center}
\mbox{\epsfysize=10.cm\epsfbox{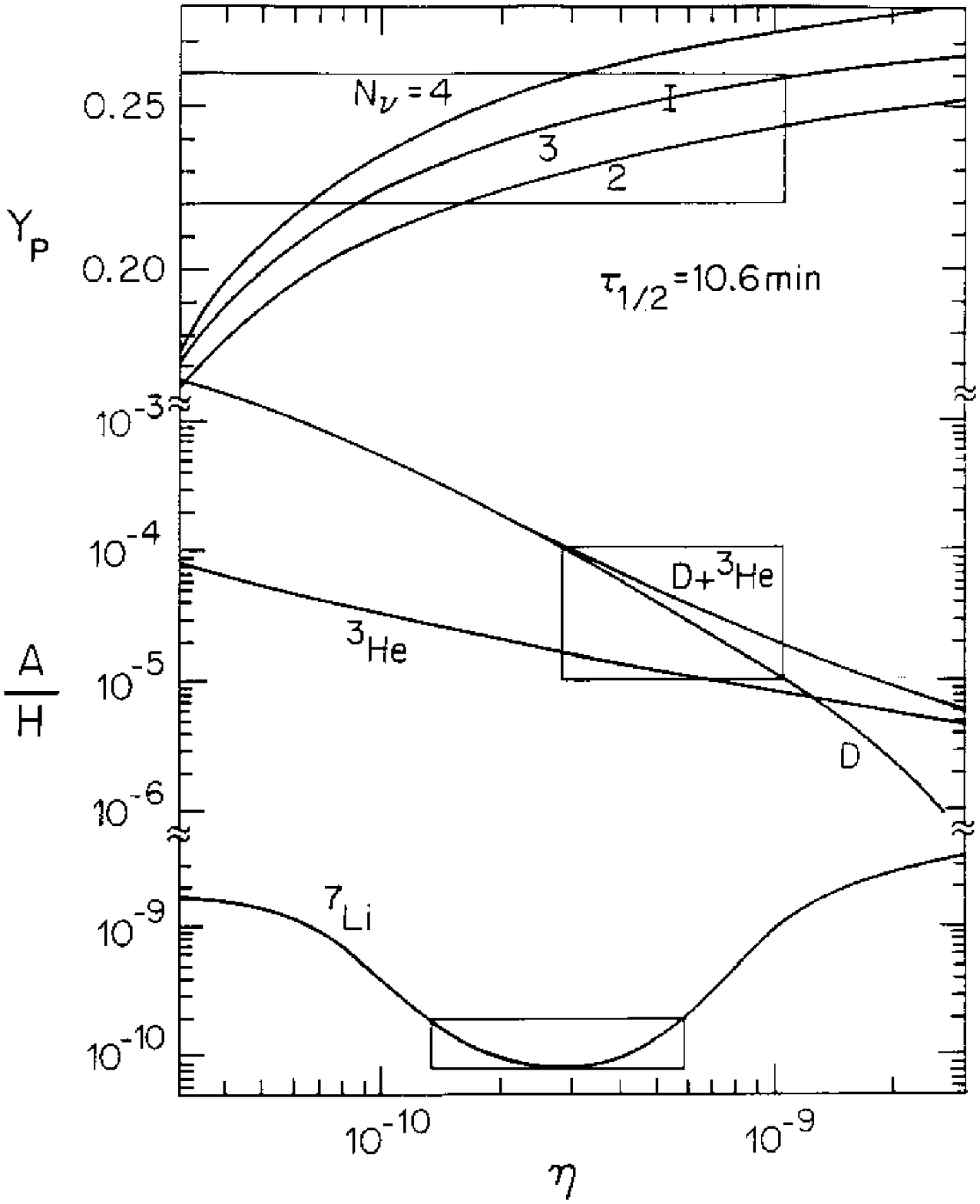}}
%   HNS   E
\caption{Big Bang nucleosynthesis predictions for the primordial abundance 
of the light elements as function of the primordial ratio $\eta$ of baryons
and photons.  From ref. [49]. }
%  HNS  A
\label{f52}
\end{center}
%  HNS  E
\end{figure}
 An expanding universe will cool down, so at the beginning the
 universe might have been hot. The remnants of the radiation of
 such a hot universe can still be observed today as microwave
 background radiation corresponding to a temperature of a few degrees.
 This radiation
was first predicted by Gamow, but accidentally observed in 1963 by
 Penzia and Wilson from Bell Laboratories\footnote{They were
 awarded the Nobel prize for this discovery in 1978.} as noise
 in microwave
 antennas used for communication with early satellites.
Such an antenna is only sensitive to a single frequency. Recently,
the whole spectrum was measured by the 
COBE\footnote{Cosmic Background Explorer.} 
satellite and it was found to be indeed
describable by a black body radiation,  as shown in fig. \ref{f51}
(from ref. \cite{borner}).
 The deviation from perfect
isotropy, if one ignores the dipole anisotropy from the  Doppler
shift caused by the movement of the earth through the microwave
background, is a few times $10^{-6}$. This  has strong implications for
theories concerning the clustering of galaxies, since this radiation
was released soon after the ``bang'' and hardly interacted afterwards, 
so the inhomogeneities in this radiation are proportional to  the
density fluctuations in the early universe!
These density fluctuations are the seeds for the final formation of galaxies.
As will be discussed
later, such small anisotropies have strong implications for the
models trying to understand the formation of galaxies and the nature of the 
dark matter in the universe. Direct evidence that the temperature from the
microwave background is indeed the temperature of the universe came
from the measurement of the temperature of gas clouds deep in space. As it 
happens,
the rotational energy levels of cyanogen $(CN)$ are such that the 3K background
radiation can excite these molecules.  From the detection of the relative 
population of the
groundstate and the higher levels the excitation temperature was determined to
be $T_{CN}=2.729^{+0.023}_{-0.031}$ K\cite{CN}, which is in excellent 
agreement with the
direct measurement of the microwave background of $2.726\pm0.010$ K by 
COBE\cite{cobe}.

Other evidence that the universe was indeed very hot at the beginning
 came from the measurement of the natural
abundance of
 the light elements: the  universe consists for
 74\% out of hydrogen, 24\% helium and  1\% for the remaining elements.
Both the rarity of heavy elements and the large abundance of helium
 are hard to explain, unless one assumes a hot universe at the beginning.

The reason for the low abundance of the heavier elements  in a hot 
universe is simple:
they are cracked by the intense radiation around. The abundance of the
light elements is plotted in fig. \ref{f52}
as function of the ratio $\eta$ of primordial
baryons and photons (from ref. \cite{olive}). Agreement with
experimental observations can only
be obtained for
 $\eta$ in the range $3-7\cdot 10^{-10}$.

The very heavy elements can be produced only at much lower temperatures,
but high pressure,
so it is usually assumed that
        the heavy elements on earth and in our bodies were cooked
 by the high pressure inside the cores of collapsing stars, which
 exploded as supernovae
 and put large quantities of these elements into the
 heavens. They clustered into galaxies under the influence of gravity.

\begin{figure}[bt]
%\begin{center}
%\vspace{-9.cm}
%\mbox{\epsfysize=12.cm\epsfxsize=10.cm\epsfbox{f53.eps}}
\epsfysize=10.cm
\epsfxsize=12.cm
\hspace{1.5cm}
\epsffile[5 70 2205 2205]{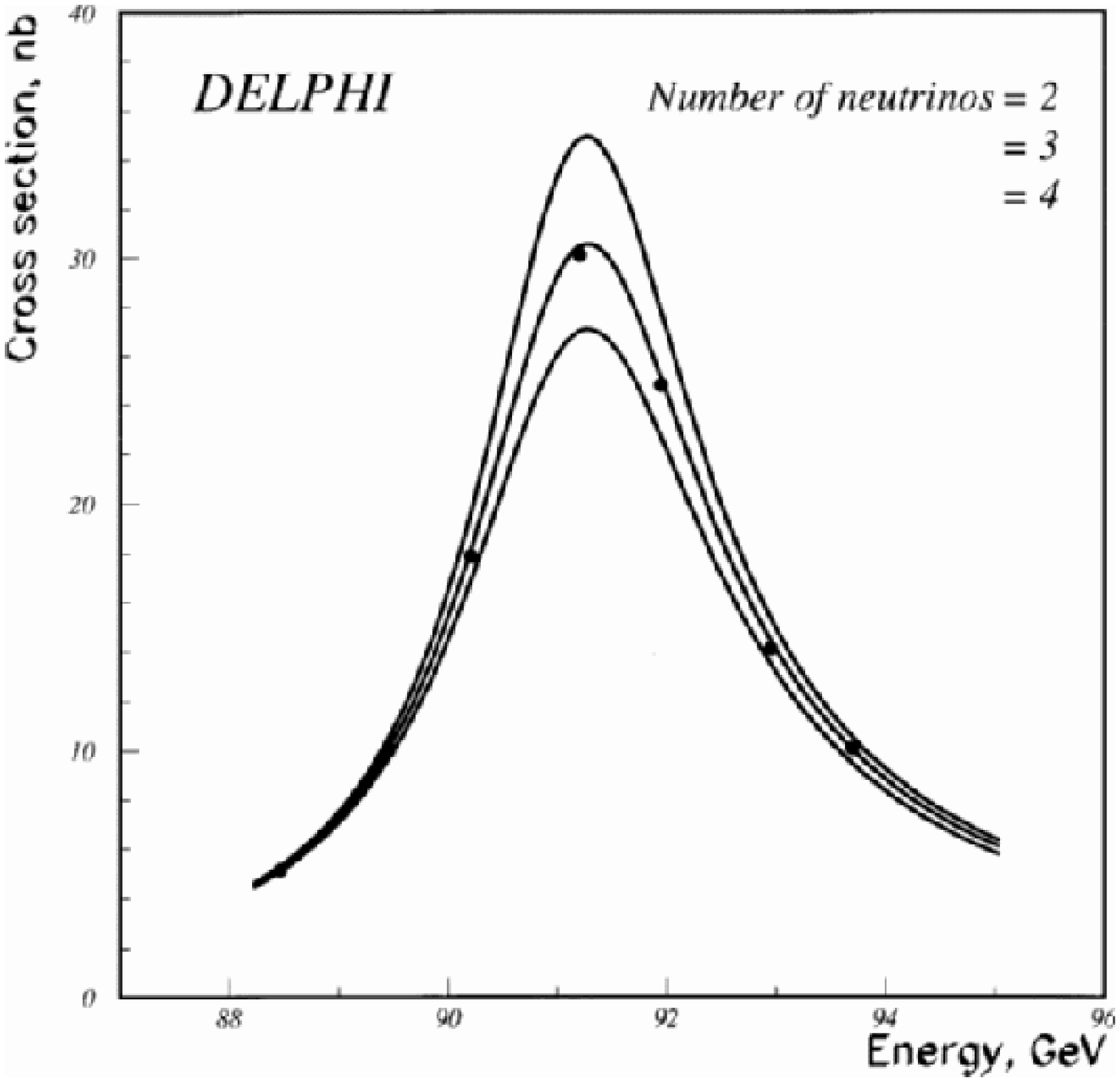}
\caption{The $Z^0$ lineshape for different number of neutrino types. 
The data (black points) exclude more than three types with
mass below $M_Z/2\approx 45$ GeV.}
\label{f53}
%\end{center}
\end{figure}

The ratio of helium and hydrogen is determined by
the number of
neutrons available for fusion into
deuterium and subsequently into helium 
at the freeze-out
temperature of about 1 MeV
 or $10^{10}$ K. At these high temperatures
no complex nuclei can exist, only free 
protons and neutrons. They can be converted
into each other via   charged weak
interactions like 
$ep\leftrightarrow\nu_en$ and $\overline{e}n\leftrightarrow\overline{\nu}_ep$. 
Note that this is the same interaction  which is responsible for 
the decay of a free neutron  into a proton,
electron and antineutrino.
The weak interactions maintain thermal
equilibrium between the protons and neutrons as long as the density and 
temperature are high enough, thus
leading to a Boltzmann distribution:
\bq \frac{n}{p}=e^{-Q/kT} ,\eq
where $Q=(m_n-m_p)c^2=1.29$ MeV is the energy
difference between the states.
Thermal equilibrium is not guaranteed
anymore if the weak interaction  rates  $\Gamma$
are slower than the expansion rate of the
universe given by the Hubble constant, i.e. freeze-out occurs when     
$\Gamma<H(t)$.
This happens by the time the temperature
is about $10^9$ K or 0.1 MeV.
Then the ratio $n/p$ is about 1/7.
Since the photon energies at these temperatures are too low to crack the
heavier nuclei, nuclei can form
 through reactions like
$n+p\leftrightarrow ^2H+\gamma$, 
$^2H+p\leftrightarrow ^3H+\gamma$  and
$^2H+n\leftrightarrow ^3H+\gamma$,
which in turn react to form $^4He$.
The latter is a very stable nucleus, which
hardly can be cracked, so the chain
essentially stops till all neutrons
are bound inside $^4He$!
Heavier nuclei are hardly produced at this stage,  since
there are no stable elements with 5 or 8 
nuclei, so as soon as $^4He$ catches another nuclei it will decay.
Consequently the $n/p$ ratio 1/7, as 
determined from the Boltzmann distribution,
yields a $^4He$ mass fraction $Y_{He}$
\bq Y_{he}=\frac{2n/p}{n/p+1}\approx \frac{1}{4}. \eq
Experimentally the mass fraction $Y_P$ of
$^4He$ is $23\pm1\%$ \cite{schramm}!
 The     mass fractions of deuterium, $^3He$, and  $^7 Li$ are many 
orders of magnitude
smaller\cite{borner,kolb}. The concentration of the latter elements is a 
strong function of the primordial baryon density (see fig. \ref{f52}), 
since at high enough density all the deuterium will fuse into $^4He$, 
thus eliminating the ``components'' for $^3He$ and $^7 Li$.
  
As said above, freeze-out occurs, if $\Gamma<H(t)$.
Thus the expansion rate $H(t)$ around
$T\approx 1$ MeV determines the $^4He$
abundance. The expansion rate in turn is determined by the fraction of 
relativistic particles, like neutrinos,
light photinos etc.
Roughly for each additional species
the primordial $^4He$ abundance increases by 1\%, as shown in  fig. \ref{f52} for
a neutron half-life time of 10.6 minutes
(from ref. \cite{olive})\footnote{The
presently accepted value is $10.27\pm 0.024$
 minutes\cite{gribbon}.}.
The neutron lifetime is not negligible on the scale of the first three 
minutes, so it has to be taken into account.
If one wants to reconcile the abundance of all light elements, there can   
only be three neutrino species (see fig. \ref{f52}) with practically no 
room for other weakly interacting relativistic particles like light photinos!
Present collider data  confirm
that there are indeed only three light
neutrinos\cite{pdb}:
\bq N_\nu=2.99\pm 0.04.\eq
The strongest constraint comes from the $Z^0$ resonance data at LEP\cite{lepc}. 
An example of the quality of the data is shown in  fig. \ref{f53}.
Note that collider data  limit the number of neutrino generations, while 
nucleosynthesis is sensitive to all kinds of light particles, where light 
means about  one MeV or less. Happily enough  the collider data require
the lightest neutralino to be  above 18.4 GeV\cite{pdb}, so there is no 
conflict between cosmology and supersymmetry.

Note that this is a beautiful example
 of the interactions between cosmology and elementary particle
 physics: from the Big Bang Theory the number of relativistic
 neutrinos is restricted to three (or four if one takes the  more  
conservative upper limit on the
$^4He$ abundance to be 0.25) and at the LEP accelerator one
 observes that the number of light neutrinos is indeed three!
 Alternatively, one can combine the
 accelerator data and the abundance
  of the light elements to ``postdict'' the
primordial helium  abundance to be 24\%
and use it to obtain
an upper limit on the baryonic density\cite{schramm}:   
\bq\rho_b\le 0.1 \rho_c\label{omegab},\eq  where $\rho_c$ is the critical
density needed for a flat universe.
The critical density will be calculated
in section  \ref{s53}.
Thus LEP data in combination with
baryogenesis  strengthens the argument
that we need non-baryonic dark matter in a flat universe, 
for which $\rho=\rho_c$.
Other arguments for dark matter will be discussed in section \ref{s511}.

In spite of the marvelous successes of this  model
 of the universe, many questions and problems remain, as mentioned in
the Introduction. However, GUT's can provide amazingly simple
solutions, at least in principle, since many details are still open.

These problems will be discussed more quantitatively in the next sections,
 starting
 with Einsteins equations in a homogeneous and isotropic universe,
conditions which
 have been well verified in the present universe and which make the
solutions
 to Einstein's equations particularly simple. Especially, it is easy
to see that
a phase transition can  lead to inflation, the key in all present
cosmological
theories.
\section{Predictions from General Relativity }\label{s51}
At large
   distances the universe is homogeneous, i.e. one finds the
same mass density everywhere in the universe, typically
\bq
\rho_{univ}=(4- 16)~10^{-27} ~{\rm kg/m^3},\eq
which corresponds to  2.5 -10  hydrogen atoms per cubic meter.
(In comparison, an extremely good vacuum of $10^{-9} N/m^2$ at 300 K
contains about $2~10^{11}$ molecules per cubic meter.
Of course, the  volume to be averaged over
should be chosen to be much larger than the size of clusters of
galaxies. Furthermore, the same density and temperature is observed
in all directions, i.e. the universe is very isotropic.
If no point and no direction is preferred in the universe, the possible
geometry of the universe becomes very simple:
            the curvature has to be the same everywhere, i.e.
instead of a curvature tensor one needs only a single number,
usually written as $K(t)=k/R^2(t)$, where $R(t)$ is the so-called
scale factor. This factor can be used to define dimensionless 
time-independent (comoving) coordinates in an expanding universe:
the proper (or real) distance $D(t)$ between two galaxies scales as
\bq D(t)=R(t) d\label{d},\eq where $d$ is the
distance at a given   time $t_0$.
The factor $k$ introduced above  defines the sign of the curvature:
                          $k=0$ implies no curvature, i.e. a flat
  universe, while $k=+1(-1)$ corresponds to a space with a positive
  curvature (spherical) and $k=-1$ corresponds to a space with a negative
  curvature (hyperbolic).

The movement of a galaxy in a homogeneous  universe can be compared to
the molecules in a gas; the stars are just the atoms of a molecule
and the molecules are homogeneously distributed.
Differentiating   equation \ref{d} results in
\bq v=\dot{R}(t)d\label{v},\eq or substituting $d$ from eq. \ref{d}
 results in the famous relation
between the velocity and the distance of two galaxies:
\bq v = \frac{\dot{R}(t)}{R(t)}D(t) \equiv H(t)D(t) \label{hub}, \eq
where
$H(t)$ is the famous Hubble constant.

This relation between the velocity and the distance of the galaxies was first 
observed
experimentally by Hubble in the 1920's. He observed that all
neighbouring galaxies showed a redshift in the spectral lines of the
light emitted by specific elements  and the redshift was roughly proportional 
to the distance. So this was the first evidence
that we are living in an expanding universe, which might have been
created by a ``Big Bang''.

The Hubble relation \ref{hub} is a direct consequence of the
homogeneity and isotropy of the universe, since the scale factor cannot
be a constant in that case. This follows directly from Einstein's
field equations  of general relativity, which  can be written as:
\bqa
\ddot{R}(t)  & = & -\frac{4\pi~G}{3c^2}(u(t)+3p(t))R(t),
\label{einI}
\eqa
\bqa
\frac{\dot{R}(t)^2}{R^2(t)}-\frac{8\pi~G}{3c^2}u(t) & = & -\frac{k c^2}{R^2},
\label{einII}
\eqa
where $G=6.67 \cdot10^{-11} {\rm Nm^2/kg^2}$  is the gravitational constant,
$\dot{R}$ and $\ddot{R}$ are the derivatives of  $R$
with respect to time, $p$ is the pressure
and $u$ is the  energy density.

A static universe, in which the derivatives
 and pressure are zero, implies $u(t)=0$,  so a static
universe cannot exist unless one introduces  additional potential energy in 
the universe, e.g. Einstein's
cosmological constant. At present there
is no experimental evidence for
such a term\cite{coscon}.

\section{Interpretation in terms of Newtonian Mechanics }\label{s52}
Since the energy density, and correspondingly the curvature, is
small in our present
universe, relativistic effects can be neglected and the
field equations \ref{einI} and \ref{einII} have a simple interpretation in 
terms of
Newtonian mechanics. Consider a spherical
shell with radius R and mass $m$. The mass inside this sphere
can be related the average density $\rho$:
 \bq M= \frac{4}{3} \pi R^3\rho.\label{rho} \eq
For an expanding universe the total mechanical energy of the mass shell
can be written as the sum of the kinetic and potential energy:
\bqa E_{tot} &=&\frac{1}{2}m\dot{R}^2-\frac{GMm}{R}\nn  
&=&\frac{1}{2}mR^2\left[\frac{\dot{R}^2}{R^2}-\frac{8}{3}\pi
G\rho\right].\label{etot}\eqa

The expression in brackets is just        the left hand side of
eq.  \ref{einII}, so the sum of kinetic -- and potential energy
determines the sign of the curvature $k$.
If $k= 1$, then $E_{tot} < 0$, implying that the universe
will      recollapse under the influence of gravity  (``Big Crunch''),
 just like a rocket
which is launched with a speed below the escape velocity, will
return to the earth; $k=-1   $, on the other hand, implies that the
universe will expand and cool forever (``Big Chill'').
In case of $k=0$ the total energy equals zero (flat Euclidean space),
in which case the gravitational energy, or equivalently the
mass density, is sufficient to halt the expansion.

The first field equation (eq. \ref{einI})
follows from the second equation (eq. \ref{einII}) by differentiation
and taking into account that in an expanding universe
energy is  converted into
gravitational potential energy: when the volume increases by
an infinitesimal amount $\Delta V$, then the remaining energy
in the gas decreases by an amount $p\Delta V_{phys}$, where $p$ 
denotes the pressure.
Therefore \bq \dot{E}(t) =-p(t)\dot{V}_{phys}(t)
=-3\frac{\dot{R}(t) }{R(t)}p(t)V_{phys}(t).
\label{e1}\eq
Here we used $V_{phys}=V_0 R^3(t)$, where $V_0$ is the volume in 
comoving time-independent coordinates analogous to eq. \ref{d}.
On the other hand follows from $E(t)=u(t) V_{phys}(t)$
 \bq \dot{E}(t) =\dot{u(t)}V_{phys}(t)
+3\frac{\dot{R}(t) }{R(t)}u(t) V_{phys}(t).
\label{e2}\eq
Combining eqs. \ref{e1} and \ref{e2} results in:
\bq \dot{u}(t) =-3\frac{\dot{R}(t)}{R(t)}
(u(t)+p(t)).\label{dotr}\eq
Substituting this equation
for $\dot{u}(t)$ after differentiation of
eq. \ref{einII} yields eq. \ref{einI}.
\section{Time Evolution of the Universe }
\label{s53}
To find out how the universe will evolve in time,
one needs to know the equation of state,
which relates the energy density to pressure.
Usually energy and pressure are proportional, i.e. $p=\alpha\rho c^2$,
where $\alpha=0$ for cold non-relativistic matter ($p=0)$ and $\alpha=
1/3$ for a relativistic hot gas, as follows from elementary Thermodynamics.
From eq. \ref{dotr} given above, it follows immediately that
\bq \rho \propto R^{-3(1+\alpha)}\label{rhot}\eq
and substituting  this into eq. \ref{einII} results in:
\bq R \propto t^{\frac{2}{3(1+\alpha)}},\label{rt}\eq
if we neglect the curvature term, i.e. either $R$ is large
or $k$ small. As we will see, both are true after the inflationary
phase of the universe.
\begin{table}[htb]
\begin{center}
\begin{math}
\begin{array}{|c|c|c|c|}  \hline
hot ~relativistic &
t>t_1  & R\propto t^{1/2}  & \rho=D_1/R^4 \\
\hline
inflation &
t_1>t>t_2  & R\propto e^{Ht}   & \rho=const.\\  \hline
hot ~relativistic &
t_2>t>t_3  & R\propto t^{1/2}  & \rho=D_2/R^4 \\  \hline
cold ~non-relativ. &
t>t_3 & R\propto t^{2/3}  &
 \rho=D_3/R^3 \\  \hline
\end{array}
\end{math}
\end{center}
\caption{The time dependence of the scale factor and energy density 
during  various stages in the evolution of the universe. Typically, 
$t_1\approx 10^{-43}$ s, $t_2\approx 10^{-35}$ s, and $t_3\approx 10^{5}$ yrs.
The constants $D_i$ are integration constants. }
\label{t51}
\end{table}

\begin{figure}[tb]
\begin{center}
\mbox{\epsfysize=6.cm\epsfbox{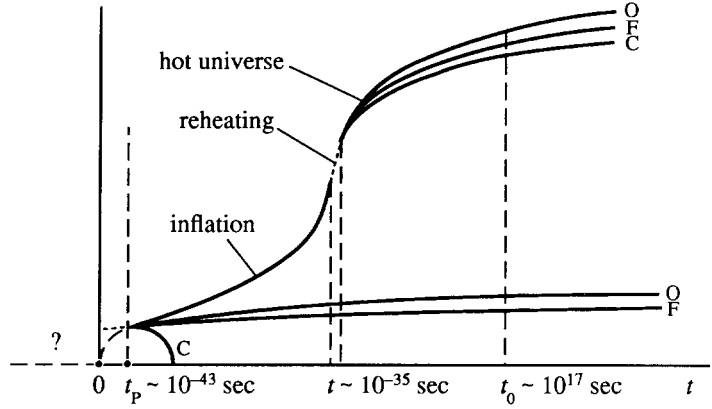}}
\caption{Evolution of the radius of the universe for a closed (C), 
flat (F) or open (O) universe with and without inflation.
From ref. [3].}
\label{f54}
\end{center}
\end{figure}

    The time dependence has been summarized in table \ref{t51}
for various stages of the universe. The inflationary period will be 
discussed in the next section.
One observes that the scale factor vanishes at some time $t=0$
and the energy density becomes infinite at that time.
This singularity explains the popular name ``Big Bang'' theory
for the evolution of the universe.
The solutions for $R(t)$ are
shown graphically in fig. \ref{f54} 
for three cases: a flat universe ($k=0$),
 an open universe ($k=-1$) and
a closed  universe ($k=1$) (from ref.\cite{linde}).

An open or flat universe will expand forever, since the kinetic energy is   
larger than the gravitational attraction. 
A closed universe will recollapse.  The lifetime
of a closed universe with $p>-\rho/3$ and a total mass M of
cold non-relativistic matter is\cite{linde}:
\bq t_c=\frac{4MG}{3}\approx \frac{M}{M_P}~10^{-43} s, \label{tc}\eq
so the present lifetime of the universe 
of at least $10^{10}$ yrs 
gives a strong upper limit  on the density of the universe.

The lifetime   can  easily be calculated, if we assume a flat universe:
from table \ref{t51} it follows that
$R(t)\propto t^{2/3}$ for most of the time.
 Substituting this and its time
derivative into the definition of the Hubble constant (eq. \ref{hub} )
results in:
\bq H(t)=\frac{2}{3t}.\label{ht}\eq
With the presently accepted value of the measured Hubble constant:
\bq H=100 ~h_0~(\frac{km}{s~Mpc})\approx h_0~(3~10^{17})^{-1} s^{-1}
\approx h_0~10^{-10}~ yrs^{-1}, \label{hub0}\eq
where $h$ indicates the experimental uncertainty ($0.4\le h_0 \le 1$),
  one finds for the age of the universe:
\bq t_{universe}=2/3H=2/(3h_0)\cdot  10^{10}
{\rm yrs} \label{tuni}.\eq
The  critical density, which is the density corresponding to a flat universe, 
can be calculated from eqn.
\ref{etot} 
  by requiring $E_{tot}=0$ (or equivalently $k=0$) and substituting for 
$\dot{R}/R$
the Hubble constant (see eq. \ref{hub}):
\bq \rho_c=\frac{3H^2}{8\pi G}=2\cdot 10^{-26}~ h_0^2 ~  kg/  m^{3} ,\label{rhoc}
\eq
where the numerical value of $H$ from eq. \ref{hub0}  was used.

The size of the observable universe, the horizon distance $D_h$,
can be calculated in the following way:
the proper distance  between two points is $R(t)d$ (eq. \ref{d}), where $d$
is the distance in comoving coordinates.
Light propagates on the light-cone.
This can be studied most easily by considering the time $\eta$ in comoving 
coordinates with
\bq dt=R(t)d\eta. \eq

In comoving coordinates the distance $d_h$
light can propagate is $cd\eta$, so
$ d_h\equiv c\int  d\eta=c\int  dt/R(t)$
or the proper distance $D_h=R(t)d_h$
equals:
\bq D_h=
cR(t)\int_0^{t'} \frac{dt'}{R(t')}.
\label{dh}\eq

For most of the time $R(t)=a t^{2/3}$ (see table \ref{t51}). 
Substituting this into
eq. \ref{dh} yields:
\bq D_h=3ct=2c/H(t)=0.9 h_0^{-1} \cdot 10^{26} ~{\rm m}, \label{hor} \eq
where for the lifetime $t$ of the universe
eqs. \ref{ht} and \ref{hub0} were used.
\section{Temperature  Evolution of the Universe }\label{s54}
In the previous section the scale factor and the
energy density were
calculated as function of time.
The energy density has two components:
the energy density from the photon radiation
in the microwave background $\rho_{rad}$ and
the energy density of the non-relativistic
matter $\rho_{matter}.$
At present $\rho_{rad}$ is negligible, but at
the beginning of the universe it was the
dominating energy.
Assuming this radiation to be in  thermal equilibrium with matter implies a 
black body
radiation with a frequency distribution given
by Planck's law and an energy density
\bq \rho_{rad}=aT^4, \label{planck} \eq
where $a=7.57\cdot 10^{-16} ~J ~m^{-3} K^{-4}.$
Since $\rho_{rad}
\propto 1/R^4$ (see table \ref{t51})  one finds from eq. \ref{planck}:
\bq T\propto 1/R(t), \label{rt1}
\eq
from which  follows immediately: $\dot{R}/R=-\dot{T}/T$
and $R^{-2}\propto T^2$. Substituting these expressions and eq. \ref{planck}
into the second field equation (eq. \ref{einII}) leads to:
\bq \left(\frac{\dot{T}}{T}\right)^2=
\frac{8\pi aG}{3c^2}T^4,\label{tt} \eq
since the term with $kc^2$ is only proportional
 to $T^2$, so it can be neglected
at high temperatures. Integrating eq. \ref{tt}  yields:
\bq T=\left(\frac{3c^2}{32\pi a G}\right)^{1/4}\cdot \frac{1}{\sqrt{t}}=
1.5\cdot 10^{10}~K\cdot \sqrt{\frac{1~s}{t}}=
1.3 ~MeV\sqrt{\frac{1~s}{t}},
\label{tt1} \eq
so the temperature drops as $ 1/\sqrt{t}.$

From this equation one observes   that
 about one microsecond after the Big Bang the
temperature has dropped from a value above the
 Planck temperature, corresponding to an energy of
$10^{19}$ GeV to a temperature of about one GeV,
so after about one microsecond the temperature
is already too low to generate protons
and after about one second the lightest matter particles, the electrons, 
are ``frozen'' out.
After about three minutes the temperature
is so low that the light elements become stable,
and after about $10^5$ years atoms can form.

At this moment all matter becomes neutral and
the photons can escape. These are the photons,
which are still around in the form of the microwave background radiation!

%\begin{figure}[tb]
%\begin{center}
%\mbox{\epsfysize=6.cm\epsfbox{f54.eps}}
%\caption{Evolution of the radius of the universe 
%for a closed (C), flat (F) or open 
%(O) universe with and without inflation.}
%\label{f54}
%\end{center}
%\end{figure}
\begin{figure}[tb]
%   HNS   A
\begin{center}
\mbox{\epsfysize=7.cm\epsfbox{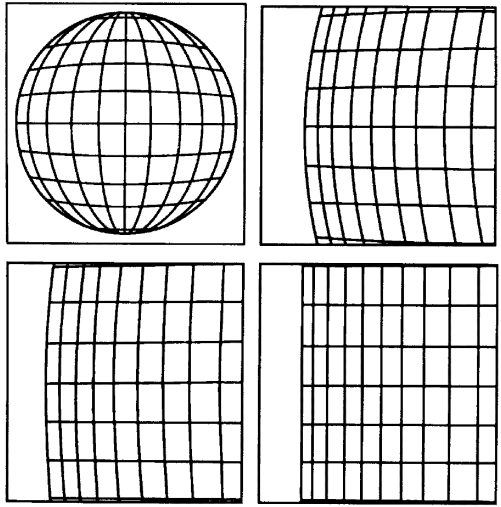}}
%   HNS  E
\caption{The flatness of the universe after inflation is easily
understood if one thinks about the inflation of a balloon.}
\label{f55}
%   HNS  A
\end{center}
%   HNS  E
\end{figure}
\begin{figure}[tb]
% HNS  A
\begin{center}
\epsfysize=6.cm
\epsfxsize=10.cm
\hspace{1.5cm}
\epsffile[25 25 550 335]{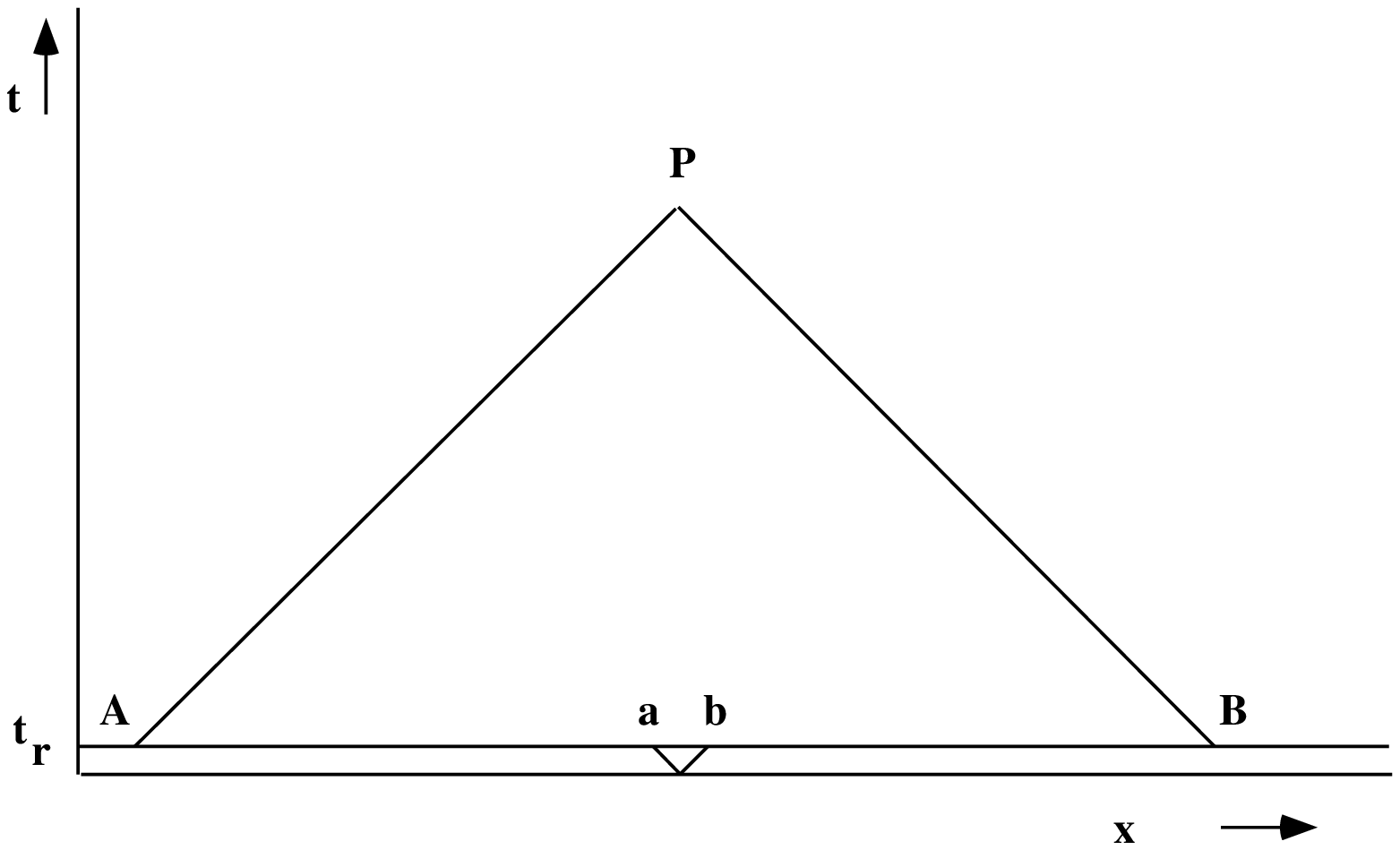}
%   HNS   E
\caption{The space-time diagram of the microwave background radiation, 
which was released at the time $t_r$. Photons observed in point P from 
opposite directions traveled some $10^{10}$ yrs with hardly any 
interactions from the points A and B, respectively. The distance 
light could have traveled between the Big Bang and $t_r$ is only 
$ab$, which is much smaller than the distance $AB$. Consequently 
the points A and B could never have been in causal contact 
with each other. Nevertheless, the radiation from A
 and B have the $same$ temperature, although the horizon $ab$ 
is much smaller (horizon problem). The problem can be solved 
if one assumes the region of causal contact was much larger than 
$ab$ through inflation  of space-time via a phase transition.}
%  HNS  A
\label{f56}
\end{center}
\end{figure}

After the discovery of the microwave background radiation, the
Big Bang theory gained widespread acceptance.
Nevertheless, the simplest model as formulated here, has several
serious problems, which can only be solved by the so-called
inflationary models. These models have the bizarre
property that the expansion of the universe goes faster than the
speed of light!
This is not a contradiction of special relativity, since
these regions are causally disconnected, so no
information will be transmitted. Special relativity does not restrict
the velocities of causally disconnected objects.
However, such inflationary scenarios require the introduction of
a scalar field, e.g. the Higgs field discussed in the previous chapter.
For certain conditions of the potential of this field, the
gravitational force becomes repulsive, as can be derived
directly from the Einstein equations given above.
This will be discussed   in more detail after a short summary of the
main problems and questions of the simple Big Bang theory.
\section{Flatness Problem }\label{s55}
At present we do not know if the universe is open or closed, but
experimentally the ratio of the actual density to the critical density is bound
 as follows\cite{kolb}: \bq
 0.1\le \Omega=\rho/\rho_c \le 2\label{omega}.\eq
The luminous matter contributes only
 about 1\% to $\Omega$, but from the
dynamics of the galaxies one estimates
that the galaxies contribute between
0.1 and 0.3, so the lower  limit on $\Omega$ stems from these observations.
The upper limit is obtained from the
lower limit on the lifetime of the universe. From the dating of the oldest 
stars and the elements one knows that
the universe is at least $10^{10}$ years old, which gives an upper limit on
the Hubble constant (eq. \ref{ht})
and consequently on the  density\cite{kolb}.
This does look like a perfectly acceptable number and the universe
might even be perfectly flat, since $\Omega=1$ is not excluded.
However, it can be shown easily, that $\Omega-1$  grows with time as
 $t^{2/3}$ and for the present lifetime $t\approx 10^{17} s$ this
number becomes big, unless very special initial conditions limit
the proportionality constant to be exceedingly small.
This constant can be calculated easily:
From eq. \ref{einII} and the definition of $\rho_c$ (eq. \ref{rhoc})
one finds:
\bq \Omega(t)-1=\frac{kc^2}{R^2(t) H^2(t)} \label{ome}\eq
Since $R\propto t^{2/3}$ for the longest time of the universe
(see table \ref{t51}) and $H(t)\propto 1/t$ (eq. \ref{tuni}) one
observes that $\Omega-1 \propto kt^{2/3}$. For this number to come
out close to zero for $t$ very large implies that $k$ must have been
very close to zero, right from the beginning. Remember that $k$ is 
proportional to the sum of potential and kinetic energy in the 
non-relativistic approximation.
One can show\cite{linde} that in order for $\Omega$ to lie in the
range close to 1 now, implies that in the early universe
$|\Omega-1|\le 10^{-59} M_P^2/T^2$, or for $T\approx M_P$,
\bq \frac{|\Omega-1|}{\Omega}\le 10^{-59}.\eq
In other words, if the density of the initial universe was
{\it above} the critical density say by $10^{-55}\rho_c$, the universe
would have collapsed long ago! On the other hand, would the density
 have been {\it below} the critical density by a similar amount, the
present density in the universe would have been negligible small
and life could not exist!

\section{Horizon Problem }
\label{s56}
Since the horizon increases linearly with time but the
expansion only with $t^{2/3}$, most of the presently
visible universe was causally disconnected at the time $t=10^5$ years,
when the microwave background was released.
Nevertheless, the temperature of the microwave background radiation
is the same in all directions! How did these photons thermalize
after being emitted some $10^{10}$ years ago?
One should realize that the density in the universe is exceedingly low,
so photons from opposite directions  have traveled some $2\cdot 10^{10}$
lightyears without interactions. Since the distance scales as $t^{2/3}$,
these regions were about $10^7$ light years apart at the time  they were 
released, i.e. the distance $AB$ in fig. \ref{f56},
which is two orders of magnitude larger than the horizon of the universe at that
time (distance $ab$), so no signal could have been transmitted.
Nevertheless, the temperature difference
 $\Delta T/T$ between  these regions is less than $10^{-5}$ as shown by
the recent COBE data.
As with the flatness problem, one can  impose an accidental
temperature isotropy in the universe as an initial condition,
but with all the hefty fluctuations during the Big Bang, this	is
a very unsatisfactory explanation.
As we will see later, inflation solves both problems in a very elegant
way.
\section{Magnetic Monopole Problem }
\label{s57}
Magnetic monopoles are predicted by GUT's as topological
defects in the Higgs field: after spontaneous symmetry breaking the vacuum 
obtains a non-zero vacuum expectation
value in a given region. Different regions may have different
orientations of the phases of the Higgs field and the borderlines
of these regions have the properties expected for magnetic 
monopoles\cite{monop}. Unfortunately the magnetic monopole
density is very small, if not zero. Their absence
has to be explained in any theory based on GUT's with SSB.
 The first attack was made by Alan Guth, who invented
inflation for this problem. Although the original
model did not solve the monopole problem, it
provided a perfectly reasonable solution for the
horizon and flatness problem.
An alternative version of inflation, the so-called {\it new}
inflation, which was invented by A.D. Linde\cite{linde}
and independently by Albrecht and Steinhardt\cite{guth},
provided also a solution of the monopole problem,
as will be discussed in    section \ref{s59}.

\section{The smoothness Problem}
\label{s58}
Our universe has density inhomogeneities in the form of
galaxies. On a large scale the spectrum of inhomogeneities is
approximately scale-invariant, which can be understood
in the inflationary scenario as follows:
 the inflation smoothens out any inhomogeneities
which might have been present in the initial
conditions. Then in the course of the phase transition
inhomogeneities are generated by the quantum
fluctuations of the Higgs field on a very small scale
of length, namely the scale where quantum effects are
important. These density fluctuations are then enlarged to an astronomical scale
by inflation and they stay scale invariant
as is obvious if one thinks about a little circle
on a balloon, which stays a circle after inflation, but just on a larger scale.

\section{Inflation }\label{s59}
The deceleration in the universe is given by
eq. \ref{einI}.
In case the energy density only consists
of kinetic and gravitational energy, the
sign of $\ddot{R}$ is negative, since both
the pressure and energy density are positive.
However,
the situation can change drastically, if the
universe undergoes a first-order phase transition.
In Grand Unified Theories such  phase transitions
are expected: e.g. the highly symmetric phase
 might  have been an SU(5) symmetric state, while
the less symmetric state corresponds to the
$\su$ symmetry
of the Standard Model.
The description of the spontaneous symmetry
breaking by the Higgs mechanism leads to
a specific picture of this phase-transition:
the Higgs field $\phi$ is a scalar field, which
fills the vacuum with a potential energy $V(\phi)$.
The value of the potential is temperature dependent: at high 
temperatures the minimum
occurs for $\phi=0$, but for temperatures below
the critical temperature, the ground state, i.e.
the state with the lowest energy, is reached
for a value of the field $\phi \ne 0$.
This is completely analogous to other phase
transitions, e.g. in superconductivity
the scalar field corresponds to the density
of spin 0 Cooper pairs or in ferromagnetism it
would be the magnetization.

If during the expansion of the universe
the energy density falls below the
energy density of this scalar field,
something dramatic can happen:
the deceleration can become an acceleration,
leading to a rapid expansion of the universe,
usually called ``inflation''.

This can be understood as follows:
if the vacuum is filled with this potential
energy of the scalar field with an energy density
$\rho_{vac}$,
 the work $W$ done  during the expansion
is  $p\Delta V$. However, the gain in energy is $\rho_{vac} \Delta V$, 
since the potential energy of the vacuum does not change (a ``void''
stays a ``void'' as long as no phase transition takes place), so an increase
in volume implies an increase in energy.
Since no external energy is supplied, the
total energy of the system must stay
constant, i.e.
$ p\Delta V + \rho_{vac}\Delta V =0,$
or
\bq p = -\rho_{vac}.\eq
This is the {\it famous equation of state}
in case of a potential dominated vacuum.
In this case  eq. \ref{einII} reduces to:
\bq   \ddot{R}= \frac{8}{3}\pi GR(t){\rho}.\eq
This equation has the solution:
\bq        R(t) \propto e^{t/\tau},\eq
where \bq \tau=\sqrt{\frac{3}{8\pi G\rho}}. \label{tau}\eq
As we have seen in the previous chapter, the symmetry breaking
of a GUT happens at an energy of $10^{16}$ GeV.
The energy density at this energy is extremely high:
\bq u=\rho c^2=\frac{E^4_{GUT}}{(\hbar c)^3}=10^{100} ~{\rm Jm^{-3}}\label{ugut},
\eq
where the powers are derived
from dimensional analysis. Inserting this result into eq. \ref{tau}  yields:
\bq \tau=10^{-37} s.\eq
Thus the universe inflates extremely rapidly: its diameter
doubles every $\tau \ln 2$ s!
The behaviour of the Hubble constant during the inflationary era is then
determined by eq. \ref{hub}, which yields:
\bq H(t)=\frac{1}{\tau}. \eq

Clearly, the inflationary scenario provides in an elegant way
solutions for many of the shortcomings  of the BBT:
\begin{itemize}
\item
The rapid expansion removes all curvature in space-time,
thus providing a solution for the flatness problem.
\item
After expansion the universe reheats
because of the quantum fluctuations around
the new minimum of the vacuum, thus
thermalizing a region much larger than
the visible region at that time. This
explains why
 all visible regions in the present universe
were in causal contact during the time 
the 2.7 K background microwave radiation was released.
\item
The rapid expansion explains the absence of magnetic monopoles, since
after sufficiently large inflation the monopoles
 are diluted to
a negligible level.
\end{itemize}
However, the whole idea of inflation only works
 if one assumes
 the inflation to go smoothly from a single homogenous
region to a large homogenous region many
times the size of our universe (so-called {\it new}
 inflation). This requires rather special conditions
for the shape of the potential, as was pointed out
by Linde\cite{linde} and independently by
Albrecht and Steinhardt\cite{guth} after
the original introduction of inflation by Guth\cite{guth}.
The problem is that
 the corresponding scalar fields providing the
potential energy of the vacuum have to be weakly interacting, since otherwise the
phase transition will involve  only microscopic distances
 according to  the uncertainty relation.
The Higgs fields providing spontaneous symmetry breaking
are interacting too strongly, so one has to introduce
additional weakly interacting scalar fields.
It is non-trivial to combine the requirement of
a negligible small cosmological constant, which represents the potential 
energy of the vacuum, with a vacuum filled with
Higgs fields to generate masses. This requires large cancellations
of positive and negative contributions, which occur e.g.
in unbroken supersymmetric theories. However, these theories,
if they describe the real world, have to be broken.
Details on these problems are discussed   by Olive\cite{olive} and in the
recent text books by B\"orner\cite{borner} and Kolb and Turner\cite{kolb}.
 In spite of these problems the arguments in favour of inflation are so 
strong, that it has become the only acceptable paradigm of present cosmology. 

\section{Origin of Matter }\label{s510}
As discussed above, matter in our universe consists
largely of hydrogen (75\%) and helium (24\%) (typically $10^{78}$ nucleons).
The absence of antimatter can be explained
if one has phase transitions and
among others CP-violation, as  discussed
in chapter \ref{ch3}. 
At present the nuclei dominate the
energy density in the universe in contrast
to the first $10^5$ years, when the
energy density of the radiation
dominated.  The reason for this change
is simply the fact   that    the energy density of the 
many photons around decreases $\propto T^4$,
while the energy density of the nuclei
decreases $\propto T^3$  (see eq. \ref{rhot}).

Large amounts of matter can be created
from the energy release
during the inflationary phase of the  universe. This can be easily estimated
as follows.   
After inflation  the universe has a  macroscopic size, typically 
the size of a    football or larger. The   large energy density   
in this volume (see eq. \ref{ugut})
yields a total energy many times the 
mass in our present universe.

 Thus within
 the inflationary scenario  the universe could originate as a  quantum
fluctuation, starting from absolute ``nothing'', i.e. a state devoid of
space, time and matter with a total energy equal to zero. Most matter was
created after the inflationary phase
from the decay of the field quanta
of the fields responsable for the
inflation. 
Of course, a quantum description of space-time can be discussed only
in the context of quantum gravity, so these ideas
must be considered speculative until  a theory of quantum
gravity is formulated and proven by experiment.
Nevertheless, it is fascinating to contemplate that
physical laws          may determine not only the evolution
of our universe, but they may remove  also the need for
assumptions      about the initial conditions.
\begin{figure}[tb]
% HNS  A
\begin{center}
\epsfysize=8.0cm
\epsfxsize=12.cm
\hspace{.4cm}
\epsffile[0 0 1400 1010]{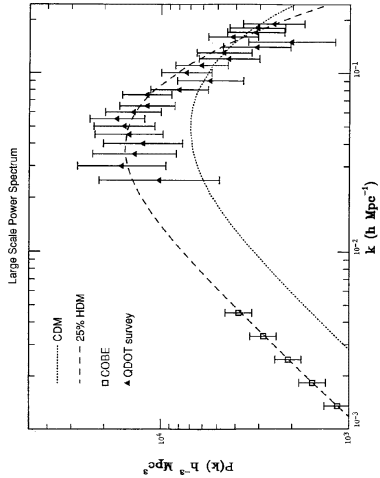}
%   HNS   E
\caption{Models including 25\% hot dark matter (HDM) can describe the large 
scale structure of the universe, as probed by the galaxy surveys (QDOT) and 
COBE temperature anisotropy, better than models with only dark matter (DM).   
From Schaefer and Shafi [57].}
%  HNS  A
\label{f57}
\end{center}
\end{figure}
\section{Dark Matter }\label{s511}
The visible matter is clustered in large galaxies,
which are themselves clustered in clusters and
superclusters  with immense voids in          between.
From the movements of the galaxies one is forced to
conclude that there must be much more matter than
the observed visible matter, if
we want to stick to Newtonian mechanics.
The most impressive evidence for the {\it dark}, i.e. not visible matter 
comes from the so-called {\it flat} rotation curves \cite{rot}:
the orbital velocities of luminous matter around the central of spiral 
galaxies remain  constant out to the far edges of the galaxies in apparent 
contradiction to velocity distributions expected from Keppler's law:
\bq v^2(r)=G\frac{M(r)}{r}, \eq
where $r$ is the radial distance to the centre of the galaxy, $M(r)$ the 
mass of the galaxy inside a sphere with radius $r$,
and $G$ the gravitational constant.
From this law one expects the velocities to decrease with $1/\sqrt{r}$, if 
the mass is concentrated in the centre, which is certainly the case for the 
visible matter.
Velocities   independent of $r$ imply
$M(r)/r$ to be constant or $M(r)\propto r$!
Such a behaviour is expected for {\it weakly} interacting matter, like neutrinos,
gravitinos or photinos, since strongly interacting matter would be attracted 
to the centre by gravity, interact, loose energy, and concentrate in the 
centre, just like the visible matter does.
Also the dynamical properties of galaxies in large clusters require large 
amounts of dark matter.
To make the speeds work out consistently, one has
to assume that the total density of 'dark' matter
is an order of magnitude more than the visible matter. Recent reviews 
for the experimental evidence of dark matter can be found in ref. \cite{tyson}.

From the concentration of light elements, as shown
in fig. \ref{f52}, one has to conclude that the
total baryonic density is only about 10\%
of the critical density (see   eq. \ref{omegab}).
The critical density is the density for a flat universe, which naturally
occurs after  inflation. Consequently,
in the inflationary scenario the dark matter
makes up about 90\% of the total mass in the
universe and it {\it has} to be non-baryonic.

Possible candidates for dark matter are
the MACHO's\footnote{Massive Compact Halo Objects.}, which have been 
observed recently through
their microlensing effect on the light of stars behind them\cite{macho}. 
But since they cluster in
heavy compact objects, they are likely to be remnants
of collapsed stars or light dwarfs, which have
too little mass to start nuclear burning.
In these cases they would be baryonic and make
up   10\% of the critical density, required
by baryogenesis. Note that
the visible baryonic matter in stars represents at most 1\% of the critical 
density.

So one needs additional dark matter, if one believes in inflation and takes 
the value of $\Omega_b=0.1$
from baryogenesis. Candidates for this additional
dark matter are
neutrinos with a mass in the $eV$ range\cite{turner}:
\bq 9<m_\nu<35~ {\rm eV}.\eq
Such small masses are experimentally
not excluded\cite{pdb}, but they would be
relativistic or ``hot''.
Unfortunately, {\it all} dark matter cannot be
relativistic, since this is inconsistent with
  the extremely small anisotropy in the microwave background as
observed by the COBE satellite \cite{coldm}.

 This anisotropy in the temperature is proportional to the anisotropy in the 
mass density at the time of
release of this radiation shortly after the Big Bang.
Through gravity the galaxies were formed around
these fluctuations (``seeds'') in the mass density.
So the present structure of the universe has to
follow from the spectrum of fluctuations
in the early universe, which can be probed by
the microwave background anisotropy.
The best fit is obtained for a mixture of 75\%
cold and 25\% hot dark matter\cite{coldm}, as shown in fig. \ref{f57} 
(from ref. \cite{shafi}).

The lightest supersymmetric particles
are ideal  candidates for cold dark matter, provided they are not too
numerous and too heavy. Otherwise they would
provide a density above the critical  density\cite{relic,roskane,rob}
in which case the universe would be closed and the lifetime
would be very short (see fig. \ref{f54}).
The LSP can  annihilate sufficiently rapidly into fermion-antifermion pairs, if
the masses of the SUSY particles are not
too heavy, as will be discussed in chapter
\ref{ch6}.
\section{Summary}\label{s512}
The Big Bang theory is remarkably successful
in explaining the basic observations
of the universe, i.e. the Hubble expansian,
the microwave background radiation and
the abundance of the elements.
From the measured Hubble constant one
can derive such basic  quantities as the
the size and the age of the universe.
Nevertheless many questions remain unanswered. They can be answered by 
postulating
 phase transitions during the evolution of the universe
from the Planck temperature of $10^{32}$ K to the 2.7 K observed today.
Among the questions:
\begin{itemize}
\item {\bf The Matter-Antimatter Asymmetry in our Universe \\}
As first spelled out by Sakharov\cite{sak}, any theory trying
to explain the preponderance of matter in our universe must
necessarily implement:
\begin{itemize}
\item Baryon- and Lepton number violation;
\item C- and CP- violation;
\item Thermal non-equilibrium conditions,
as expected  after phase transitions.
\end{itemize}
\item {\bf The Dominance of Photons over Baryons \\}
If the excess of matter originates from small CP-violation  effects,
most matter and antimatter will have enough time to annihilate
into photons thus providing an explanation why the number
of photons as observed in the 3K microwave background radiation
is about $10^9$ to $10^{10}$  times as high as the number of baryons
in our universe.
\item {\bf Inflation\\}
An inflationary phase, i.e. a rapid expansion generated
by a potential energy term, which, according to Einstein's equations of 
General Relativity, provides  a repulsive instead of attractive gravitational 
force,  is the only viable explanation to solve the following problems:
\begin{itemize}
\item{\bf Horizon Problem \\}
The fact that the observed temperature of the 2.7 K microwave background
radiation is to a very high degree the same in all directions
can only be explained if we assumes that all regions were in causal
contact with each other at the beginning. However, the size
of our universe is larger than the ``horizon'', i.e. the distance
light could have traveled since the beginning. Therefore, one
can only explain the temperature isotropy, if one assumes
that all regions were in causal contact at the beginning and that
space expanded faster than the speed of light.
This is indeed the case in the inflationary scenario.
\item {\bf Flatness Problem \\}
Experimentally the observed density in our universe is close
to the so-called critical density, which is the density,
where the total energy of the universe is zero, i.e. the
kinetic energy of the expanding universe is just compensated
by the gravitational potential energy.
This corresponds to a ``flat'' universe, i.e. zero curvature.
The inflationary scenario naturally explains, why the universe is so
flat: the rapid expansion by more than 50 orders of magnitude drives
all curvature to zero\footnote{Just like blowing up a balloon
removes all wrinkles and curvature from the
surface.}.
\item {\bf Magnetic Monopole Problem } \\
Magnetic monopoles are predicted by GUT's.
 Their absence in our universe is explained
by the inflationary models, if one assumes
 the inflation to go smoothly from a single homogeneous
region to a large homogeneous region many
times the size of our universe (so-called {\it new}
 inflation).
\item {\bf The Smoothness Problem}\\
Experimentally the cosmic background radiation shows the features in accord with
the Harrison-Zel'dovich scale-invariant
spectrum (n=1)\cite{cobe}, which is the
spectrum expected after inflation.
The scale invariance can be understood
as follows:
 the inflation smoothens out out any inhomogeneities
which might have been present in the initial
conditions. Then in the course of the phase transition
inhomogeneities are generated by the quantum
fluctuations of the Higgs field on a very small scale
of length, namely the scale where quantum effects are
important. These density fluctuations are then enlarged to an astronomical scale
by inflation and they stay scale invariant
as is obvious if one thinks about e.g. a little circle
on a balloon, which stays a circle after inflation, but just on a larger scale.
\end{itemize}
\end{itemize}
In Grand Unified Theories the  conditions needed for the inflationary Big 
Bang Theory are naturally met:
at least two phase transitions, which generate mass and thus provide
non-equilibrium conditions, are expected:
 one at the unification scale of $10^{16}$ GeV,
 i.e. at   temperature of about $10^{28}$ K and one at the electroweak scale,
 i.e. a temperature of about $10^{16}$ K.
Furthermore, a potential energy term in the vacuum is  expected from the 
scalar fields in these theories, which are needed to generate particle 
masses in a gauge-invariant way. In the minimal model at least 29 scalar 
fields are required. Unfortunately,
none have been discovered so far, so little is known about the scalar
sector. 

Nevertheless, the arguments in favour of inflation are so strong, that it has 
become the only acceptable paradigm of present cosmology. Experimental 
observation of scalar fields would provide
a great boost in the acceptance of the role of scalar fields, both in 
cosmology and particle physics.

\chapter{Comparison of GUT's with Experimental Data}\label{ch6}
In this chapter the various low energy GUT predictions are
compared with data. The most  restrictive constraints are
the coupling constant unification combined with the lower
limits on the proton lifetime. They exclude the 
SM\cite{el2,abf,lalu} as well
as many other models\cite{abf,abfI,yana} with 
either a more complicated Higgs sector
or models, in which one searches for the minimum
number of new particles  required to fulfil the
constraints mentioned above.
From the many models tried, only a few yielded
unification at the required energies, but these
models have particles introduced ad-hoc without
the appealing properties of Supersymmetry.
Therefore we will concentrate here on the 
supersymmetric models
and ask if the predictions of the simplest, i.e. {\it minimal}
models\cite{su5susy} are consistent with
all the constraints described in the previous chapters.
%   Other non-minimal supersymmetric
% models have been studied too\cite{nanop}.
 The relevant RG equations for the running of the
couplings and the masses are given in the appendix.
Assuming soft symmetry breaking at the GUT scale, 
all SUSY masses can be
expressed in terms of 5 parameters and the masses 
at low energies are then
determined by the well known Renormalization Group 
(RG) equations. So many
parameters cannot be derived from the unification 
condition alone,
 However, further constraints can be considered:
\begin{itemize}
\item $M_Z$ predicted from electroweak  symmetry
breaking\cite{ewbr,rrb,ir,roskane,loopewbr}.
\item b-quark mass predicted from the unification of Yukawa
couplings\cite{bmas1,ara1,ara2}.
\item Constraints from the lower limit on the proton
lifetime~\cite{arn,relictst,lanpol}.
\item Constraints on the relic density in the
universe~\cite{relictst,roskane}.
\item Constraints on the top mass ~\cite{mtmax,rrb,ir,ara2}.
\item Experimental lower limits on SUSY masses~\cite{pdb,higgslim}.
\item Constraints from $b\rightarrow s\gamma$ 
decays\cite{bsgamma,bsgamm1,bsgamm2,roskane}.
\end{itemize}
Of course, in many of the references given
above, several constraints are studied
simultaneously, since
considering   one constraint at a time yields
only one relation between parameters.  Trying to 
find complete solutions with only a few constraints 
requires then
additional assumptions, like naturalness, no-scale 
models, fixed ratios
for gaugino- and scalar masses or a fixed ratio  
for the Higgs mixing
parameter  and the scalar mass,  or
combinations of these assumptions.

Several ways to study the constraints simultaneously
have been pursued. One can either sample the whole 
parameter space in a systematic or
random way and check the regions which are allowed 
by the experimental constraints.

Alternatively, one can try a statistical
analysis, in which all the constraints are
implemented in a $\chi^2$ definition and try to 
find the most probable region of the parameter 
space by minimizing the $\chi^2$ function.

In the first case one has to ask: which weight 
should one give to the
various regions of parameter space and how large 
is the parameter space? Some sample
the space only logarithmically, thus emphasizing 
the low energy regions\cite{roskane}, others provide a linear 
sampling\cite{car,nanop,bor,ez}.
In the second case one is faced with the difficulty, 
that the function to be  minimized is not monotonous,  
because of the experimental  limits  on the particle 
masses,   proton lifetime, relic density and so on.
At the transitions where these constraints become
effective, the derivative of the $\chi^2$ function 
is not defined. Fortunately, good
minimizing programs in multidimensional parameter 
space, which do not rely on the derivatives, exist\cite{minuit}.
The advantage of such a statistical analysis
is that one obtains probabilities for the
allowed regions of the parameter space and
can calculate confidence levels.
The results of such an analysis\cite{bek} will be 
presented after a short description
of the experimental input values.
Other analysis have obtained similar mass spectra 
for the predicted particles in the 
MSSM\cite{rrb,arnnat,roskane,bor,ram1} or extended
versions of the MSSM\cite{nanopo}.
\section{Unification of the Couplings}
\label{s61}
In the SM based on the group $\rm SU(3)\times 
SU(2)\times U(1)$   the
 couplings are defined as:
\bq\label{SMcoup}{\matrix{
\alpha_1&=&(5/3)g^{\prime2}/(4\pi)&=&5\alpha/(3\cos^2\theta_W)\cr
\alpha_2&=&\hfill g^2/(4\pi)&=&\alpha/\sin^2\theta_W\hfill\cr
\alpha_3&=&\hfill g_s^2/(4\pi)\cr}}
\eq%
where $g'~,g$ and $g_s$ are the $U(1)$, $SU(2)$ and $SU(3)$ coupling constants;
the first two coupling constants are related to the fine structure constant by
(see fig. \ref{f25}):
\bq \label{SW}
e = \sqrt{4\pi\alpha}=g\sin\theta_W=g'\cos \theta_W.\eq
   The factor of 
$5/3$ in the
definition of $\alpha_1$ has been included for 
the proper normalization at
the unification point (see eq. \ref{c53}).
 The couplings, when defined as
effective values including loop corrections in 
the gauge boson propagators,
become energy dependent (``running'').  A running coupling requires
the specification of a renormalization prescription, for which one usually
uses the modified minimal subtraction ($\MSbar$) scheme\cite{msb}.

In this scheme the world averaged values of the couplings at the
Z$^0$ energy are
\bqa\label{worave}
\alpha^{-1}(\MZ)&=&127.9\pm0.1\\
\sin^2\theta_\MSbar&=&0.2324\pm0.0005\\
\alpha_3&=&0.123\pm0.006.
\eqa
The value of $\alpha^{-1}$ is given in ref. \cite{dfs}\ and the
    value of $\sin^2\theta_\MSbar$ has been  been taken from a detailed
 analysis of all available data by
 Langacker and Polonsky\cite{sinms2}, which agrees
 with the latest analysis of the LEP data\cite{lep}. 
The error includes the uncertainty
from the top quark. We have not used the smaller 
error of 0.003 for a given value of $\mt$, since 
the fit was only done within the SM, not the MSSM, 
so we prefer to use the more conservative error including 
the uncertainty from $\mt$.

The $\alpha_3$ value
corresponds to the value at \mz\ as determined 
from quantities calculated in the 
``Next to Leading Log Approximation''\cite{bet}.
These quantities are less sensitive to the
  renormalization scale, which is
an indicator of the unknown higher order 
corrections; they are the dominant uncertainties 
in quantities relying   on second order QCD 
calculations. This $\alpha_s$ value is in excellent agreement with a
preliminary value of $0.120\pm 0.006$
   from a fit to the $\zz$ cross sections and 
asymmetries measured at LEP\cite{lep}, for which   the
third order QCD corrections have been calculated too;     
the renormalization scale uncertainty is correspondingly small.

The top quark mass was simultaneously fitted and found to be\cite{lep}:
\bq M_{top}=166^{+17~+19}_{-19~-22}~{\rm GeV},\label{mtop}\eq
where the first error is statistical and the 
second error corresponds to a variation
of the Higgs mass between 60 and 1000 GeV.
The central value  corresponds to a Higgs mass of 300 GeV.

For SUSY models, the dimensional reduction $\DRbar$ scheme is a more
appropriate renormalization scheme\cite{akt}.  
This scheme also has the advantage
that all thresholds can be treated by simple step 
approximations.  Thus
unification occurs in the $\DRbar$ scheme if all 
three $\aii(\mu)$ meet
exactly at one point.
                     This crossing point then gives 
the mass of the heavy
gauge bosons.  The $\MSbar$ and $\DRbar$ couplings 
differ by a small offset
\bq\label{MSDR}{{1\over\alpha_i^\DRbar}=
{1\over\alpha_i^\MSbar}-{C_i\over\strut12\pi}
}\eq
where the $C_i$ are the quadratic Casimir coefficients 
of the group ($C_i=N$
for SU($N$) and 0 for U(1) so $\alpha_1$ stays the same). 
 Throughout the
following, we use the $\DRbar$ scheme for the MSSM.

\section{\mz~ Constraint  from Electroweak Symmetry 
Breaking }\label{s62}
As discussed in chapter \ref{ch4}  
the electroweak breaking in the MSSM 
is triggered by the large negative corrections 
to the mass of one of the Higgs doublets\cite{ewbr}.
After including  the one-loop corrections to the Higgs 
potential\cite{erz,loopewbr},
 the following expression for \mz~can be found (see appendix):
\bqa \label{mzloop}M^2_Z&=&2\frac{\displaystyle m^2_1-m^2_2 \tan^2\beta -
 \Delta^2_Z}{\tan^2\beta -1}, \\
 \Delta^2_Z&=&\frac{3g^2}{32\pi^2}\frac{m^2_t}{M^2_W}\left[
 f(\tilde{m}^2_{t1})+f(\tilde{m}^2_{t2})+2m^2_t +
(A^2_tm_0^2-\mu^2\cot^2\beta )
 \frac{f(\tilde{m}^2_{t1})-f(\tilde{m}^2_{t2})}{\tilde{m}^2_{t1}-
 \tilde{m}^2_{t2}}\right]
 \eqa
where
$m_1$ and $m_2$ are the mass parameters in the Higgs
potential, $\tan\beta$ is the mixing angle
between the Higgs doublets and the function
$f$ has been defined in the appendix.
The corrections $\Delta_Z$ are zero if the top- and stop
quark masses are identical, i.e. if supersymmetry would be exact.
They grow with the difference $\tilde{m}^2_t-\mt^2$, so these
corrections become unnaturally large for large values of the
stop masses, as will be discussed later.
In addition to relation \ref{mzloop}  one finds 
from the minimzation of the potential a relation 
between $\tan\beta$ and $m_3$ (see appendix), 
so requiring electroweak breaking effectively 
reduces the original 5  free mass parameters to only 3.
\section{Evolution of the Masses}
\label{s63}
In the soft breaking term of the Lagrangian 
\mze~ and \mha~ are the
universal masses of the gauginos and scalar 
particles at the
GUT scale, respectively and $\mu$ constrains  the masses
of the Higgsinos.
At lower energies the masses of the SUSY particles
start to differ from these  universal masses
due to the radiative corrections. E.g. the 
coloured particles
get  contributions proportional to $\as^2$ from gluon loops,
while the non-coloured ones get contributions depending on
the electroweak coupling constants only.
The evolution of the masses is given by the  renormalization group
equations\cite{rgem,ir}, which have been 
summarized in the appendix. Approximate
numerical  mass formulae for the
squarks and sleptons,  
mass mixing between the top quarks, gauginos, and
the Higgs mass parameters  are given
in chapter \ref{ch4}.% (eqns. %\ref{sqsl,blr,t12,gaugino,mh12}).
 The exact
formulae can be found in the appendix.

\section{Proton Lifetime Constraints}
\label{s64}
GUT's predict proton decay and the present lower limits
on the proton lifetime $\tau_p$ yield quite strong  constraints
on the GUT scale and the SUSY parameters.
As mentioned at the beginning, the direct decay $p\rightarrow e^+\pi^0$
via s-channel exchange requires
the GUT scale to be above $10^{15}$ GeV. This is not fulfilled
in the Standard Model (SM), but always fulfilled 
in the Minimal Supersymmetric Standard Model (MSSM). 
Therefore we do not
consider this constraint.
However, the decays via box diagrams with winos and Higgsinos
predict much shorter lifetimes, especially in the 
preferred mode
 $p\rightarrow \overline{\nu} K^+$.
 From the present experimental lower limit of  
$10^{32}$ yr\cite{pdb} for
this decay mode Arnowitt and Nath\cite{arn}\  
deduce an upper limit
on the parameter B, which is proportional
to $1/\tau_p$:  \bq B<293\pm 42 (M_{H_3}/3\mgut)\ {\rm GeV}^{-1}.\eq
Here $M_{H_3}$ is the Higgsino mass, which is expected to be
 above $\mgut$; else it would induce
too rapid proton decay.
If $M_{H_3}$ would become much larger than
$\mgut$, one would enter the non-perturbative regime. Arnowitt and 
Nath\cite{arn1} give the following
acceptable range:
\bq 3<M_{H_3}/\mgut <10 \label{mh3}\eq
 To obtain a conservative upper limit
on $B$, we allow  $M_{H_3}$ to become an order 
of magnitude heavier
than $\mgut$, so we require \bq B< 977\pm 140\ 
{\rm GeV}^{-1}.\label{B}\eq

 The   uncertainties from the unknown heavy Higgs mass  are large
compared with the contributions from the first and third generation,
which contribute through the mixing in the CKM matrix.
Therefore   we only consider the second order generation
contribution, which can be written as\cite{arn}\ :
\bq B=-2(\alpha_2/(\alpha_3  \sin(2\beta))(m_{\tilde{g}}/
          m^2_{\tilde{q}}) ~10^6~{\rm GeV}^{-1}.\eq
One observes that the upper limit on $B$ favours small
 gluino masses  $m_{\tilde{g}}$, large squark masses
         $ m_{\tilde{q}} $, and small values of \tb.
         To fulfil   this constraint requires
                    \bq\tb < 10\label{tanb}\eq for the 
whole parameter space and requires a minimal
value  of the parameter  $\mze$ in case $\mha$ is not 
too large, since  $m_{\tilde{g}}\approx 2.7 \mha$ and  
$m_{\tilde{q}}^2\approx \mze^2+7\mha^2$ (see eq. \ref{msq}). 
The constraint can always be fulfilled for very large 
values of $\mha$.
However,  the finetuning constraint   \ref{susest}   
 implies $  m_{\tilde{g}} \le 1000$ GeV or  $\mha 
\le 350$ GeV. In this case
eq. \ref{B} requires $\mze$ to be  above a few 
hundred GeV, if $\mha$ becomes
of the order of 100 GeV or below, as will be discussed below.

\section{Top Mass Constraints}
\label{s65}
The top mass can be expressed as:
\bq \mt^2=(4\pi)^2\ Y_t(t)\ v^2\ \sin^2(\beta), \label{mt}\eq
where the running of the Yukawa coupling
as function of $t=log(\frac{M_X^2}{Q^2})$ is given by\cite{rgem}:
\bq Y_t(t)=\frac{\displaystyle Y_t(0)E(t)}{\displaystyle 1+6Y_t(0)F(t)}.
\label{ytt}\eq
One observes that $Y_t(t)$ becomes independent
of $Y_t(0)$ for large values of $Y_t(0)$, implying 
an upper limit on the top mass\cite{ir,mtmax}.
 Requiring electroweak symmetry breaking
implies a minimal value of the top Yukawa
coupling, typically $Y_t(0)\ge {\cal O}(10^{-2})$. 
In this case the term
   $6Y_t(0)F(t)$ in the denominator of \ref{ytt}
is much larger than one, since $F(t)\approx 290$ at 
the weak scale, where  $t\approx 66$.
 In this case $Y_t(t)=E(t)/6F(t)$, so from eq. \ref{mt} it follows:
\bq m_t^{2}=\frac{(4\pi)^2\ E(t)}{6F(t)}\ v^2\ \sin^2(\beta)\approx 
(190~{\rm GeV})^2\sin^2(\beta),\eq
where $E$ and $F$ are functions of the couplings only (see appendix).
The physical (pole) mass is about 6\% larger
than the running mass\cite{gas,gray}:
\bq M_{t}^{pole}=m_t \left(1+\frac{4}{3}\frac{\as}
{\pi}\right)\approx (200~{\rm GeV})\sin\beta,\label{topm}.\eq

The electroweak breaking conditions
require $\pi/4<\beta<\pi/2$ (eq. \ref{beta}); hence 
the equation above implies for the MSSM approximately:
\bq 145 < M_{t}^{pole} < 200 ~{\rm GeV},\label{mtlim} \eq
which is consistent with
the experimental value of 166 GeV as determined at 
LEP (see eq. \ref{mtop}).
Although  the latter  value was determined
from a fit using the SM, one does not expect shifts
outside the errors, if the fit would be made  for the MSSM.

 As will be shown in chapter \ref{ch6}  
for such large top masses, the b-quark 
mass becomes a sensitive function of $\mt$ 
and of the starting values of the gauge
couplings at $\mgut$. 

\section{b-quark Mass Constraint}
\label{s66}
As discussed in chapter \ref{ch3}  
the masses of the up-type quarks are  
arbitrary  in the $SU(5)$ model, but the masses
of the down-type quarks are related to 
the lepton masses within a generation, if
one assumes 
unification of the Yukawa couplings  at the
GUT scale. 
This does not work for the light quarks, but the
ratio of b-quark and $\tau$-lepton masses 
can be correctly
predicted by the radiative corrections to 
the masses\cite{bmas,bmas1}.

 To calculate the
experimentally observed mass ratio the 
second order
renormalization group equations for
the running masses have to be used. 
These equations are integrated between 
the value of the physical mass and $\mgut$.

 For the running mass of the b-quark we used\cite{gas}:
\bq m_b=4.25\pm0.3~ {\rm GeV}.\label{bmas}\eq
This mass depends on the choice of scale 
  and the value of $\as(m_b)$.
Consequently, we have assigned a rather
conservative error of 0.3 GeV instead of 
the proposed value of 0.1 GeV\cite{gas}.
Note that the running mass (in the 
$\overline{MS}$ scheme) is related to the
physical (pole) mass $M_b^{pole}$ by\cite{gray}:
\bq m_b=M_b^{pole}\left(1-\frac{4}{3}\frac{\as}
{\pi}-12.4(\frac{\as}{\pi})^2\right)\approx 0.825 \;M_b^{pole}, \eq
so $m_b=4.25$ corresponds to $M_b^{pole}\approx 5$ GeV.
We ignore the running of $m_\tau $ below 
$m_b$ and use for the pole mass:
$M_\tau=1.7771\pm 0.0005$ GeV\cite{taumas}.

 \section{Dark Matter Constraint }\label{s67}
As discussed in chapter \ref{ch5}
there is abundant evidence for the existence
of non-relativistic, neutral, non-baryonic
dark matter in our universe.
The lightest supersymmetric particle (LSP) is supposedly stable
and would be  an ideal candidate for dark  matter.

The present lifetime of the universe is at least $10^{10}$ years,
which implies an upper limit on the expansion rate (see eq.
\ref{ht}) and correspondingly on the total relic abundance
 (compare eq. \ref{rhoc}).
Assuming $h_0>0.4$ one finds that for each relic particle 
species $\chi$\cite{kolb}:
  \bq\Omega_\chi h^2_0<1.\eq
This bound can only be obeyed, if
most of the LSP's   annihilated into 
fermion-antifermion pairs, which in turn would
annihilate into  photons again.
As will be shown below, the LSP is most likely
a gaugino-like neutralino $\chi_0$.
 In this case the annihilation rate 
$\chi_0\chi_0\rightarrow f\overline{f}$ depends most
sensitively on the mass of the lightest 
(t-channel) exchanged  sfermion:
$\Omega_\chi h^2_0\propto m_f^4/m_\chi^2$\cite{rosdm}.
Consequently, the upper limit on the relic 
density implies an upper limit on the sfermion mass.
However, as discussed in chapter  \ref{ch4},
the neutralinos are mixtures of gauginos and 
higgsinos. The higgsino component also allows
s-channel exchange of the $Z^0$ and Higgs bosons. 
The size of the Higgsino component
depends on   the relative sizes of the elements 
in the mixing matrix \ref{neutmat}, especially  
on the mixing angle $\tb$ and the size of the 
parameter $\mu$ in comparison to $M_1\approx 
0.4m_{1/2}$ and $M_2\approx 0.8 m_{1/2}$.
Consequently, the relic density is a complicated 
function of the SUSY parameters,
especially if one takes into account the
resonances and thresholds in the annihilation cross 
sections\cite{griest}, but in
general one finds a large region in parameter
space where the universe is not 
overclosed\cite{relictst}. In the  preferred
gaugino-like neutralino region the relic
density constraint
translates into  an upper bound of about 
1000 GeV on  $\mze$~\cite{roskane}, except 
for large $\mha$, where some SUSY masses become much
larger than 1 TeV and are therefore
disfavoured by the fine-tuning criterion 
(see eq. \ref{susest}).

\section{Experimental lower Limits on SUSY 
Masses}\label{s68}
SUSY particles have not been found so far 
and from the searches
at LEP one knows that the lower limit on the 
charged leptons and charginos is
about half the \Z~ mass (45 GeV)\cite{pdb} 
and the Higgs mass has to be above
62 GeV\cite{higgslim}. The lower limit on 
the lightest neutralino is 18.4 GeV\cite{pdb},
while the sneutrinos have to
be above 41 GeV\cite{pdb}.
  These limits require  minimal values for the 
SUSY mass parameters.

There exist also limits on squark and gluino 
masses from the hadron colliders\cite{pdb}, but these
limits depend on the assumed decay modes.
Furthermore, if one takes the limits given above 
into account, the  constraints from the limits of all other
particles are usually fulfilled, so they
do not provide additional reductions of  the
parameter space in case of the {\it minimal} SUSY model.
\section{Decay $b\rightarrow s\gamma$}
\label{s69}
Recently    CLEO has published an upper bound 
for this transition $b\rightarrow s\gamma<5.4\cdot10^{-4}$\cite{cleo}. 
Furthermore a central value of $3.5\cdot10^{-4}$ 
and a lower limit of $1.5\cdot10^{-4}$  can be extracted from the
observed process $B\rightarrow K^*\gamma$\cite{cleo} and assuming that 
the branching ratio for this process is 15\% 
(using lattice calculations)\cite{bkgam}.

In the SM the transition $b\rightarrow s\gamma$ 
can happen through one-loop diagrams with a 
quark (charge 2/3) and a charged gauge boson. 
SUSY allows for additional loops involving a 
charged Higgs and the charginos and 
neutralinos\cite{bsgamma,bsgamm1,bsgamm3,bsgamm2}.
In SUSY large cancellations occur, since
the $W-t$ and $H^\pm-t$ loops have an opposite
sign as compared to the $\chi^\pm-\tilde{t}$ 
loop and all loops are of the same order of magnitude.

Kane et al.\cite{roskane} find acceptable
rates for the $b\rightarrow s\gamma$ ~ 
transition in the MSSM for
a large range of parameter space, even if they
include constraints from electroweak symmetry 
breaking and unification of gauge and
Yukawa couplings. They do not find as   strong 
lower limits on the charged Higgs boson masses   
as others\cite{bsgamm2}.
Similar conclusions were reached by Borzumati\cite{borz}, 
who used the
more complete calculations including the
flavour changing neutral currents\cite{bsgamm3}.
It turns out that with the present errors
the combination of all constraints
discussed above are more restrictive
than the limits on $b\rightarrow s\gamma$,
so we have not included it in the analyses
discussed below.
\section{Fit Strategy}\label{s610}
As mentioned before, given the five   parameters 
in the MSSM plus   $\agut$
and $\mgut$, all other SUSY masses, the b-quark 
mass, and $\mz $
   can be calculated
by performing the complete evolution of the
 couplings including all thresholds.

The proton lifetime prefers small  values of
$\tb$ (eq. \ref{tanb}), while all SUSY masses
are expected to be below 1 TeV from the 
fine-tuning argument (see eq. \ref{susest}).

Therefore   the following strategy  was adopted:
 \mze\ and \mha\ were varied between 0 and 1000 GeV and
$\tb$ between 1 and 10.
The trilinear coupling $A_t(0)$   at $\mgut$
was kept mostly at zero, but the large radiative
corrections to it were taken into account,
so at lower  energies it is unequal zero.
Varying $A_t(0)$ between $+3\mze$ and $-3\mze$ 
did not change the
results significantly, so   the following        
results are quoted for $A_t(0)$.

The remaining four
parameters   - $\agut,\ \mgut,\   \mu,\  $ and $Y_t(0)$ -
 were fitted with the MINUIT program\cite{minuit} by minimizing the
following $\chi^2$ function:

\begin{eqnarray*} \chi^2&=&
       {\sum_{i=1}^3\frac{(\aii(\mz)-\alpha^{-1}_{MSSM_i}(\mz))^2}{
\sigma_i^2}}      \\
 & &+\frac{(\mz-91.18)^2}{\sigma_Z^2}     \\
 & &+\frac{(\mb-4.25)^2}{\sigma_b^2}     \\
 & &+{\frac{(B  - 997)^2}{\sigma_B^2}} {(for ~B > 997)}  \\
 & &+{\frac{(D(m1m2m3))^2}{\sigma_D^2}} {(for~ D > 0)}     \\
 & &+{\frac{(\tilde{M}-\tilde{M}_{exp})^2}{\sigma_{\tilde{M}}^2}}
 {(for~\tilde{M} > \tilde{M}_{exp})}.\\
\label{chi2}
\end{eqnarray*}
The first term is the contribution of the 
difference between the three
calculated and measured gauge coupling  
constants at \mz~and  the following
two terms are the contributions from the 
\mz-mass ~and \mb-mass  constraints. 
The last three terms impose constraints from the proton
lifetime limits, from
electroweak symmetry
breaking, i.e. $D=V_H(v_1,v_2)-V_H(0,0) < 0$ 
 (see eq. \ref{m2}),
and from experimental lower limits on the SUSY masses.
The top mass, or equivalently, the top
Yukawa coupling enters sensitively into the
calculation of $\mb$ and $\mz$.
Instead of the top Yukawa coupling
one could have taken the top mass as a parameter. 
However, if the couplings are evolved from $\mgut$ 
downwards, it is more convenient to run also the 
Yukawa coupling downward, since the RGE of the 
gauge and Yukawa couplings form a set of coupled 
differential equations in second order (see appendix).
Once the Yukawa coupling is known at $\mgut$,
the top mass can be calculated at any scale.
 The top mass can be taken   as an input parameter too
 using the value from the LEP data.  
Unfortunately, the value from
the LEP data (eq. \ref{mtlep}) is not yet  
very precise  compared with the range expected
in the MSSM (eq. \ref{mtlim}), so it does not 
provide a sensitive constraint. Instead, we 
prefer  to fit the Yukawa coupling in order 
to obtain the most probable top mass in  the MSSM.
As it turns out,
the resulting parameter space from the minimization 
of this $\chi^2$ includes the space allowed by the 
dark matter   and the $b\rightarrow s\gamma$ constraints, which
have been discussed above.
% Therefore these
%have not been included in the $\chi^2$, 
%although they might have reduced the parameter 
%space slightly.

The following errors were attributed:
$\sigma_i$ are the experimental errors in the 
coupling constants,
as given above, $\sigma_b$=0.3 GeV,
 $\sigma_B$=0.14 GeV, while  $\sigma_D$ and 
$\sigma_{\tilde{M}}$ were set to 10 GeV.
 The values of the  latter errors are not   critical,
 since the corresponding terms in the numerator
 are  zero
 in case of a good fit and even for the 90\% C.L.  limits
 these constraints could be fulfilled and the $\chi^2$ was determined
 by  the other terms, for which one knows the errors.

For unification in the $\DRbar$ scheme,
 all three couplings $\aii(\mu)$ must
cross at a single unification point 
 $\mgut$\cite{drb}.  Thus in
these models
one can fit the couplings at $\mz$ by 
extrapolating from a single
 starting point at 	
 $\mgut$   back to $\mz$ for each of 
the $\alpha_i$'s and taking
into account all light thresholds. The fitting 
program\cite{minuit} will then adjust the starting values
of the four high energy  parameters ($\mgut\ ,\agut,\   \mu $ and $Y_t(0)$)
until  the five low energy values
 (three coupling constants, $\mz$ and $\mb$) are
``hit''. The fit is repeated for all values of $\mze$ and $\mha$
between 0 and 1000 GeV and $\tb$  between 1 and 10. Alternatively, 
fits were performed
in which $\mha$ was left free too.

The light thresholds are taken into account by 
changing the coefficients of the RGE at the 
value $Q=m_i$, where
the threshold masses $m_i$ are obtained from
the analytical solutions of the corresponding RGE 
(see section \ref{s44}).
These solutions depend on the integration range, 
which was chosen between $m_i$ and $\mgut$. 
However, since one does not know $m_i$ at the 
beginning, an iterative procedure has to be used: 
one  first uses $\mz$ as a lower integration limit, 
calculates $m_i$,
and uses this as lower limit in the next iteration. 
Of course, since the  coupling constants are running,
the latter have to be iterated too, so the values 
of $\alpha_i(m_i)$ have to be used for calculating 
the mass at the scale $m_i$\cite{rrb,acpz}.
Usually three iterations are enough to find a stable solution.

Following Ellis, Kelley and Nanopoulos\cite{el2} the 
possible effects from heavy thresholds
 are set to zero, since
proton lifetime   forbids Higgs triplet
masses  to be below
$\mgut$   (see eq. \ref{mh3}).
These heavy thresholds have been considered
by other authors   for different assumptions\cite{baha,sinms2,acz}.

SUSY particles influence the evolution only through their
appearance in the loops,  so they enter only in higher order.
Therefore it  is
sufficient to consider the loop corrections to the masses  
in first order,
in which case    simple analytical solutions can be found, even
if the one-loop  correction to the Higgs potential from the top
Yukawa coupling is taken into account (see appendix).
There is one exception: the  corrections to 
the bottom and tau mass are compared directly 
with data,    which implies that the second order
solutions have to be taken for the RGE predicting 
the ratio of the
bottom and tau mass. Since this ratio involves 
the top Yukawa coupling $Y_t$, the RGE for $Y_t$ 
has to be considered in second order too. These 
second order corrections are  important for the 
bottom mass, since the strong coupling constant becomes
large at the small scale of the bottom mass, i.e. 
$\alpha_s(m_b)\approx 0.2$.

In total  one has to solve a system of 18 coupled 
differential equations:  5 second order ones(for the 3 gauge
couplings, $Y_t$ and $Y_b/Y_\tau$) and 13 
first order ones (for the masses and parameters 
in the Higgs sector, see appendix).
The  second order ones are solved numerically\footnote{The program 
DDEQMR from the CERN library was used for the solution of these 
coupled second order differential equations.} taking
into account the thresholds of the light particles using
the iteration procedure discussed above.
Note that from the starting values
of all parameters at $\mgut$ one can
calculate all light thresholds from the simple first order equations before
one starts the numerical integration of the
five second order equations. 
Consequently, the program is fast in finding the optimum solution, even if
before each iteration the light thresholds
have to be recalculated.
\section{Results     }\label{s611}
The upper part of
fig. \ref{f61}  shows the evolution of the coupling constants
in the MSSM for two cases: one for the minimum value of the $\chi^2$
function given in eq. \ref{chi2} (solid lines) and 
one corresponding to
the 90\% C.L. upper limit of the thresholds    of the light
SUSY particles (dashed lines).
The  position of the light     thresholds 
is shown in the bottom part   as jumps 
in the first order $\beta$ coefficients, which are 
increased according to the entries in
table {\ref{ta1} as soon as a new threshold is passed. 
Also the second order coefficients
are changed correspondingly (see table \ref{ta2}), 
but their effect on the evolution
is not visible in the top figure in contrast
to the first order effects, which change the
slope of the lines considerably in the top figure.
One observes that the changes  in the coupling constants   
occur  in a rather
narrow energy regime, so qualitatively
this picture is very similar to fig. \ref{f41},
in which case all sparticles were assumed to be
degenerate at an  effective SUSY mass scale $\msusy$\cite{abf}.
Since the running of the couplings depends only
logarithmically on the sparticle masses, the
90\% C.L. upper limits are as large as several TeV, as
shown by the dashed lines in fig. \ref{f61} and more 
quantitatively
in table \ref{t61}.
In this table the initial choices of $\mze$
and $\tb$ as well as the fitted parameters 
$\agut,\ \mgut,\ \mha,\ \mu,\ Y_t(0)$ (and  the corresponding top mass
after running down $Y_t$ from $\mgut$ to $\mt$)
 are  shown at the top and given these parameters
the corresponding masses of the SUSY particles
can be calculated. Their values are given in the 
lower part of the table. Note we fitted here five parameters
with five constraints, so the $\chi^2$=0, if a good solution
can be found. This is indeed the case.
The upper and lower limits in table \ref{t61} will be discussed below. 

Only the
value of the top Yukawa coupling is given, since 
for the ratio of bottom and tau mass
only the ratio  of the Yukawa couplings enters,
not their absolute values. For the running of the 
gauge couplings and the mixing in the quark
sector, only the small contribution from the
top Yukawa coupling is taken into account, 
since for the range of $\tb$ considered,
all other Yukawa contributions are negligible.

As mentioned before, varying $A_t(0)$ between 
$+3\mze$ and $-3\mze$
does not influence the results very much, so 
its value at the unification scale was kept at 0,
but its non-zero value at lower energies  due 
to the large radiative corrections  was taken 
into account. The fits are shown for positive
values of the Higgs mixing paramter $\mu$,
but similar values are obtained for negative
values of $\mu$ with an equally good $\chi^2$
value for the fit.

The parameters  $\mze, ~\mha $ and $\mu$ are
correlated, as shown in fig. \ref{f62},
where the value of $\mu$ is shown for all
combinations of $\mze$ and $\mha$  between
100 and 1000 GeV.
 One observes that $\mu$ increases
with increasing \mze~ and $\mha$. The strong
correlation between $\mha$ and $\mu$ originates
mainly from the electroweak symmetry breaking
condition, but also from the fact that  the thresholds in
the running of the gauge couplings all have
to occur at a similar scale. For example, from
fig. \ref{f61} it is obvious that the  dashed      
lines for $1/\alpha_1$ and $1/\alpha_2$ will
not meet with the solid line of $1/\alpha_3$,
simply because the thresholds are too different;    
the thresholds in $1/\alpha_3$ are mainly
determined by $\mha$, while the thresholds
for the upper two lines include the winos
and higgsinos too, so one obtains automatically a positive
correlation between $\mu$ and $\mha$.

The $\chi^2$ value is acceptable in the whole region, 
except for the regions where either
$\mze$ or $\mha$ or both become very small, as shown 
in  fig. \ref{f63}.
The increase in this corner is completely due
to the constraint from the proton lifetime (see section \ref{s65}). 
This plot was made for $\tb=2$. For larger values
the region excluded by proton decay quickly increases; for $\tb=10$ 
practically the whole region is excluded.

One notices from fig. \ref{f62} already a strong
correlation between $\mu$ and $\mha$. This is
explicitly
     shown in fig. \ref{f64}.
 The steep walls   originate from  the experimental lower
 limits on the SUSY masses and the requirement
of radiative symmetry breaking.
In the minimum the $\chi^2$ value is zero, but one notices
 a long valley,   where the $\chi^2$ is only slowly increasing. 
Consequently, the upper limits on the sparticle masses, 
which grow with increasing values of $\mu$ and $\mha$,
become several TeV, as shown in table \ref{t61}.
The 90\% C.L. upper limits were obtained by
requiring an increase in $\chi^2$ of 1.64.
%Of course, the upper limits  depend  on the %errors, 
so  the rather
%conservative  errors  defined above were used.
 The upper limits are a sensitive function of the  
central value of $\as$: decreasing the central  value 
of $\as$ by two standard deviations (i.e.
 0.012) can increase   the thresholds of  sparticles
   several TeV. Acceptable fits are only obtained for   
input $\as$ values between 0.108 and   0.132, if the error
is kept at 0.006. Outside this range all requirements cannot
be met simultaneously any more, so the MSSM predicts $\as$ 
in this range.

As discussed previously, sparticle   masses in the 
TeV range spoil the cancellation
of the quadratic divergences.
 This can be seen explicitly in the corrections to $\mz$:
 $\Delta_Z$ is exactly zero if the masses of 
stop-- and top quarks
 are identical, but the corrections grow
quickly if the degeneracy is removed, as shown 
in fig. \ref{f65}.
 For the  SUSY masses at the minimum value of 
$\chi^2$ the
corrections to $\mz$ are    small.
If one requires that only solutions are allowed  
for which the
corrections to $\mz$ are not large compared with 
$\mz$ itself,
one has to limit the mass of the heaviest stop quark  to
about one TeV.
 The corresponding 90\% C.L. upper limits of the 
individual
sparticles masses are given in the right hand column 
of table \ref{t61}.
The correction to $\mz$ is 6 times \mz~ in this case. The limits are obtained
by scanning $\mze$ and $\mha$ till
the $\chi^2$ value increases by 1.64, while
optimizing the values of $\tb,~ \mu,~,\agut,~Y_t(0)$ and $\mgut$. 
The lower limits on the SUSY parameters are 
shown in the left column of 
table \ref{t61}. The lowest values of $\mze=45$ GeV
and $\mha$=85 GeV are required to have simultaneously a
sneutrino mass above 42 GeV and a  wino  mass above 45 GeV.
If the proton lifetime is included, the minimum value
of either $\mze$ or $\mha$ have to increase (see fig. \ref{f62}).
Since the squarks and gauginos are much more sensitive to 
$\mha$, one obtains the lower limits by increasing $\mze$.
The minimum value for $\mze$ is about 400 GeV in this case.
But in both cases the $\chi^2$ increase for the 
lower limits is  due to the b-mass,
which is predicted to be 4.6 GeV from the 
parameters determining the lower limits.

 The b-quark mass is a strong function of both 
$\tb$ and $\mt$, as  shown in fig. \ref{f66}; this dependence
originates from the $W-t$ loop to the bottom quark.
The horizontal band corresponds to the mass of the b-quark
after QCD corrections: $\mb=4.25\pm 0.3$ GeV (see eq. \ref{bmas}).
Since also $\mz$ is a strong function of the same
parameters, the requirement of gauge and Yukawa 
coupling unification together with electroweak 
symmetry breaking strongly constrains  the SUSY particle spectrum.
A typical  fit with a $\chi^2$ equal zero
is given in the central column of table \ref{t61}, 
but is should be noted that the
values in the other columns provide acceptable
fits too at the 90\% C.L..

The mass  of the lightest Higgs particle,
   called $h$ in table \ref{t61},
is a rather strong function of \mt,
as shown in fig. \ref{f67} for various choices
of $\tb$, $\mze$ and $\mha$. All other
parameters were optimized for these inputs and
after the fit the values of the Higgs and top
mass were calculated and plotted.
One observes that the mass of the lightest Higgs particle varies
between 60 and 150 GeV and the top mass between
134 and 190  GeV. Furthermore, it is evident
that $\tb$ almost uniquely determines the value
of $\mt$, since even if $\mha$ and $\mze$ are
varied between 100 and 1000 GeV, one finds the
practically the same $\mt$ for a given  $\tb$  and
the value of $\mt$ varies between 134 and 190 GeV, if $\tb$ is varied between
1.2 and 5. 
 This range is in excellent agreement with   
the estimates given in eq. \ref{mtlim},
if one takes into account that $M_t^{pole}\approx 1.06 \mt$ 
     (see eq. \ref{topm}).

Note the strong correlation between 
$\tb$ and $\mt$ in fig. \ref{f67}:
for a given value of $\tb$ $\mt$ is constrained
to a vary narrow range almost independent
of $\mha$ and $\mze$. Furthermore one observes a rather strong positive 
correlation between $m_higgs$ and all
other parameters ($\tb,~ \mze,~$ and $\mha$)  
originating from the loop corrections to the potential.

In summary, the following parameter
ranges are allowed (if  the limit on proton 
lifetime is obeyed and  extreme finetuning
is to be avoided, i.e. $\tilde{m}_{t2}< 1$ TeV):
\begin{eqnarray*}
   400~ <& ~m_0~ &<~ 1000~ \mbox{GeV}           \\
   80~ <& ~m_{1/2}~ &<~  475 ~\mbox{GeV}    \\
  330~ <& ~\mu~  &<~ 1100 ~\mbox{GeV}    \\
    1~ <& ~\tan \beta~  &<~ 10                        \\
  134~ <& ~m_t~  &<~ 190~\mbox{GeV~(from fig. \ref{f67}.)}            \\
0.108~ <& ~\alpha_s~  &<~0.132                      \\
\end{eqnarray*}

The corresponding constraints on the
SUSY masses are (see table \ref{t61} for details):
\begin{eqnarray*}
 25~<&   \chi^0_1(\tilde{\gamma})      &<~ 202 ~\mbox{GeV}  \\
 48~<&   \chi^0_2(\tilde{Z}) ,
    \chi^{\pm}_1(\tilde{W})            &<~  386 ~\mbox{GeV} \\
 217~<& \tilde{g}                     &<~ 1104 ~\mbox{GeV} \\
 440~<& \tilde{q}                     &<~ 1070  ~\mbox{GeV} \\
 240~<& \tilde{t}_1                   &<~  725 ~\mbox{GeV} \\
 414~<& \tilde{t}_2                   &<~ 1000  ~\mbox{GeV} \\
 406~<& \tilde{e}_L                   &<~  521 ~\mbox{GeV} \\
 401~<& \tilde{e}_R                   &<~  440  ~\mbox{GeV} \\
400~<& \tilde{\nu}_L                   &<~  516 ~\mbox{GeV} \\
 291~<&  \chi^0_3(\tilde{H}_1)    &<~ 
 799 ~\mbox{GeV} \\
313~<&  \chi^0_4(\tilde{H}_2)    &<~ 
 812 ~\mbox{GeV} \\
315~<&  \chi^\pm_2(\tilde{H}^\pm)    &<~ 
 831 ~\mbox{GeV} \\
527~<&   H^\pm     &<~ 
 1034 ~\mbox{GeV} \\
 523~<&       {H}                     &<~  1033  ~\mbox{GeV} \\
521~<&       {A}                     &<~  1031  ~\mbox{GeV} \\
  60~<&       {h}                     &<~  150  
                                     ~\mbox{GeV ~(from fig. \ref{f67}.)} \\
\end{eqnarray*}

The lower limits will all increase as soon as
the LEP limits on sneutrinos, winos and the lightest Higgs increase.

The lightest Higgs particle  is certainly within reach of
experiments at present or
future accelerators\cite{lep200,higgshunter}. Its observation in the predicted 
mass range
of 60 to 150 GeV   would
be a strong case   in support of
    this minimal version of a supersymmetric grand unified theory.
\begin{figure}
 \begin{center}
  \leavevmode
  \epsfxsize=15cm
  \epsfysize=18cm
  \epsffile{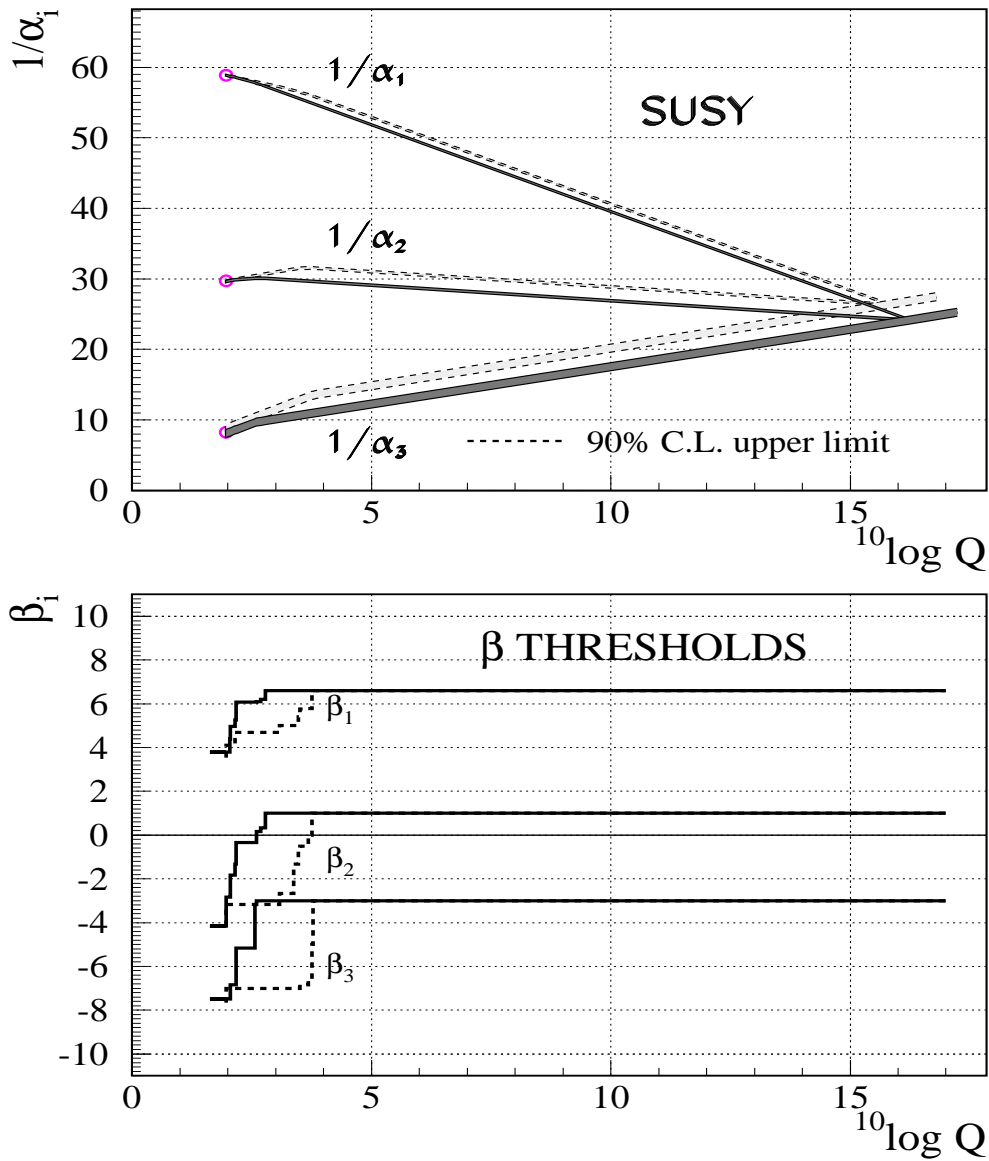}
 \end{center}
 \caption{\label{f61}
 Evolution of the inverse of the three couplings in the
 MSSM.                   The line  above $\MG $ follows the prediction
 from the supersymmetric SU(5) model.
The SUSY thresholds have been indicated in the lower part  of the curve:
they are treated as step functions in the
     first order $\beta$ coefficients in the  renormalization group
equations, which correspond to a change in slope in the evolution
of the couplings in the top figure.
The dashed lines correspond to the 90\% C.L. upper limit for the
SUSY thresholds.
}
\end{figure}

\clearpage
\begin{figure}
 \begin{center}
  \leavevmode
  \epsfxsize=15cm
  \epsfysize=18cm
  \epsffile{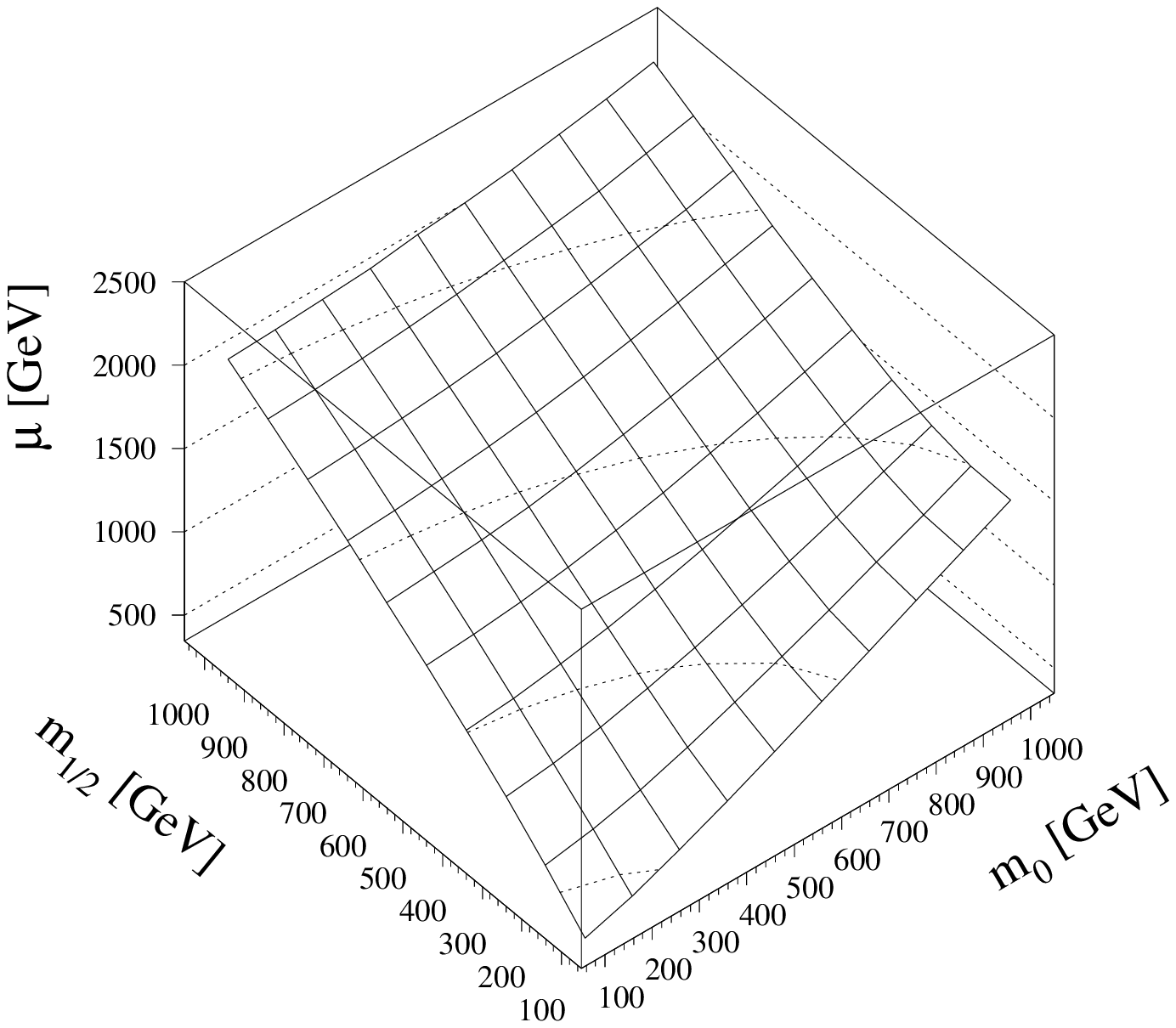}
 \end{center}
\caption{\label{f62}
   The fitted MSSM parameter    $\mu$
  as function of  $\mze$ and $\mha$ for $\tb=2$.
  }
\end{figure}

\clearpage
\begin{figure}
 \begin{center}
  \leavevmode
  \epsfxsize=15cm
  \epsfysize=18cm
  \epsffile{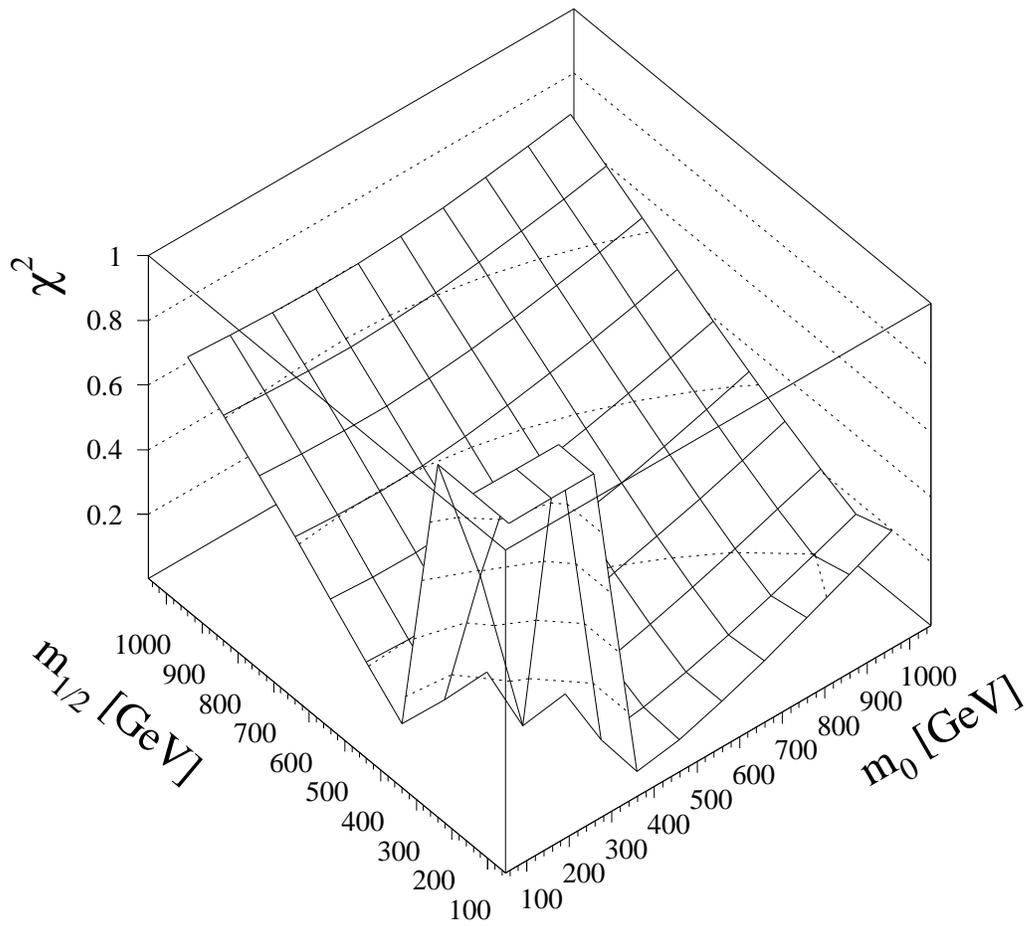}
 \end{center}
\caption{\label{f63}
   The  $\chi^2$ of the fit
  as function of  $\mze$ and $\mha$ for   $\tb=2$. The sharp increase 
in $\chi^2$ in the corner is caused by the lower limit on the proton lifetime.
  }
\end{figure}

\begin{figure}
 \begin{center}
  \leavevmode
  \epsfxsize=15cm
  \epsfysize=18cm
  \epsffile{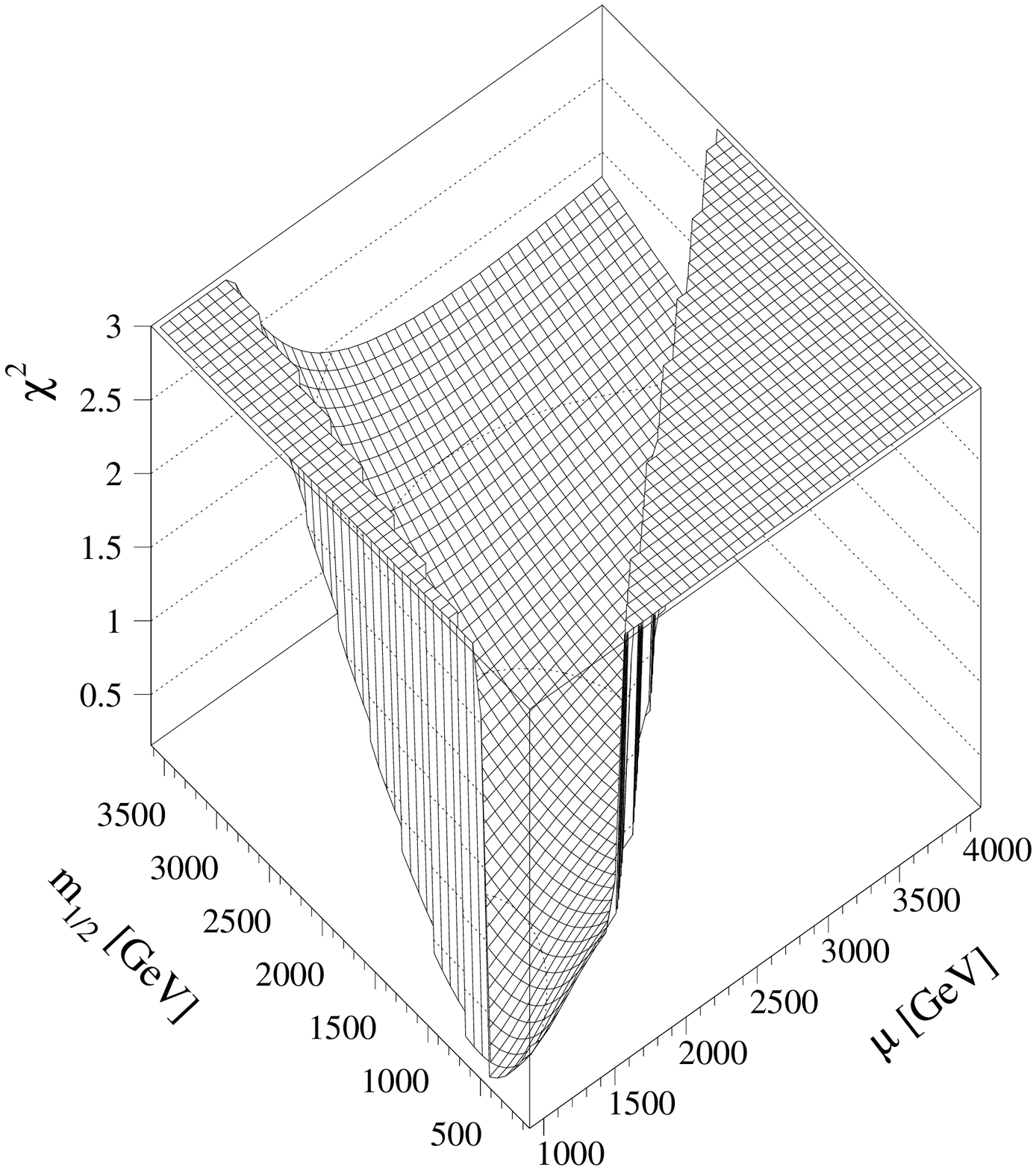}
 \end{center}
 %\vspace{18cm}
  \caption{\label{f64}
   The correlation between
   $\mha$ and $\mu$ for $\mze$=500 GeV.     }
\end{figure}

\clearpage
\begin{figure}
 \begin{center}
  \leavevmode
  \epsfxsize=15cm
  \epsfysize=18cm
  \epsffile{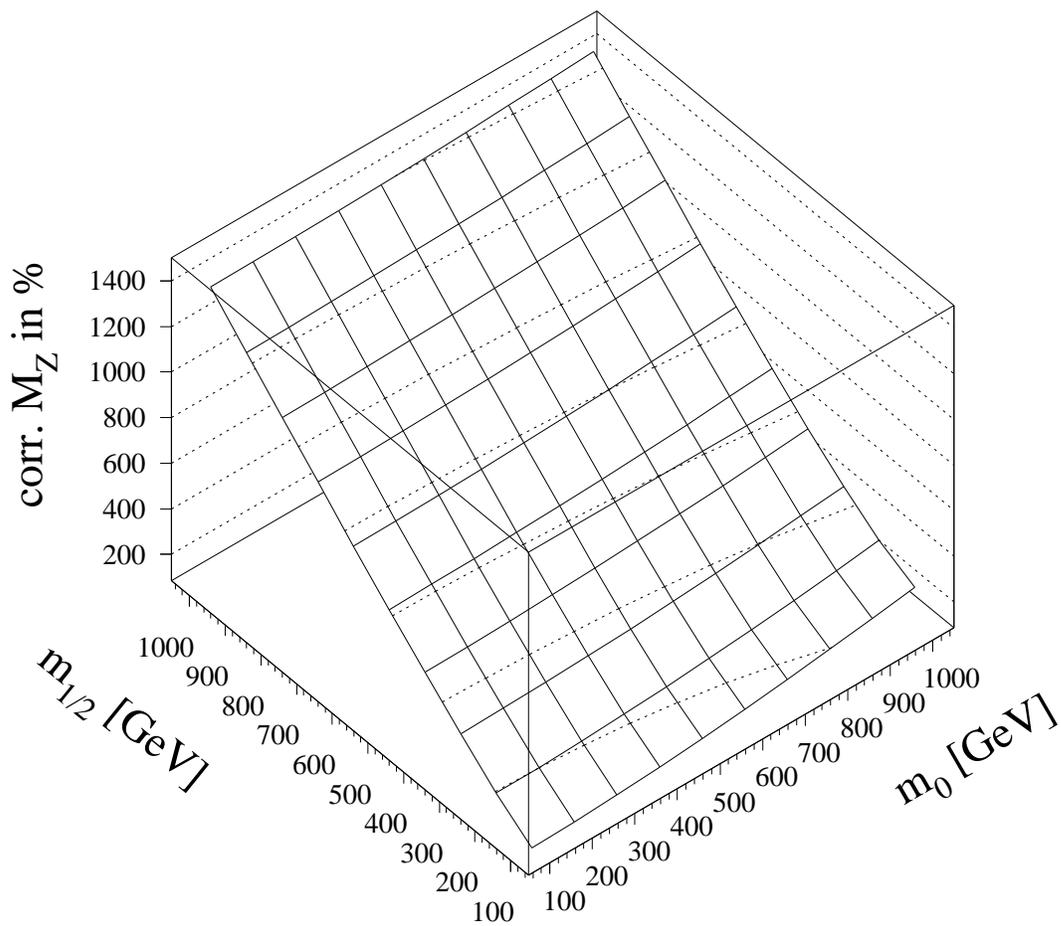}
 \end{center}
\caption{\label{f65}
   The  one-loop correction factor to $\mz$
  as function of  $\mze$ and $\mha$.
  }
\end{figure}
\clearpage

\begin{figure}
 \begin{center}
  \leavevmode
  \epsfxsize=15cm
  \epsfysize=18cm
  \epsffile{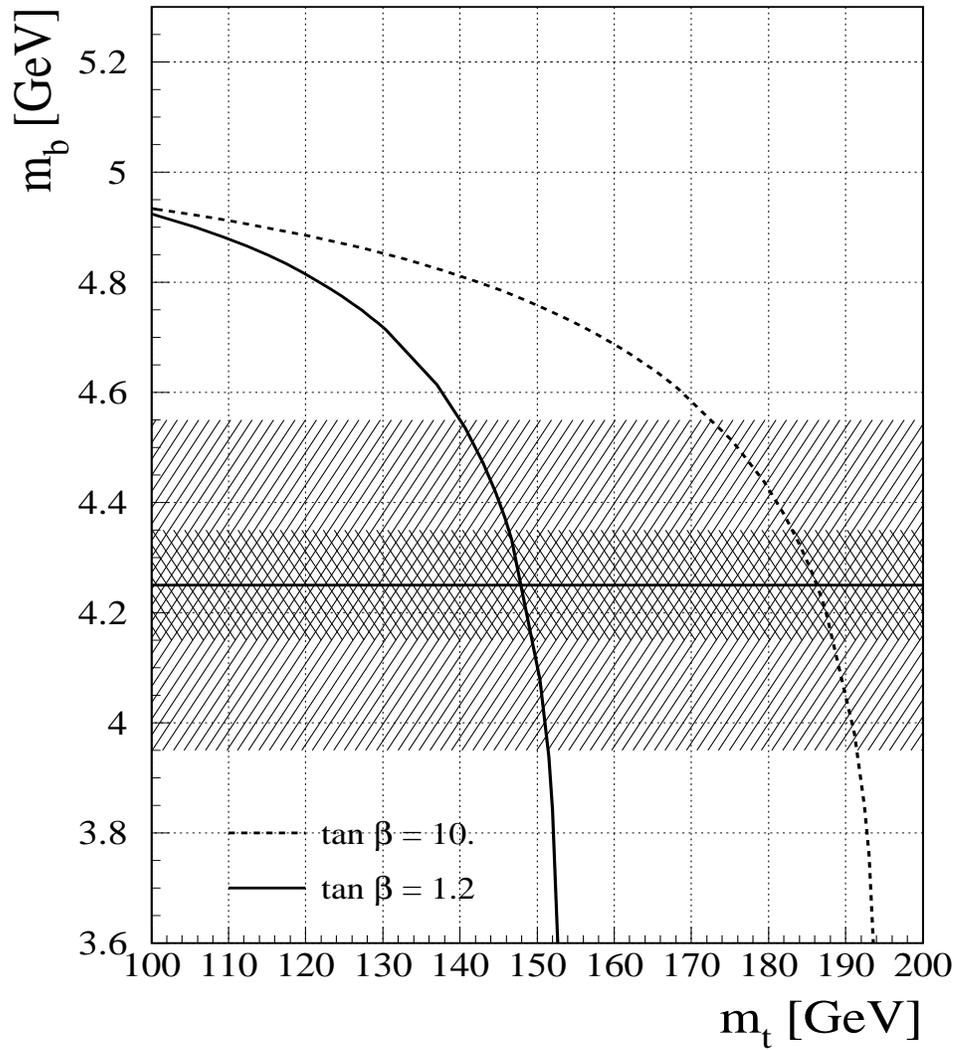}
 \end{center}
 \caption{\label{f66}
  The correlation between
  $\mb$                      and $\mt$ for $\mze$=400 GeV and two values 
of $\tb$.
  The hatched area indicates the experimental value for \mb.
           }
\end{figure}
\clearpage

\begin{figure}
 \begin{center}
  \leavevmode
  \epsfxsize=13cm
  \epsfysize=16cm
  \epsffile{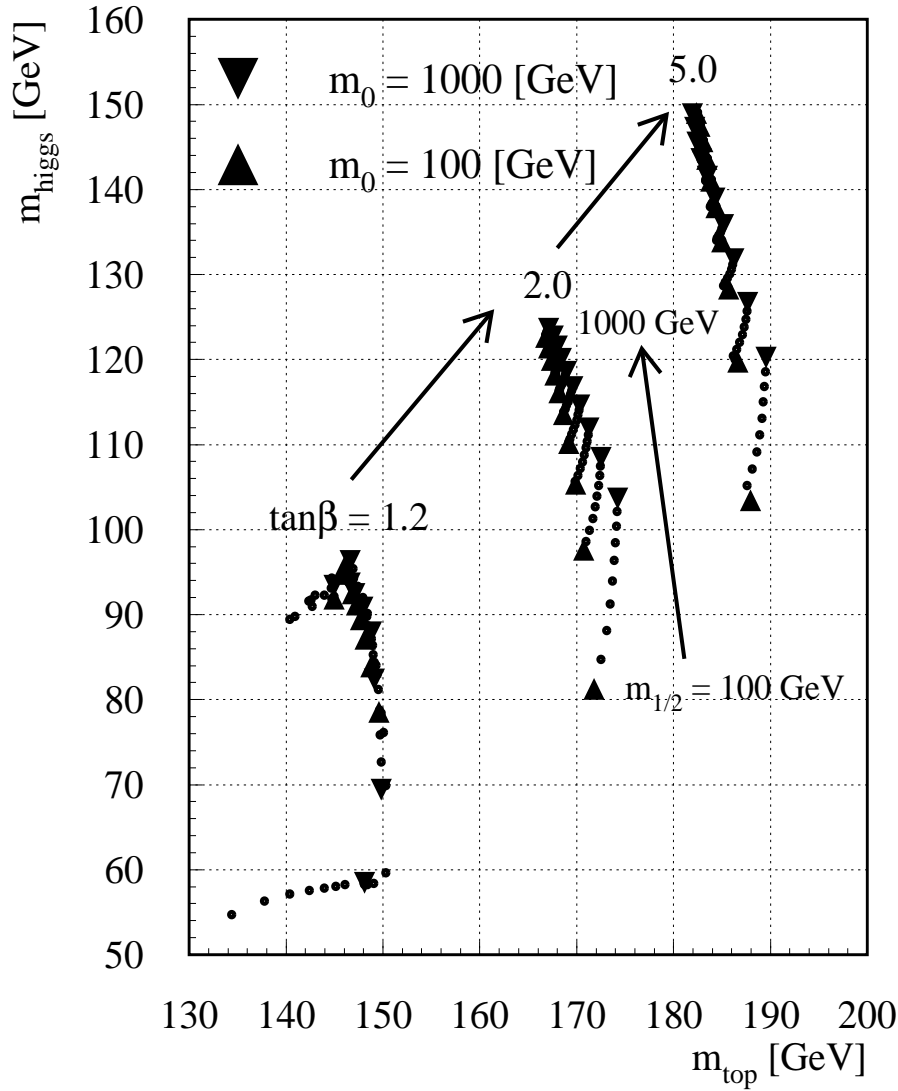}
 \end{center}
 \caption{The mass of the lightest Higgs particle as function of the top quark
  mass for   values of $\tb$ between 1.2 and 5   and values of $\mze$
and $\mha$ between 100 and 1000 GeV.
 The parameters  of $\mu, ~\mgut,~ \agut$ and $Y_t(0)$~are optimized
  for each choice of these parameters;    the corresponding values of the 
top and lightest Higgs mass  are shown as symbols.
 For small values of $\mha$ the Higgs mass increases with $\mze$, as shown 
for a ``string'' of points, each representing
a step of 100 GeV in $\mze$ for a given value of $\mha$, which is increasing 
in steps of 100 GeV,
starting with the low values for the lowest strings.
At high values of $\mha$ the value of $\mze$ becomes irrelevant and the 
``string''   shrinks to a point.
Note the strong positive correlation
between $m_{higgs}$ and all other parameters:
the highest
value of the Higgs mass corresponds to the maximum values of the input 
parameters, i.e.
$\tb=5$, $\mze=\mha=1000$ GeV; this
value does not correspond to the minimum $\chi^2$. More likely values 
correspond to $m_{higgs}\approx 92$ GeV for $\mha=100$, $\mze=400$ and $\tb=2$. }
\label{f67}
\end{figure}

\clearpage
\renewcommand{\arraystretch}{1.30}
\renewcommand{\rb}[1]{\raisebox{1.75ex}[-1.75ex]{#1}}
{\small
\begin{table}[t]
\begin{center}
\begin{tabular}{|c|r|r||r||r|r|}
\hline
Symbol& \multicolumn{2}{|c||}{ \makebox[4.6cm]{Lower limits }}&
      \makebox[2.3cm]{{\bf Typical fit}} &
       \multicolumn{2}{|c|}{ \makebox[4.6cm]{90\% C.L. Upper limits  }} \\
\hline
\hline
Constraints & \makebox[2.3cm]{ GEY } &\makebox[2.3cm]{ GEY+P } &
 \makebox[2.3cm]{ {\bf  GEY+(PF) }}&  
 \makebox[2.3cm]{ GEY+ (P)  }      &
 \makebox[2.3cm]{ GEY+(P)+F  }\\
\hline
\hline
 \multicolumn{6}{|c|}{ Fitted SUSY parameters } \\
\hline
 $m_0$   &  45 &400&{\bf 400} &  400 & 400               \\
\hline
 $m_{1/2}$  &85 &80 &{\bf 111} & 1600 & 475               \\
\hline
 $\mu$  &170 &330&{\bf 633} & 1842 & 1101             \\
\hline
 $\tan\beta$  &20. &3.0&{\bf 2.3} &  8.5& 2.9              \\
\hline
 $Y_t(0)$   &0.0047 &0.0035&{\bf 0.0140} & 0.0023 & 0.0084             \\
\hline
 $m_t$   &184 & 172&{\bf 177} &  168 & 178               \\
\hline
 $1/\alpha_{GUT}$  &24.0&24.3&{\bf 24.5} & 25.9 & 25.2
\\ \hline
$M_{GUT}$  &$2.0\;10^{16}$&$2.0\;10^{16}$& $
{\bf 2.0\;10^{16}}$ & $0.8\;10^{16}$ & $1.3\;10^{16}$ \\ \hline
\hline
 \multicolumn{6}{|c|}{SUSY masses in [GeV]} \\
\hline
\hline
  $\chi^0_1(\tilde{\gamma})$   & 28 &25 &{\bf  40}& 720  &  202  \\
\hline
  $\chi^0_2(\tilde{Z})$  &52&52&{\bf  78}  & 1346  &  386   \\
\hline
  $\chi^{\pm}_1(\tilde{W})$    &  49&48&{\bf  76}  & 1347 &  386   \\
\hline
  $\tilde{g}$    &  235 &217&{\bf 293}& 3377 & 1105   \\
\hline  \hline
  $\tilde{e}_L$      &  90 &406&{\bf 410}& 1160  &  521   \\
\hline
  $\tilde{e}_R$    &  56&401&{\bf 402} & 729  &  440   \\
\hline
  $\tilde{\nu}_L$     &  42&400&{\bf404}  & 1157  &  516   \\
\hline  \hline
  $\tilde{q}_L$   &  221&443&{\bf 477} & 3030  & 1071   \\
\hline
  $\tilde{q}_R$   &  213&440&{\bf 471} & 2872  & 1030   \\
\hline
 $\tilde{b}_L$   &  200&352&{\bf 370} & 2610  & 903  \\
\hline
  $\tilde{b}_R$   &  215&440&{\bf 471} & 2862  & 1027  \\
\hline
  $\tilde{t}_1$    &181&240&{\bf 213} & 2333  &  725  \\
\hline
  $\tilde{t}_2$  & 311 & 414&{\bf 450} & 2817  & 1008   \\
\hline        \hline
  $ \chi^0_3(\tilde{H}_1)$  &  157  & 292&{\bf  404}& 1771  &  799  \\
\hline
  $\chi^0_4(\tilde{H}_2)$   &  181&313&{\bf 423}& 1780  &  812  \\
\hline
  $\chi^{\pm}_2(\tilde{H}^{\pm})$& 186&315&{\bf 429}& 1816  &  831   \\
\hline   \hline
  $       h $    &   105&96&{\bf  97}&  146 &  127   \\
\hline
  $       H $   & 145&523&{\bf 629}& 2218  &  1033   \\
\hline
  $       A $   & 145&521&{\bf 627}& 2217  &  1031   \\
\hline
  $       H ^{\pm}$    &165&527&{\bf 631 }& 2219  &  1034  \\
\hline
 \end{tabular} \end{center}
 \caption{\label{t61} Values of SUSY masses and parameters for various 
constraints: G=gauge coupling unification;
 E=electroweak symmetry breaking;
 Y=Yukawa coupling unification;
 P=Proton lifetime constraint;
 F=finetuning constraint. Constraints in brackets indicate that they are 
fulfilled but not required. The value of the lightest Higgs $h$ can be 
lower than indicated (see text).
 }
\end{table}
}

\chapter{Summary.}\label{ch7}
Many of the questions posed by cosmology
suggest phase transitions during the evolution of the universe
from the Planck temperature of $10^{32}$ K to the 2.7 K observed today.
Among them the baryon asymmetry in our universe and
inflation, which is the only viable solution to explain the
horizon problem,
 the flatness problem, the
magnetic monopole problem, and
the smoothness problem (see chapter \ref{ch5}).

In Grand Unified Theories (GUT) phase transitions  are expected:
 one at the unification scale of $10^{16}$ GeV,
 i.e. at  a temperature of about $10^{28}$ K and one at the electroweak scale,
 i.e. at a temperature of about $10^{14}$ K.
Furthermore, scalar fields, which are a prerequisite for inflation, are 
included in GUT's. In the minimal model at least 29 scalar fields are 
required. Unfortunately,
none have been discovered so far, so little is known about the scalar
sector, although the verification of the relation between the couplings
and the masses of the electroweak gauge bosons indeed are
indirect evidence that their mass is generated by the interaction with a 
scalar field.
 Experimental observation of these scalar fields would provide
a great boost for cosmology and particle physics.
First estimates of the required mass spectra of the scalar fields 
can be obtained by comparing the experimental consequences of 
  Grand Unified Theories   (GUT) with low energy phenomenology.

 One of the interesting
``discoveries'' of LEP was the fact that within 
the Standard Model (SM) unification of the gauge 
couplings could be excluded (see fig. \ref{f41}). 
In contrast, the minimal supersymmetric extension of 
the SM (MSSM) provided perfect unification. This observation 
boosted the interest in Supersymmetry
enormously, especially since
 the MSSM was  not ``designed'' to provide unification, 
but it was invented many years
ago  and turned out to have  very interesting properties:
\begin{itemize}
\item
Supersymmetry automatically provides gravitational interactions,
thus paving the road for a ``Theory Of Everything''.
\item
The symmetry between bosons and fermions alleviates the divergences 
in the radiative corrections, in which case
these corrections can be made responsible for the
electroweak symmetry breaking at a much lower scale than
the GUT scale. 
%Unfortunately, a symmetry
%between fermions and bosons can only be 
%realized by assuming for each particle
%of the SM with spin $j$ there exists
%a supersymmetric partner with spin $j-1/2$,
%but these predicted superpartners have not been observed sofar.
%Consequently, the original enthousiasm for
%SUSY faded away in the last decennia
%after the ``spartner'' were not discovered
%despite intensive searches.
\item The lightest supersymmetric partner (LSP) is a natural candidate
for non-relativistic dark matter in our universe.
\end{itemize}
Other non-supersymmetric models can yield unification too,%\cite{abfI,yana}, 
but they do not exhibit the
elegant symmetry properties of supersymmetry, they offer no
explanation for dark matter and no explanation for the electroweak
symmetry breaking. Furthermore the quadratic divergences in
the radiative corrections do not cancel.

   The Minimal Supersymmetric Standard Model (MSSM) model has many 
predictions, which can be compared
with experiment, even in the energy range where the predicted
SUSY particles are out of reach. Among these predictions:
\begin{itemize}
\item
\mz.
\item
\mb.
\item
Proton decay.
\item
Dark matter.
\end{itemize}
It is surprising, that in addition to the unification of the coupling c
onstants the {\it minimal} supersymmetric model
can fulfil  all experimental constraints from these predictions.
As far as we know, supersymmetric   models are the only ones, which
are  consistent
 with all these observations simultaneously.
\begin{figure}
%   HNS  A
\begin{center}
\mbox{\epsfysize=15.cm\epsfxsize=10.cm\epsfbox{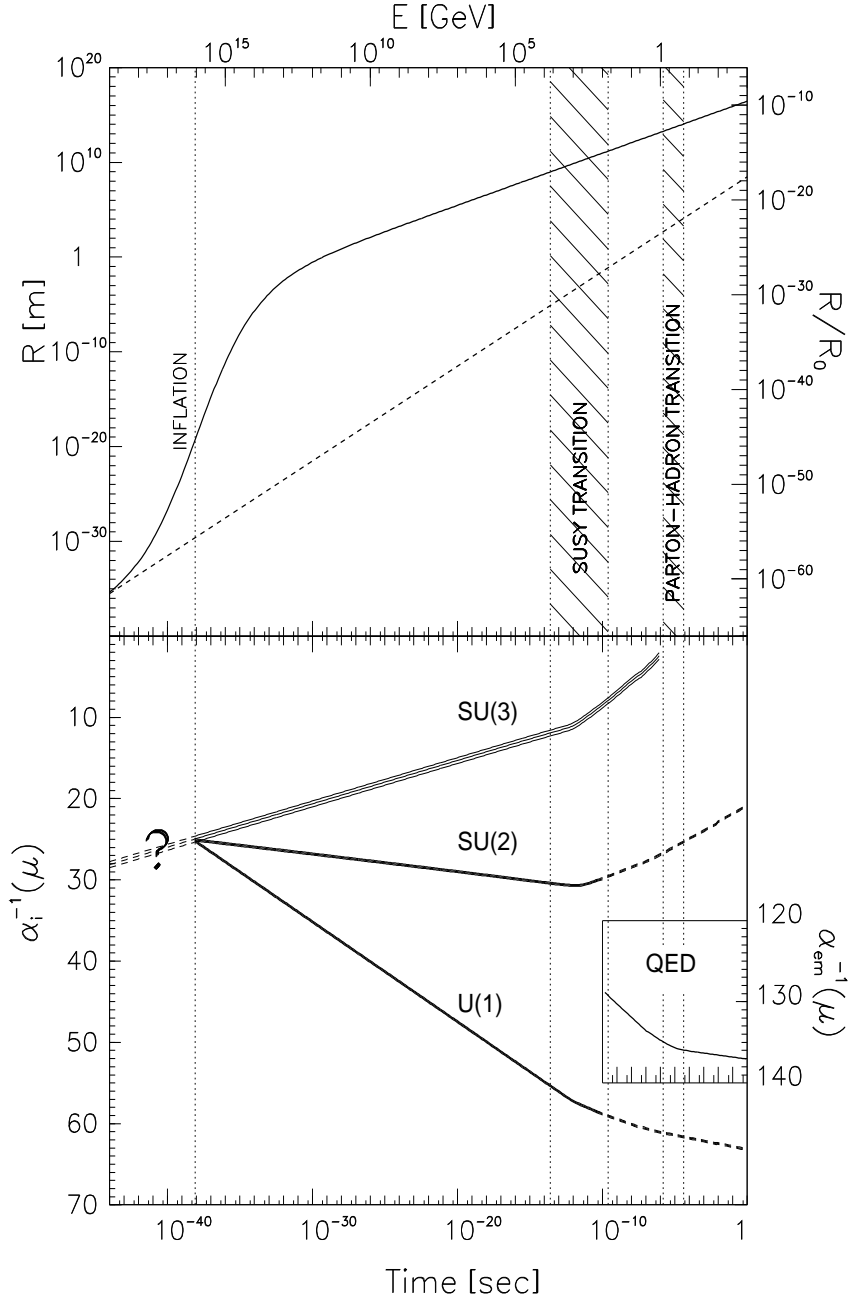}}
\vspace{1.cm}
%    HNS  E
\caption{Possible evolution of the   radius of the universe and the coupling
constants. Before $t=10^{-38}$ s spontaneous symmetry breaking 
occurs, which breaks the symmetry of the GUT into the well known 
symmetries at low energies. In the mean time the universe inflates 
to a size far above the distance light could have traveled as  
indicated by the  dashed line.  From [114].} 
\label{f71}
\end{center}
\end{figure}
Within the MSSM the evolution of the universe can be traced 
back to about $10^{-38}$ seconds after the `bang', as sketched 
in fig. \ref{f71}. If we believe in
 the inflationary scenario even the actual creation
of the universe is describable by physical laws.
In this view the universe would originate as a  quantum
fluctuation, starting from absolute ``nothing'', i.e. a state devoid of
space, time and matter with a total energy equal to zero.
Indeed, estimates of the total positive
non-gravitational energy and negative
potential energy are about equal
in our universe, i.e. according to this view
the universe
is the ultimate ``free lunch''.
All this mass  was generated from the
potential energy of the vacuum, which also caused  the inflationary phase.
%\footnote{ A classical analogy for
%the energy release during a phase %transition
%from a symmetric phase to a non-symmetric %phase is the
%freezing of water: during the transition %of the high temperature
%phase  with rotational symmetry to the low %temperature phase
%without this symmetry,  heat  is 
%released.}.

Of course, a quantum description of space-time can be discussed only
in the context of quantum gravity, so these ideas
must be considered speculative until  a   renormalizable theory of quantum
gravity is formulated and proven by experiment.
Nevertheless, it is fascinating to contemplate that
physical laws          may determine not only the evolution
of our universe, but they may remove  also the need for
assumptions      about the initial conditions.

From the experimental constraints  at
low energies the
 mass spectra  for the  SUSY particles can be predicted 
(see table \ref{t61}  in the previous chapter).
The lightest Higgs particle  is certainly within reach of
experiments at present or
future accelerators. Its observation in the predicted mass range
of 60 to 150 GeV   would
be a strong case   in support of
    this minimal version of the supersymmetric grand unified theory. 
Discovering also the heavier SUSY particles  implies  that the
known strong, electromagnetic and weak forces were all unified
 into a single ``primeval'' force during the birth of our universe.
 Future experiments will tell!

\newpage
{\bf\noindent Acknowledgments.}\\
\vspace{0.2cm}

I want to thank sincerely
Ugo Amaldi,    Hermann F\"urstenau, Ralf Ehret and Dmitri Kazakov for
their close collaboration in this exciting field. Without their
enthusiasm, work  and sharing of ideas many of our common results 
presented in this review would not have been available.
Furthermore, I thank John Ellis, Gian Giudice, Howie Haber, Gordy Kane, 
Stavros Katsanevas, Sergey Kovalenko, Hans K\"uhn,
Jorge Lopez, Dimitri Nanopoulos, Pran Nath, Dick Roberts, 
Leszek Roszkowski, Mikhail Shaposhnikov, William Trischuk, and Fabio Zwirner
for   helpful discussions and/or commenting parts of the manuscript.
% and         
%educating an ignorant newcomer in this 
%field.

Last, but not least, I want to thank Prof. Faessler for inviting me
to a seminar in T\"ubingen and his encouragement to write down
the results presented there in this review.

\addcontentsline{toc}{chapter}{Appendix A }
\setcounter{equation}{0}
\setcounter{chapter}{0}
\setcounter{table}{0}
\setcounter{section}{0} \renewcommand{\theequation}{A.\arabic{equation}}
\renewcommand{\thetable}{A.\arabic{table}}
\renewcommand{\thechapter}{A.\arabic{}}
\renewcommand{\thesection}{A.\arabic{section}}
\newpage
{\huge\bf Appendix A }
\section{Introduction }
In this appendix all the Renormalization Group Equations (RGE)
for the evolution of the masses and the couplings are given.
  SUSY particles influence the evolution only through their
appearance in the loops, so
  they enter only in higher order.
Therefore it  is
sufficient to consider the loop corrections to the masses only
in first order,
in which case  a simple analytical solution can be found, even
if the one-loop  correction to the Higgs potential from the top
Yukawa coupling is taken into account.
There is one exception: the  corrections to the bottom and tau mass are 
compared directly with data,    which implies that the second order
solutions have to be taken for the RGE predicting the ratio of the
bottom and tau mass. Since this ratio involves the top Yukawa coupling $Y_t$, 
the RGE for $Y_t$ has to be considered in second order too. These second 
order corrections are  important for the bottom mass, since the strong 
coupling constant becomes
large at the small scale of the bottom mass, i.e. $\alpha_s(m_b)\approx 0.2$.

So in total  one has to solve a system of 18 coupled differential 
equations (5 second order, 13 first order):
\begin{itemize}
\item 3 second order equations for the running of the gauge coupling  
constants $\alpha_i,~i=1,3$;
\item 2 second order equations for the running of the top Yukawa 
coupling $Y_t$ and the ratio of bottom and tau Yukawa coupling $R_{b\tau}$;
\item 1 first order equation for the masses of the left-handed   doublet of
an u-type and d-type squark pair $Q$;
\item 1 first order equation for the masses of the right-handed up-type  
squarks $U$;
\item 1 first order equation for the masses of the right-handed down-type  
squarks $D$;
\item 1 first order equation for the masses of the left-handed doublet of 
sleptons $L$;
\item 1 first order equation for the masses of the right-handed singlet of  
a charged lepton $E$;
\item 4 first order equations for the 4 mass parameters of the
Higgs potential ($m_1, m_2, m_3,$ and $\mu$);
\item 3 first order  equations for the (Majorana) masses of the 
gauginos ($M_1,M_2$ and $M_3$).
\item 1 first order equation for the trilinear coupling between
left- and right handed squarks and the Higgs field  $A_t$, where
 the subscript indicates that one only considers this coupling
for the third generation.
\end{itemize}
Note that   the absolute values of the bottom and tau Yukawa couplings  
need not to be known, if one neglects
their small contribution to the running of the gauge couplings.   
If one wants to include these, one
has to  integrate the RGE  for $Y_b$ and $Y_\tau$
separately (they are given below too) instead of the RGE for their ratio only.
Integrating the ratio has the advantage, that the
boundary condition at $\mgut$ is known to be one,
if one assumes Yukawa coupling unification.

The particle masses are related
directly to the Yukawa couplings:
\bqa Y_t(m_t)&=&\frac{h_t^2}{(4\pi)^2}; \ \ \ \ \
 m_{t} = h_t(m_t)\ v\ \sin\beta   \\
 Y_b(m_b)&=&\frac{h_b^2}{(4\pi)^2};\ \ \ \ \ \
 m_{b}=h_b(m_b)\ v\ \cos\beta  \\
Y_\tau(m_\tau)&=&\frac{h_\tau^2}{(4\pi)^2};\ \ \ \ \ \ 
 m_\tau=h_\tau (m_\tau)\ v\ \cos\beta.
\eqa
It follows that
\bq \frac{Y_t}{Y_b}=\frac{m_t^2}{m_b^2}\frac{1}{\tan^2 \beta}. \label{ytb}\eq
For  $\tb<10$, as required by proton decay limits, one observes that
 $Y_b$  is at least an order of magnitude
smaller than $Y_t$ for the values of $m_t$
considered. Hence its contribution is
indeed negligible in the running of the gauge
couplings, so below we will only consider
the contribution of $Y_b$ in the ratio 
of $Y_b/Y_{\tau}$, which is independent
of the absolute value of $Y_b(0)$.

Below we collect all the RGE's in a coherent notation and consider
the coefficients for the various threshold regions, i.e. virtual particles 
with mass $m_i$ are considered to contribute to the running of the gauge 
coupling constants effectively only for $Q$ values above $m_i$. Thus the 
thresholds are treated as simple step functions
in the coefficients of the RGE.

Furthermore,  all first order solutions are given in an analytical form, 
including the corrections from the top Yukawa coupling\cite{bek1}. Note 
that a more demanding analysis requires
a numerical solution of the first five second order equations, in which
the coefficients are changed according to the thresholds   found as 
analytical solutions of the  first order equations for the evolution of 
the masses. The results given in chapter \ref{ch6}
all use the numerical solution of these second order 
equation\footnote{The program DDEQMR from the CERN library was used for 
the solution of these coupled second order differential equations.}.

Using the supergravity inspired
breaking terms, which assume a common mass
$m_{1/2}$ for the gauginos and another common mass $m_0$
for the scalars, leads to the following  breaking term in the Lagrangian:
\begin{eqnarray} {\cal L}_{Breaking} & = &
-m_0^2\sum_{i}^{}|\varphi_i|^2-m_{1/2}\sum_{\alpha}^{}\lambda_\alpha
 \lambda_\alpha \label{2} \\ & - & 
                     Am_0\left[h^u_{ab}Q_aU^c_bH_2+h^d_{ab}Q_aD^c_bH_1+
 h^e_{ab}L_aE^c_bH_1\right] - Bm_0\left[\mu H_1H_2\right].
    \end{eqnarray}

Here

\begin{tabular}{ll}
$h^{u,d,e}_{ab}$ & are the Yukawa couplings, \ $a,b =1,2,3$ run over the
generations \\
 $Q_a$ & are the SU(2) doublet quark fields \\
$U_a^c$ & are the SU(2) singlet charge-conjugated up-quark fields \\
$D_b^c$ & are the SU(2) singlet charge-conjugated down-quark fields \\
$L_a$ & are the SU(2) doublet lepton fields \\
$E_a^c$ & are the SU(2) singlet charge-conjugated lepton fields \\
$H_{1,2}$ & are the SU(2) doublet Higgs fields \\ $\varphi_i$ & are all
scalar fields \\ $\lambda_\alpha $ & are the gaugino fields \end{tabular}

The last two terms in ${\cal{L}}_{Breaking}$ originate from the
cubic and quadratic terms in the superpotential with
A, B and $\mu$ as free parameters.
In total we  now have three couplings $\alpha_i$ and five mass parameters:
$$m_0,~m_{1/2},~ \mu(t),~A(t),~B(t).$$
with the following boundary conditions at $\mgut$ 
$(t=0)$:
\bqa
 {\rm scalars:}&&
 \tilde{m}^2_Q=\tilde{m}^2_U=
\tilde{m}^2_D=\tilde{m}^2_L=\tilde{m}^2_E=
m_0^2;\\
 {\rm gauginos:}&&
 M_i=m_{1/2}, \ \ \ i=1,2,3;\\
 {\rm couplings:}&&
 \tilde{\alpha}_i(0)=\tilde{\alpha}_{GUT},\ \ \ i=1,2,3 .\eqa
 Here $M_1$, $M_2$, and $ M_3$ are the  gauginos masses of 
the $U(1)$, $SU(2)$ and $SU(3)$ groups.In $N=1$ supergravity 
one expects at the Planck scale $B=A-1$.

With these parameters and the initial conditions
at the GUT scale the masses of all SUSY
particles can be calculated via the renormalization
group equations.

\section{Gauge Couplings }
The following definitions are used:
\bqa     \tilde{\alpha}_i&=&\frac{\alpha_i}{4\pi}\\
 t&=&\ln (\frac{\mguts}{Q^2})     \\
\beta_i&=&b_i~\tilde{\alpha}_{GUT} \\
f_i(t)&=&\frac{1}{\beta_i}\left(1-\frac{1}{(1+\beta_it)^2}\right)  \\
h_i(t)&=&\frac{t}{(1+\beta_it)}, \label{def}  \eqa
 where $\alpha_i$ (i=1,3)  denote the three gauge coupling constants
of $U(1)$, $SU(2)  and  SU(3)$, respectively,
$\alpha_{GUT}$ is the common gauge coupling at the GUT scale $\mgut$ 
and $b_i$ are the coefficients of the RGE,
as defined below.

The  second order RGE's for the gauge couplings including the effect of 
the Yukawa couplings are\cite{rgem,ir,rge1}:
\begin{eqnarray}
\frac{d\tilde{\alpha}_i}{dt} &
= & -b_i\tilde{\alpha}_i^2 -\tilde{\alpha}_i^2\left(
\sum_j b_{ij}\tilde{\alpha}_j-a_i Y_t\right),
\label{rge}
\end{eqnarray}
where $a_1=\frac{26}{5}, a_2=6,a_3=4$  for  SUSY and
$a_1=\frac{17}{10}, a_2=\frac{3}{2},a_3=2$  for  the SM.

                    The first order  coefficients for the SM are\cite{einjon}:
\begin{equation}
b_i=\left( \begin{array}{r} b_1 \\ b_2 \\b_3 \end{array} \right)
   =
\left( \begin{array}{r}           0    \\
                             - 22 / 3  \\
                                -11    \end{array} \right) +N_{Fam}
\left( \begin{array}{r}         4 / 3  \\
                                4 / 3  \\
                                4 / 3  \end{array} \right) + N_{Higgs}
\left( \begin{array}{r}         1 / 10 \\
                                1 / 6  \\
                                  0    \end{array} \right) ,
\label{smb1}
\end{equation}
while for the supersymmetric extension
 of the SM   (to be called MSSM in the following)\cite{einjon}:
\begin{equation}
b_i=\left( \begin{array}{r} b_1 \\ b_2 \\b_3 \end{array} \right)
   =
\left( \begin{array}{r}          0     \\
                                -6     \\
                                -9      \end{array} \right) +N_{Fam}
\left( \begin{array}{r}          2     \\
                                 2     \\
                                 2      \end{array} \right) +N_{Higgs}
\left( \begin{array}{r}          3/10  \\
                                 1/2   \\
                                 0      \end{array} \right)     ,
\label{susyb1}
\end{equation}
Here  $N_{Fam}$ is the number of families of matter supermultiplets
and $N_{Higgs}$ is the number of Higgs doublets.
We use  $N_{Fam}=3$ and $N_{Higgs}=1$ or 2, which corresponds to the
minimal SM or minimal SUSY model, respectively.

The second order coefficients are:

 \begin{equation}      b_{ij}=
\left(\begin{array}{rrr}
\rule{0cm}{0.5cm}
0&            0&            0\\
\rule{0cm}{0.5cm}
                                   0&-\frac{136}{3}&            0\\
\rule{0cm}{0.5cm}
                                   0&            0&         -102
\end{array}\right)   + N_{Fam}
\left(\begin{array}{rrr}
\rule{0cm}{0.5cm}
\frac{19}{15}&\frac{3}{5}  &\frac{44}{15}\\
\rule{0cm}{0.5cm}
                        \frac{1}{5}  &\frac{49}{3} &    4        \\
\rule{0cm}{0.5cm}
                        \frac{11}{30}&\frac{3}{2}  &\frac{76}{3}
\end{array}\right)   + N_{Higgs}
\left(\begin{array}{rrr}
\rule{0cm}{0.5cm}
\frac{ 9}{50}&\frac{9}{10} &    0        \\
\rule{0cm}{0.5cm}
                        \frac{3}{10} &\frac{13}{6} &    0        \\
\rule{0cm}{0.5cm}
                              0      &      0      &    0
\end{array}\right).\label{smb2}
\end{equation}
      For the         SUSY    model they become: % \cite{ein}:
\begin{equation}   b_{ij}=
\left(\begin{array}{rrr}
\rule{0cm}{0.5cm}
           0&            0&            0\\
\rule{0cm}{0.5cm}
                                   0&          -24&            0\\
\rule{0cm}{0.5cm}
                                   0&            0&          -54
\end{array}\right)   + N_{Fam}
\left(\begin{array}{rrr}
\rule{0cm}{0.5cm}
\frac{38}{15}&\frac{6}{5}  &\frac{88}{15}\\
\rule{0cm}{0.5cm}
                        \frac{2 }{5 }&     14      &    8        \\
\rule{0cm}{0.5cm}
                        \frac{11}{15}&      3      &\frac{68}{3}
\end{array}\right)   + N_{Higgs}
\left(\begin{array}{rrr}
\rule{0cm}{0.5cm}
\frac{ 9}{50}&\frac{9}{10} &    0        \\
\rule{0cm}{0.5cm}
                        \frac{3}{10 }&\frac{7}{2}  &    0        \\
\rule{0cm}{0.5cm}
                              0      &      0      &    0
\end{array}\right)   .
\label{susyb2}
\end{equation}

The contributions for the individual thresholds to $b_i$  and $b_{ij}$ are 
listed in tables \ref{ta1} (from ref. \cite{el2}) and \ref{ta2}, respectively.

The running of each $\alpha_i$ depends
on the values of the two other     coupling constants,
if the second order effects are taken into account.
However, these  effects are small, because the
           $b_{ij}$'s               are multiplied by
$     {\alpha_j}/{4\pi}\leq 0.01$. Higher orders are presumably
even smaller.

If the small Yukawa couplings are neglected, the RGE's
 \ref{rge} can be solved  by integration to obtain
$\alpha_i^\prime(\mu^{\prime  } )$ at a scale
$\mu^\prime$  for a given
$\alpha_i(\mu  )$:
\begin{equation}
\alpha_i^\prime(\mu^\prime) =
\left[ \beta_0 \cdot\ln  \frac{\mu^{\prime 2}}{\mu^2}+
\frac{1}{\alpha_i(\mu)}
+\frac{\beta_1}{\beta_0}
\ln\left(\frac{1/\alpha_i^{\prime}(\mu^\prime)+\beta_1/\beta_0}
              {1/\alpha_i (\mu)        +\beta_1/\beta_0}
   \right)\right]^{-1}
\label{itera}
\end{equation}
with
\begin{eqnarray}
\beta_0&=&   \frac{-1}{2\pi}
              \left( b_i + \frac{b_{ij}}{4\pi} \alpha_j(\mu)+
                           \frac{b_{ik}}{4\pi} \alpha_k(\mu)  \right) \\
\beta_1&=&  \frac{-2\cdot b_{ii}}{(4\pi)^2} .
\end{eqnarray}

This exact solution to the second order renormalization group equation
can be used to calculate the coupling constants at an arbitrary energy,
if they have been measured at a given energy,
i.e. one calculates
$\alpha_i(\mu^\prime)$ from a given $\alpha_i(\mu)$.
This transcendental equation is most easily solved numerically
by iteration. If the Yukawa couplings are included, their
running has to be considered too and one can solve the coupled
equations of gauge couplings and Yukawa couplings only numerically.
\section{Yukawa Couplings }
In order  to calculate the evolution of the Yukawa coupling for the b 
quark in the region between $m_b$ and $\mgut$, one has to consider four  
different  threshold regions:
\begin{itemize}
\item Region I between the typical sparticle masses $\msusy$ and the GUT scale.
\item Region II between $\msusy$ and the top mass $m_t$.
\item Region III between $m_t$ and $M_Z$.
\item Region IV between $M_Z$ and $m_b$.
\end{itemize}
\subsection{RGE  for Yukawa Couplings in Region I }
The second order RGE for the three Yukawa couplings of the third generation   
in the regions between $\msusy$ and $\mgut$ are\cite{bjo}:
%Second order RGE: SUSY from J.E. Bj\" orkman and D.R.I. Jones,
%Nucl. Phys. B259 (1985) 533;
\bqa
\frac{dY_t}{dt} & = & Y_t\left(\frac{16}{3}\tilde{\alpha}_3
+         3  \tilde{\alpha}_2 +
\frac{13}{15}\tilde{\alpha}_1-         6 Y_t\right.\nn
& &-(\frac{16}{3}b_3+\frac{128}{9})\tilde{\alpha}_3^2
   -(          3 b_2+\frac{9}{2})\tilde{\alpha}_2^2
   -(\frac{13}{15}b_1+\frac{169}{450})\tilde{\alpha}_1^2
-           8 \tilde{\alpha}_3\tilde{\alpha}_2
-\frac{136}{45}\tilde{\alpha}_3\tilde{\alpha}_1
- \tilde{\alpha}_2\tilde{\alpha}_1   \nn
& &\left. -16\tilde{\alpha}_3 Y_t
   -6\tilde{\alpha}_2 Y_t
   -\frac{6}{5}\tilde{\alpha}_1 Y_t   +22Y_t^2\right) \\
\frac{dY_b}{dt} & = & Y_b\left(\frac{16}{3}\tilde{\alpha}_3
+           3\tilde{\alpha}_2 +
\frac{7}{15}\tilde{\alpha}_1- Y_t\right. \nn
& &-(\frac{16}{3}b_3+\frac{128}{9})\tilde{\alpha}_3^2
   -(          3 b_2+\frac{9}{2})\tilde{\alpha}_2^2
   -(\frac{ 7}{15}b_1+\frac{49}{450})\tilde{\alpha}_1^2
-           8 \tilde{\alpha}_3\tilde{\alpha}_2
-\frac{8}{9}\tilde{\alpha}_3\tilde{\alpha}_1
- \tilde{\alpha}_2\tilde{\alpha}_1   \nn
& &\left.
   -\frac{4}{5}\tilde{\alpha}_1 Y_t   +         5 Y_t^2\right) \\
\frac{dY_\tau}{dt} & = & Y_\tau\left(
+           3\tilde{\alpha}_2 +
\frac{9}{5}\tilde{\alpha}_1     \right. \nn
& &\left. -(          3 b_2+\frac{9}{2})\tilde{\alpha}_2^2
   -(\frac{9}{5}b_1+\frac{81}{50})\tilde{\alpha}_1^2
- \frac{9}{5}\tilde{\alpha}_2\tilde{\alpha}_1\right)
\eqa
If one assumes Yukawa coupling unification for particles belonging
to the same multiplet, i.e. $Y_b=Y_\tau$ at the GUT scale,
one can calculate easily the RGE for the ratio  
$R_{b\tau}(t) = m_b /m_\tau =\sqrt{Y_b(t)/Y_\tau(t)}$:
\bqa
\frac{dR_{b\tau}}{dt} & = & R_{b\tau}\left(\frac{8}{3}\tilde{\alpha}_3
-\frac{2}{3}\tilde{\alpha}_1- \frac{1}{2}Y_t\right.\nn
& &-(\frac{8}{3}b_3+\frac{64}{9})\tilde{\alpha}_3^2
   +(\frac{ 2}{3}b_1+\frac{34}{45})\tilde{\alpha}_1^2
-           4 \tilde{\alpha}_3\tilde{\alpha}_2
-\frac{4}{9}\tilde{\alpha}_3\tilde{\alpha}_1
+\frac{2}{5} \tilde{\alpha}_2\tilde{\alpha}_1   \nn
& &\left.
   -\frac{2}{5}\tilde{\alpha}_1 Y_t   + \frac{5}{2}Y_t^2\right)
\eqa

\subsection{RGE  for Yukawa Couplings in Region II }
%from Fischler and J. Oliensis, Phys. Lett.  119 B %(1982) 385\\
For the region between $M_{SUSY}$ and $m_t$
one finds\cite{fis}:
\begin{eqnarray}
\frac{dY_t}{dt} & = & Y_t\left(           8\tilde{\alpha}_3
+ \frac{9}{4}\tilde{\alpha}_2 +
\frac{17}{20}\tilde{\alpha}_1-\frac{9}{2}Y_t\right.\nn
& &+108\tilde{\alpha}_3^2 +\frac{23}{4}\tilde{\alpha}_2^2
-\frac{1187}{600}\tilde{\alpha}_1^2 -9\tilde{\alpha}_3\tilde{\alpha}_2
-\frac{19}{15}\tilde{\alpha}_3\tilde{\alpha}_1+
 \frac{9}{20}\tilde{\alpha}_2\tilde{\alpha}_1   \nn
& &\left. -36\tilde{\alpha}_3 Y_t
   -\frac{225}{16}\tilde{\alpha}_2 Y_t
   -\frac{393}{80}\tilde{\alpha}_1 Y_t   +12Y_t^2\right) \\
\frac{dY_b}{dt} & = & Y_b\left(           8\tilde{\alpha}_3
+ \frac{9}{4}\tilde{\alpha}_2 +
\frac{1}{4}\tilde{\alpha}_1-\frac{3}{2}Y_t\right. \nn
& &+108\tilde{\alpha}_3^2 +\frac{23}{4}\tilde{\alpha}_2^2
+\frac{127}{600}\tilde{\alpha}_1^2 -9\tilde{\alpha}_3\tilde{\alpha}_2
-\frac{31}{15}\tilde{\alpha}_3\tilde{\alpha}_1+
 \frac{27}{20}\tilde{\alpha}_2\tilde{\alpha}_1   \nn
& &\left. - 4\tilde{\alpha}_3 Y_t
   -\frac{99}{16}\tilde{\alpha}_2 Y_t
   -\frac{91}{80}\tilde{\alpha}_1 Y_t   +\frac{1}{4}Y_t^2\right) \\
\frac{dY_\tau}{dt} & = & Y_\tau\left(
  \frac{9}{4}\tilde{\alpha}_2 +
\frac{9}{4}\tilde{\alpha}_1-3Y_t\right. \nn
& & +\frac{23}{4}\tilde{\alpha}_2^2
- \frac{1371}{200}\tilde{\alpha}_1^2
-\frac{27}{20}\tilde{\alpha}_2\tilde{\alpha}_1   \nn
& &\left. -20\tilde{\alpha}_3 Y_t
   -\frac{45}{8}\tilde{\alpha}_2 Y_t
   -\frac{17}{8}\tilde{\alpha}_1 Y_t   +\frac{27}{4}Y_t^2\right) \\
\frac{dR_{b\tau}}{dt} & = & R_{b\tau}\left(         4 \tilde{\alpha}_3
-\tilde{\alpha}_1+ \frac{3}{4}Y_t\right. \nn
& &+                           54\tilde{\alpha}_3^2
   + \frac{53}{15}\tilde{\alpha}_1^2
-\frac{9}{2}       \tilde{\alpha}_3\tilde{\alpha}_2
-\frac{31}{30}\tilde{\alpha}_3\tilde{\alpha}_1
+\frac{27}{20} \tilde{\alpha}_2\tilde{\alpha}_1   \nn
& &\left.
   +          8\tilde{\alpha}_3 Y_t
   -\frac{9}{32}\tilde{\alpha}_2 Y_t
   +\frac{79}{160}\tilde{\alpha}_1 Y_t - \frac{13}{4}Y_t^2\right)
\end{eqnarray}

\subsection{RGE  for Yukawa Couplings in Region III }
%WRONG!!!!!!!!!!! Still to be figured out!!
%$M_{SUSY}$ and $m_t$
%from Fischler and J. Oliensis, Phys. Lett.  119 B %(1982) 385\\
For the region between $\mt$ and $\mz$ one finds:
\begin{eqnarray}
\frac{dY_b}{dt} & = & Y_b\left(           8\tilde{\alpha}_3
+ \frac{9}{4}\tilde{\alpha}_2 +
\frac{1}{4}\tilde{\alpha}_1\right. \nn
& &\left. +108\tilde{\alpha}_3^2 +\frac{23}{4}\tilde{\alpha}_2^2
+\frac{127}{600}\tilde{\alpha}_1^2 -9\tilde{\alpha}_3\tilde{\alpha}_2
-\frac{31}{15}\tilde{\alpha}_3\tilde{\alpha}_1+
 \frac{27}{20}\tilde{\alpha}_2\tilde{\alpha}_1   \right) \\
\frac{dY_\tau}{dt} & = & Y_\tau\left(
  \frac{9}{4}\tilde{\alpha}_2 +
\frac{9}{4}\tilde{\alpha}_1 \right. \nn
& & \left. +\frac{23}{4}\tilde{\alpha}_2^2
- \frac{1371}{200}\tilde{\alpha}_1^2
-\frac{27}{20}\tilde{\alpha}_2\tilde{\alpha}_1  \right)  \\
\frac{dR_{b\tau}}{dt} & = & R_{b\tau}\left(         4 \tilde{\alpha}_3
-\tilde{\alpha}_1\right. \nn
& &\left.+                           54\tilde{\alpha}_3^2
   + \frac{53}{15}\tilde{\alpha}_1^2
-\frac{9}{2}       \tilde{\alpha}_3\tilde{\alpha}_2
-\frac{31}{30}\tilde{\alpha}_3\tilde{\alpha}_1
+\frac{27}{20} \tilde{\alpha}_2\tilde{\alpha}_1   \right)
\end{eqnarray}
\subsection{RGE  for Yukawa Couplings in Region IV }
\begin{eqnarray}
\frac{dR_{b\tau}}{dt} & = & R_{b\tau}\left(         4 \tilde{\alpha}_3
-\tilde{\alpha}_1
   + 54\tilde{\alpha}_3^2
   + \frac{53}{15}\tilde{\alpha}_1^2
-\frac{31}{30}\tilde{\alpha}_3\tilde{\alpha}_1
  \right)
  \end{eqnarray}
\section{Squark and Slepton Masses }
%The RGE for the sparticle masses are\cite{iba}:
%notation  of
%Ibanez et al, Nucl Phys. B256 (1985) 218)
  Using the notation introduced at the beginning, the
RGE equations for the squarks and sleptons can be written as\cite{rgem}:
\bqa
\frac{d\tilde{m}^2_L}{dt} & = & \left(
 3\tilde{\alpha}_2M^2_2 + \frac{3}{5}\tilde{\alpha}_1M^2_1\right) \\
\frac{d\tilde{m}^2_E}{dt} & = & (
 \frac{12}{5}\tilde{\alpha}_1M^2_1) \\
\frac{d\tilde{m}^2_Q}{dt} & = & (\frac{16}{3}\tilde{\alpha}_3M^2_3
+ 3\tilde{\alpha}_2M^2_2 + \frac{1}{15}\tilde{\alpha}_1M^2_1)
-\delta_{i3}Y_t(
\tilde{m}^2_Q+\tilde{m}^2_U+m^2_2+A^2_tm_0^2-\mu^2) \nn & & \\
\frac{d\tilde{m}^2_U}{dt} & = & \left(\frac{16}{3}\tilde{\alpha}_3M^2_3
+\frac{16}{15}\tilde{\alpha}_1M^2_1\right)
-\delta_{i3}2Y_t(
\tilde{m}^2_Q+\tilde{m}^2_U+m^2_2+A^2_tm_0^2-\mu^2) \\
\frac{d\tilde{m}^2_D}{dt} & =
 & \left(\frac{16}{3}\tilde{\alpha}_3M^2_3
+ \frac{4}{15}\tilde{\alpha}_1M^2_1\right) 
 \eqa
The $\delta_{i3}$ factor ensures that this term
is only included for the third generation.
\subsection{Solutions for the squark and slepton masses.}
The solutions for the RGE given above are \cite{ir}:
%\underline{Squarks and Sleptons masses }

\begin{eqnarray}
\tilde{m}^2_{E_{L}}&=&m^2_0+m^2_{1/2}\tilde{\alpha}_{GUT}
\left(\frac{3}{2}
f_2(t)+\frac{3}{10}f_1(t)\right)
-\cos(2\beta)M_Z^2(\frac{1}{2}-\sin^2\theta_W)\\
\tilde{m}^2_{\nu_{L}}
                   &=&m^2_0+m^2_{1/2}\tilde{\alpha}_{GUT}
\left(\frac{3}{2}
f_2(t)+\frac{3}{10}f_1(t)\right)
+\cos(2\beta)      \frac{1}{2}M_Z^2\\
   \tilde{m}^2_{E_{R}}
&=&m^2_0+m^2_{1/2} \tilde{\alpha}_{GUT}\left(\frac{6}{5}f_1(t)\right)
    -\cos(2\beta)M_Z^2\sin^2\theta_W  \\
\tilde{m}^2_{U_{L}}
&=&m^2_0+m^2_{1/2}\tilde{\alpha}_{GUT}
\left(\frac{8}{3}f_3(t)
+\frac{3}{2}f_2(t)
+\frac{1}{30}f_1(t)\right)
-\cos(2\beta)M_Z^2(-\frac{1}{2}
+\frac{2}{3}\sin^2\theta_W) \nn & &\\
\tilde{m}^2_{D_{L}}
&=&m^2_0+m^2_{1/2}\tilde{\alpha}_{GUT}
\left(\frac{8}{3}f_3(t)
+\frac{3}{2}f_2(t)
+\frac{1}{30}f_1(t)\right)
-\cos(2\beta)M_Z^2(\frac{1}{2}
-\frac{1}{3}\sin^2\theta_W)  \nn & & \\
\tilde{m}^2_{U_{R}}
&=&m^2_0+m^2_{1/2}\tilde{\alpha}_{GUT}
\left(\frac{8}{3}f_3(t)
+\frac{8}{15}f_1(t)\right)+\cos(2\beta)M_Z^2(
\frac{2}{3}\sin^2\theta_W)\\
\tilde{m}^2_{D_{R}}
&=&m^2_0+m^2_{1/2}\tilde{\alpha}_{GUT}
\left(\frac{8}{3}f_3(t)
+\frac{2}{15}f_1(t)\right)-\cos(2\beta)M_Z^2(
\frac{1}{3}\sin^2\theta_W)\\
\end{eqnarray}
For the third generation the effect of the top
Yukawa coupling needs to be taken into account, in which case 
the solution given  above are changed to\cite{bek1}:
\begin{eqnarray}
\tilde{m}^2_{b_{R}}&=&\tilde{m}^2_{D_{R}} \\
\tilde{m}^2_{b_{L}}&=&\tilde{m}^2_{D_{L}}+
\left[\frac{1}{3}(m^2_2-\mu^2-m^2_0)
-\frac{1}{2}\tilde{\alpha}_{GUT}
\left(f_2(t)+\frac{1}{5}
f_1(t)\right)m^2_{1/2}\right]  \\
\tilde{m}^2_{t_{R}}&=&\tilde{m}^2_{U_{R}}+2
\left[\frac{1}{3}(m^2_2-\mu^2-m^2_0)
-\frac{1}{2}\tilde{\alpha}_{GUT}\left(
f_2(t)+\frac{1}{5}
f_1(t)\right)m^2_{1/2}\right] +m_t^2    \\
\tilde{m}^2_{t_{L}}&=&\tilde{m}^2_{U_{L}}+
\left[\frac{1}{3}(m^2_2-\mu^2-m^2_0)
-\frac{1}{2}\tilde{\alpha}_{GUT}
\left(f_2(t)+\frac{1}{5}
f_1(t)\right)m^2_{1/2}\right]+m_t^2
\end{eqnarray}
A non-negligible Yukawa coupling causes a mixing
between the weak interaction eigenstates. 
The mass matrix is\cite{rgem}:
\bq \left(\begin{array}{cc}
\tilde{m}^2_{t_{R}}&
 -h_t(A_t\ m_0\ |H_2^0|+\mu  |H_1^0|)\\
  - h_t(A_t\ m_0\ |H_2^0|+\mu  |H_1^0|)&
\tilde{m}^2_{t_{L}}
\end{array}
   \right) \label{tlr} \eq
and the mass
eigenstates  are:
\begin{eqnarray}
\tilde{m}^2_{t_{1,2}}&=&
\frac{1}{2}\left[\tilde{m}^2_{t_{L}}+\tilde{m}^2_{t_{R}} \pm
\sqrt{(\tilde{m}^2_{t_{L}}-\tilde{m}^2_{t_{R}}
)^2+4m^2_t(A_t m_0 + \mu\cot\beta )^2}
\right]\label{mt1t2}
\end{eqnarray}
\section{Higgs Sector }
\subsection{Higgs Scalar Potential}

\vspace{1cm}

The MSSM has two Higgs doublets ($Q=T_3+Y_W/2$):
$$ H_1(1,2,-1)= \left(\begin{array}{c}H^0_1 \\
H^-_1\end{array}\right), \ \ \
 H_2(1,2,1)= \left(\begin{array}{c}H^+_2 \\
H^0_2\end{array}\right),$$

The tree level potential for the neutral sector can
be written as:
 \bq 
 V(H_1^0,H_2^0)  =  m^2_1|H_1^0|^2+m^2_2|H_2^0|^2-m^2_3(H_1^0H_2^0+h.c.)+
\frac{g^2+g^{'2}}{8}(|H_1^0|^2-|H_2^0|^2)^2 \eq 
 with the following boundary conditions at
  the GUT scale $m_1^2=m^2_2=\mu^2 +m_0^2, \ m^2_3= -B\mu m_0$, 
where the value of $\mu$
is the one at the GUT scale.

The renormalization group equations for the
mass parameters in the Higgs potential can be written
as\cite{rgem}:
\bqa
\frac{d\mu^2}{dt} & = & 3(
\tilde{\alpha}_2 +\frac{1}{5}\tilde{\alpha}_1 -Y_t)\mu^2 \\
\frac{dm^2_1}{dt} & = &
3(\tilde{\alpha}_2M^2_2 +\frac{1}{5}\tilde{\alpha}_1M^2_1)+
3(\tilde{\alpha}_2 +\frac{1}{5}\tilde{\alpha}_1 -Y_t)\mu^2\\
\frac{dm^2_2}{dt} & = &
3(\tilde{\alpha}_2M^2_2 +\frac{1}{5}\tilde{\alpha}_1M^2_1)+
3(\tilde{\alpha}_2 +\frac{1}{5}\tilde{\alpha}_1)\mu^2-3Y_t(
\tilde{m}^2_Q+\tilde{m}^2_U+m^2_2
+A^2_tm_0^2) \nn & & \\
\frac{dm^2_3}{dt} & = & \frac{3}{2}(
\tilde{\alpha}_2+\frac{1}{5}\tilde{\alpha}_1-Y_t)m^2_3+3\mu m_0Y_t A_t-
3\mu(\tilde{\alpha}_2M_2 +\frac{1}{5}\tilde{\alpha}_1M_1)
\eqa

\subsection{Solutions for the Mass Parameters in the Higgs Potential }
The solutions for the RGE given above are \cite{rgem}:
 \begin{eqnarray}
\mu^2(t)&=& {\displaystyle q(t)^2\mu^2(0)}\\
m_1^2(t)&=&m_0^2+\mu^2(t)+
m^2_{1/2}
\tilde{\alpha}_{GUT}(\frac{3}{2}f_2(t)+
\frac{3}{10}f_1(t))\label{mit1}\\
m^2_2(t)&=&q(t)^2\mu^2(0)+m^2_{1/2}e(t)+A_t(0)m_0m_{1/2}f(t)+
m_0^2(h(t)-k(t)A_t(0)^2) \label
{m12} \\
m^2_3(t)&=&q(t)m^2_3(0)+r(t)\mu(0)m_{1/2}+s(t)A_t(0)m_0\mu(0)
\label{mit2} \end{eqnarray}
where
 \begin{eqnarray*}
q(t)&=&\frac{\displaystyle 1}{\displaystyle
(1+6Y_t(0)F(t))^{1/4}}(1+\beta_2t)^{3/(2b_2)}(1+\beta_1t)^{3/(10b_1)}\\
h(t)&=&\frac{1}{2}(\frac{3}{D(t)}-1) \\
k(t)&=&\frac{3Y_t(0)F(t)}{D^2(t)} \\
f(t)&=&-\frac{6Y_t(0)H_3(t)}{D^2(t)} \\
D(t)&=&1+6Y_t(0)F(t) \\
e(t)&=&\frac{3}{2}\left[\frac{G_1(t)+Y_t(0)G_2(t)}{D(t)}+
\frac{(H_2(t)+6Y_t(0)H_4(t))^2}{3D^2(t)}+H_8(t)\right] \\
s(t)&=&\frac{3Y_t(0)F(t)}{D(t)}q(t)\\
r(t)&=&\left(\frac{3Y_t(0)H_3(t)}{D(t)}-H_7(t)\right)q(t)\\
E(t)&=&(1+\beta_3t)^{16/(3b_3)}(1+\beta_2t)^{3/b_2}(1+\beta_1t)^{13/(15b_1)}\\
F(t)&=&\int\limits_{0}^{t}E(t')dt' \\
H_2(t)&=&\tilde{\alpha}_{GUT}(\frac{16}{3}h_3(t)
+3h_2(t)+\frac{13}{15}h_1(t)) \\
H_3(t)&=& tE(t)-F(t) \\
H_4(t)&=&F(t)H_2(t)-H_3(t) \\
H_5(t)&=&\tilde{\alpha}_{GUT}(-\frac{16}{3}f_3(t)+6f_2(t)
-\frac{22}{15}f_1(t)) \\
H_6(t)&=&\int\limits_{0}^{t}H_2^2(t')E(t')dt' \\
H_7(t)&=&\tilde{\alpha}_{GUT}(3h_2(t)+\frac{3}{5}h_1(t)) \\
H_8(t)&=&\tilde{\alpha}_{GUT}(-\frac{8}{3}f_3(t)+f_2(t)
-\frac{1}{3}f_1(t)) \\
G_1(t)&=&F_2(t)-\frac{1}{3}H^2_2(t) \\
G_2(t)&=&6F_3(t)-F_4(t)-4H_2(t)H_4(t)+2F(t)H^2_2(t)-2H_6(t) \\
F_2(t)&=&\tilde{\alpha}_{GUT}(\frac{8}{3}f_3(t)+\frac{8}{15}f_1(t))\\
F_3(t)&=&F(t)F_2(t)-\int\limits_{0}^{t}E(t')F_2(t')dt' \\
F_4(t)&=&\int\limits_{0}^{t}E(t')H_5(t')dt'
\end{eqnarray*}
The functions $f_i$ and $h_i$ have been defined before.
The Higgs mass spectrum can be obtained from
the potential given above by diagonalizing the
mass matrix:
\bq
M_{ij}^2=\frac{1}{2}\frac{\partial^2 V_H}{\partial\phi_j\partial\phi_j}
\label{Mij}\eq
 where $\phi_i$ is a generic notation
for the real or imaginary part of the Higgs
field. Since the Higgs particles
are quantum field oscillations around the minimum, eq. \ref{Mij} has to 
be evaluated
at the minimum.
The mass terms at tree level haven been given
in the text. However,
as discovered a few years ago, the radiative
corrections to the Higgs mass spectrum are not small
and one has to take the corrections from a heavy
top quark into account. In this case
the effective potential for the neutral sector can be written as\cite{erz}:
\begin{eqnarray*}
  V(H_1^0,H_2^0) &=& m^2_1|H_1^0|^2+m^2_2|H_2^0|^2-m^2_3(H_1^0H_2^0+h.c.)+
\frac{g^2+g^{'2}}{8}(|H_1^0|^2-|H_2^0|^2)^2  \\
           &+&  \frac{3}{32\pi^2}\left[
\tilde{m}_{t1}^4 ( \ln \frac{\tilde{m}_{t1}^2}{Q^2}-\frac{3}{2})
+\tilde{m}_{t2}^4 ( \ln \frac{\tilde{m}_{t2}^2}{Q^2}-\frac{3}{2})
-      {m}_{t}^4 ( \ln \frac{      {m}_{t}^2}{Q^2}-\frac{3}{2})\right],
\end{eqnarray*}
where  $\tilde{m}_{ti}$ are field dependent masses, which are obtained 
from eqns. \ref{mt1t2}
 by substituting
 $       {m}_t^2=h_t^2\ H_2^2.$ 

The minimum of the potential can be found by requiring:

\begin{eqnarray}
\frac{\partial V}{\partial |H_1^0|} & = & 2m_1^2v_1-2m_3^2v_2 +
\frac{g^2+ g^{'2}}{2}(v_1^2-v_2^2)v_1 \nn
& & +\frac{3}{8\pi^2}h_t^2\mu(A_t m_0v_2+\mu v_1)
\frac{f(\tilde{m}^2_{t1})-f(\tilde{m}^2_{t2})}{\tilde{m}^2_{t1}-
\tilde{m}^2_{t2}}= 0\\
\frac{\partial V}{\partial |H_2^0|} & = & 2m_2^2v_2-2m_3^2v_1 -
\frac{g^2+ g^{'2}}{2}(v_1^2-v_2^2)v_2 \nn
& & +\frac{3}{8\pi^2}\left\{ h_t^2 A_t m_0(A_t m_0v_2+\mu v_1)
\frac{f(\tilde{m}^2_{t1})-f(\tilde{m}^2_{t2})}{\tilde{m}^2_{t1}-
\tilde{m}^2_{t2}}\right. \nn
& &\left.
+[(f(\tilde{m}^2_{t1})+f(\tilde{m}^2_{t2})-2f(m^2_t)]h_t^2v_2\right\}=0,
\end{eqnarray}
where
\bq f(m^2)=m^2(\ln\frac{m^2}{m^2_t}-1) \eq

%\vspace{1cm}
From the minimization conditions given above one obtains:
\begin{eqnarray}
  v^2&=&\frac{\displaystyle 4
       }{\displaystyle (g^2+g^{'2})(\tan^2\beta -1)}\Bigg\{
       m_1^2-m_2^2\tan^2\beta \\
  & &  -\frac{3h_t^2}{16\pi^2}\left[
[f(\tilde{m}^2_{t1})+f(\tilde{m}^2_{t2})-2f(m^2_t)]\tan^2\beta+
(A_t^2m_0^2\tan^2\beta -\mu^2)
\frac{f(\tilde{m}^2_{t1})-f(\tilde{m}^2_{t2})}{\tilde{m}^2_{t1}-
\tilde{m}^2_{t2}}\right]\Bigg\}  \nn
%
%%%%%%%%%%%%%%%%%%%%%%%%%%%%%%%%%%%
%
2m_3^2&=&(m_1^2+m_2^2)\sin 2\beta
  + \frac{3h_t^2 \sin 2\beta}{16\pi^2}\left\{
 f(\tilde{m}^2_{t1})+f(\tilde{m}^2_{t2})-2f(m^2_t)\right. \\
 & & \left. +(A_t m_0+\mu\tan\beta)
(A_t m_0 +\mu\cot\beta)
\frac{f(\tilde{m}^2_{t1})-f(\tilde{m}^2_{t2})}{\tilde{m}^2_{t1}-
\tilde{m}^2_{t2}}
\right\}  \nonumber
\end{eqnarray}

From the above equations one  can derive easily:
\begin{eqnarray}
M^2_Z&=&2\frac{\displaystyle m^2_1-m^2_2 \tan^2\beta -
\Delta^2_Z}{\tan^2\beta -1}, \\
\Delta^2_Z&=&\frac{3g^2}{32\pi^2}\frac{m^2_t}{M^2_W\cos^2\beta}
\left[
f(\tilde{m}^2_{t1})+f(\tilde{m}^2_{t2})+2m^2_t +
(A^2_t m_0^2 -\mu^2\cot^2\beta )
\frac{f(\tilde{m}^2_{t1})-f(\tilde{m}^2_{t2})}{\tilde{m}^2_{t1}-
\tilde{m}^2_{t2}}\right] \nn
\end{eqnarray}

Here all $m_i$ are evaluated at $M_Z$ using eqns \ref{mit1}-\ref{mit2}. Only the
splitting in the stop sector has been taken into account, since this splitting
depends on the large Yukawa coupling for the top quark (see the mixing matrix
(eq. \ref{tlr})). More general formulae are given in ref. \cite{loopewbr}.
The Higgs masses corresponding to this one loop potential are\cite{erz}: \\
%(J.    Ellis, G. Ridolfi and F. Zwirner, Phys. Lett. %B262(1991) 477\\
%and A. Brignole,
%J. Ellis, G. Ridolfi and F. Zwirner, Phys. Lett. %B271(1991) 123\\
\bqa
m^2_A&=&m_1^2+m_2^2+\Delta^2_A,        \\
\Delta^2_A&=& \frac{3g^2}{32\pi^2}\frac{m^2_t}{M^2_W\sin^2\beta}\left[
f(\tilde{m}^2_{t1})+f(\tilde{m}^2_{t2})+2m^2_t +(A^2_t m_0^2+\mu^2      )
\frac{f(\tilde{m}^2_{t1})-f(\tilde{m}^2_{t2})}{\tilde{m}^2_{t1}-
\tilde{m}^2_{t2}}\right]\\
m^2_{H^{\pm}}&=&m^2_A+M^2_W+\Delta^2_H, \\
\Delta^2_H&=&-\frac{3g^2}{32\pi^2}\frac{m^4_t\mu^2}{\sin^4\beta
M^2_W}\frac{h(\tilde{m}^2_{t1})-h(\tilde{m}^2_{t2})}{\tilde{m}^2_{t1}-
\tilde{m}^2_{t2}}\\
m^2_{h,H}&= &
\frac{1}{2}\left[m^2_A+M^2_Z +\Delta_{11}+\Delta_{22} \right.\nn &&\left. \pm
\sqrt{\begin{array}{ll}
(m^2_A+M_Z^2+\Delta_{11}+\Delta_{22})^2 & -4m^2_AM_Z^2\cos^22\beta
-4(\Delta_{11}\Delta_{22}-\Delta_{12}^2)\\ -4(\cos^2\beta
M^2_Z+\sin^2\beta M^2_A)\Delta_{22} & -4(\sin^2\beta M^2_Z+\cos^2\beta
M^2_A)\Delta_{11}\\ -4\sin2\beta (M^2_Z+M^2_A)\Delta_{12}& \end{array} }\right]
 \\
\Delta_{11}&=&\frac{3g^2}{16\pi^2}\frac{m^4_t}{\sin^2\beta M^2_W}
\left[\frac{\mu(A_t m_0+
\mu\cot\beta )}{\tilde{m}^2_{t1}-\tilde{m}^2_{t2}}
\right]^2d(\tilde{m}^2_{t1},\tilde{m}^2_{t2}), \\
\Delta_{22}&=&\frac{3g^2}{16\pi^2}\frac{m^4_t}{\sin^2\beta M^2_W}\left[\ln 
(\frac{\tilde{m}^2_{t1}\tilde{m}^2_{t2}}{m^4_t})+
\frac{2A_t m_0 (A_t m_0+
\mu\cot\beta )}{\tilde{m}^2_{t1}-\tilde{m}^2_{t2}}
\ln
(\frac{\tilde{m}^2_{t1}}{\tilde{m}^2_{t2}}) \right.\nn
 & & \left.+\left[
 \frac{A_t m_0(A_t m_0 +\mu\cot\beta)}
{\tilde{m}^2_{t1}-\tilde{m}^2_{t2}}\right]^2
d(\tilde{m}^2_{t1},\tilde{m}^2_{t2})\right], \\
\Delta_{12}&=&\frac{3g^2}{16\pi^2}\frac{m^4_t}{\sin^2\beta M^2_W}
\frac{\mu(A_t m_0 +
\mu\cot\beta )}{\tilde{m}^2_{t1}-\tilde{m}^2_{t2}}
\left[\ln (\frac{\tilde{m}^2_{t1}}{\tilde{m}^2_{t2}}) +
\frac{A_t m_0 (A_t m_0+\mu\cot\beta)}
{\tilde{m}^2_{t1}-\tilde{m}^2_{t2}}
d(\tilde{m}^2_{t1},\tilde{m}^2_{t2})\right], \nn
\eqa

where
% $$ f(m^2)=m^2(\ln\frac{m^2}{m^2_t}-1), \ \
$$h(m^2)=\frac{m^2}{m^2-\tilde{m}^2_{q}}\ln \frac{m^2}{\tilde{m}^2_{q}}, $$
$$ d(m^2_1,m^2_2)=2-\frac{m^2_1+m^2_2}{m^2_1-m^2_2}\ln
\frac{m^2_1}{m^2_2},$$ and
$\tilde{m}^2_q$ is the mass of a light squark.
 \section{Charginos and Neutralinos }
 The  RGE group equations for the gaugino masses  
of the $SU(3)$, $SU(2)$ and $U(1)$
groups are simple:
\bq
\frac{dM_i}{dt}=-{b_i}{\tilde{\alpha}_i}M_i
\eq
with as boundary condition at $\mgut$: $M_i(t=0)=\mha$.
The solutions are:
\bq
M_i(t)=\frac{\tilde{\alpha}_i(t)}{\tilde{\alpha}_i(0)}m_{1/2}
\eq
Since the gluinos obtain corrections from
the strong coupling constant $\alpha_3$, they grow
heavier than the gauginos of the $SU(2)$ group.
There is an additional complication to calculate  the
mass eigenstates, since both Higgsinos and gauginos
are spin 1/2 particles, so the mass eigenstates
are in general mixtures of the weak interaction eigenstates.

  The mixing of
the Higgsinos and gauginos, whose mass eigenstates
are called charginos and neutralinos for the charged
and neutral fields,
  can be parametrized by  the following Lagrangian:
 $$ {\cal L}_{Gaugino-Higgsino}=
 -\frac{1}{2}M_3\bar{\lambda}_a\lambda_a
 -\frac{1}{2}\bar{\chi}M^{(0)}\chi -(\bar{\psi}M^{(c)}\psi + h.c.)  $$
where $\lambda_a , a=1,2,\ldots ,8,$ are the Majorana gluino fields and
$$ \chi = \left(\begin{array}{c}\tilde{B} \\ \tilde{W}^3 \\
\tilde{H}^0_1 \\ \tilde{H}^0_2
\end{array}\right), \ \ \ \psi = \left( \begin{array}{c}
\tilde{W}^{+} \\ \tilde{H}^{+}
\end{array}\right),$$
are the Majorana neutralino and Dirac chargino fields,
 respectively.
Here all the terms in the Lagrangian were assembled into
matrix notation (similarly to the mass matrix
for the mixing between $B$ and $W^0$ in the SM, eq. \ref{mas}).
The mass matrices can be written as \cite{susyrev}:
\bq M^{(0)}=\left(
\begin{array}{cccc}
M_1 & 0 & -M_Z\cos\beta \sw & M_Z\sin\beta \sw \\
0 & M_2 & M_Z\cos\beta \cw   & -M_Z\sin\beta \cw  \\
-M_Z\cos\beta \sw & M_Z\cos\beta \cw  & 0 & -\mu \\
M_Z\sin\beta \sw & -M_Z\sin\beta \cw  & -\mu & 0
\end{array} \right)\eq
\bq M^{(c)}=\left(
\begin{array}{cc}
M_2 & \sqrt{2}M_W\sin\beta \\ \sqrt{2}M_W\cos\beta & \mu
\end{array} \right) \eq
The last matrix has two chargino eigenstates $\tilde{\chi}_{1,2}^{\pm}$
with mass eigenvalues
\bq M^2_{1,2}=\frac{1}{2}\left[M^2_2+\mu^2+2M^2_W \mp
\sqrt{(M^2_2-\mu^2)^2+4M^4_W\cos^22\beta +4M^2_W(M^2_2+\mu^2+2M_2\mu
\sin 2\beta )}\right]\eq
The four mass eigenstates of the neutralino
 mass matrix  are denoted by
$\tilde{\chi}_i^0(i=1,2,3,4)$ with masses
 $M_{\tilde{\chi}_1^0}\leq \cdot\cdot\cdot\leq
 M_{\tilde{\chi}_4^0}$.
 The sign of the  mass eigenvalue corresponds
to the CP quantum number of the Majorana neutralino state.

In the limiting case $M_1,M_2,\mu >>M_Z$ one can neglect the
off-diagonal elements and the mass eigenstates become:
\bq \tilde{\chi}_i^0=[\tilde{B},\tilde{W}_3,
\frac{1}{\sqrt{2}}(\tilde{H}_1-\tilde{H}_2),
\frac{1}{\sqrt{2}}(\tilde{H}_1+\tilde{H}_2)] \eq
with eigenvalues $|M_1|,|M_2|, |\mu|,$ and $|\mu|$,
respectively.
In other words, the bino and neutral wino do not
mix with each other nor with the Higgsino eigenstates
in this limiting case.
As we will see in a quantitative analysis, the data
indeed prefers $M_1,M_2,\mu > M_Z$, so the LSP is bino-like,
which has consequences for dark matter searches.

\section{RGE for the Trilinear Couplings in the
Soft Breaking Terms }
The Lagrangian for the soft breaking terms
has two free parameters $A$ and $B$
for the trilinear coupling and the
mixing between the two Higgs doublets,
respectively.

Since the $A$ parameter always occurs in conjunction
with a Yukawa coupling, we will only consider
the trilinear coupling for the third generation,
called $A_t$.
The evolutions of the $A$ and $B$ parameters are given
by the following RGE\cite{ir}:
\bqa
\frac{dA_t}{dt} & = & \left(\frac{16}{3}\tilde{\alpha}_3\frac{M_3}{m_0}
+ 3\tilde{\alpha}_2\frac{M_2}{m_0} +
\frac{13}{15}\tilde{\alpha}_1\frac{M_1}{m_0}\right)-6Y_tA_t\\
\frac{dB}{dt} & =
& 3\left(\tilde{\alpha}_2\frac{M_2}{\mze} +
\frac{1}{5}\tilde{\alpha}_1\frac{M_1}{\mze}\right)-3Y_tA_t
\eqa
The $B$ parameter can be replaced by $\tb$
through the minimization conditions of the potential.
The solution for $A_t(t)$ is:
\bqa
A_t(t)&=&\frac{\displaystyle A_t(0)}{\displaystyle 1+6Y_t(0)F(t)}
+\frac{\mha}{\mze}\left(H_2-
\frac{\displaystyle 6Y_t(0) H_3}{\displaystyle 1+6Y_t(0)F(t)}\right)
\eqa

\begin{table}[htb]
\begin{center}
$$\begin{array}{|c|c|c|c|}  \hline
Particle  & b_1 & b_2 & b_3 \\
\hline
\rule{0cm}{0.5cm}
\tilde{g} & 0 & 0 & 2  \\
\rule{0cm}{0.5cm}
\tilde{l}_l & \frac{3}{10}  & \frac{1}{2} & 0  \\
\rule{0cm}{0.5cm}
\tilde{l}_r & \frac{3}{5}  &            0& 0  \\
\rule{0cm}{0.5cm}
\tilde{w}   &            0 &  \frac{4}{3}& 0  \\
\rule{0cm}{0.5cm}
\tilde{q}-\tilde{t} & \frac{49}{60}  &            1&\frac{5}{3}  \\
\rule{0cm}{0.5cm}
\tilde{t}_l         & \frac{1}{60}  & \frac{1}{2} &\frac{1}{6}  \\
\rule{0cm}{0.5cm}
\tilde{t}_r         & \frac{4}{15}  &           0 &\frac{1}{6}  \\
\rule{0cm}{0.5cm}
\tilde{h}           & \frac{2}{5}  & \frac{2}{3} &          0  \\
\rule{0cm}{0.5cm}
           H        & \frac{1}{10}  & \frac{1}{6} &         0   \\
\rule{0cm}{0.5cm}
       t            & \frac{17}{30}  &  1        &\frac{2}{3}  \\
                   & & & \\
\hline
                   & & & \\
\rule{0cm}{0.5cm}
 Standard~Model     & \frac{41}{10}  & -\frac{19}{6} & -7    \\
\rule{0cm}{0.5cm}
 Minimal~SUSY       & \frac{33}{5}  &            1  & -3    \\
                   & & & \\
\hline
 \end{array}
$$
\end{center}
 \caption{\label{ta1} Contributions to the first order  coefficients of 
the RGE for the gauge coupling constants.
 }
\end{table}

\begin{table}[htb]
\begin{center}
$$
\begin{array}{|c|c|}
\hline
           & \\
 Particles & b_{ij}          \\
           & \\
\hline
\hline
           & \\
 \tilde{g} &
 \left(\begin{array}{rrr}
  \rule{0cm}{0.5cm}
              0&            0&    0        \\
  \rule{0cm}{0.5cm}
              0&            0&    0        \\
  \rule{0cm}{0.5cm}
             0      &      0      &    48
 \end{array}\right)   \\
          & \\
\hline
          & \\
 \tilde{w} &
 \left(\begin{array}{rrr}
  \rule{0cm}{0.5cm}
              0      &      0      &     0 \\
  \rule{0cm}{0.5cm}
              0      & \frac{64}{3}&    0        \\
  \rule{0cm}{0.5cm}
              0      &      0      &     0
 \end{array}\right) \\
          & \\
\hline
          & \\
 \tilde{q},\tilde{l} &
 \left(\begin{array}{rrr}
  \rule{0cm}{0.5cm}
  \frac{19}{15} & \frac{3}{5} & \frac{44}{15} \\
  \rule{0cm}{0.5cm}
  \frac{ 1}{5} &-\frac{7}{3} &       4       \\
  \rule{0cm}{0.5cm}
  \frac{11}{30} & \frac{3}{2} &-\frac{8}{3}
 \end{array}\right)   \\
          & \\
\hline
          & \\
 \begin{array}{l} Heavy ~Higgses \\ and ~Higgsinos
 \end{array} &

 \left(\begin{array}{rrr}
  \rule{0cm}{0.5cm}
  \frac{9}{50} & \frac{9}{10} &    0 \\
  \rule{0cm}{0.5cm}
  \frac{3}{10}    & \frac{29}{6}&    0        \\
  \rule{0cm}{0.5cm}
              0      &      0      &     0
 \end{array}\right)   \\
     &  \\
\hline
%%%%%%%%%%%%%%%%%%%%%%%%%%%%%%%%%
\end{array} $$
\end{center}
\caption{\label{ta2} Contributions to the second order coefficients of the 
RGE for the gauge couplings.  }
\end{table}

\end{document}